\documentclass[longauth]{aa}
\usepackage[varg]{txfonts}

\begin{document}

\title{Linear polarization structures in LOFAR observations of the interstellar medium in the 3C196 field} 
\titlerunning{Linear polarization structures in the 3C196 field}

\author{V. Jeli\'{c}\inst{1,2,3}\thanks{E-mail:vjelic@astro.rug.nl} \and A.~G. de Bruyn\inst{1,2}  \and V.~N. Pandey\inst{2}  
 \and M. Mevius\inst{2}  \and M. Haverkorn\inst{4,5} \and M.~A. Brentjens\inst{2} \and L.~V.~E. Koopmans\inst{1}\and S. Zaroubi\inst{1}  
 \and F.~B. Abdalla\inst{6,7} \and K.~M.~B. Asad\inst{1} \and S. Bus\inst{1} \and E. Chapman\inst{6} 
 \and B. Ciardi\inst{8} \and E.~R. Fernandez\inst{1} 
 \and A. Ghosh\inst{1} \and G. Harker\inst{6} \and I.~T. Iliev\inst{9} \and H. Jensen\inst{10} \and S. Kazemi\inst{11} 
 \and G. Mellema\inst{10} \and A.~R.~Offringa\inst{2}  \and A.~H. Patil\inst{1} \and H.~K. Vedantham\inst{1,2} \and S. Yatawatta\inst{2}}

\authorrunning{V. Jeli\'{c} et al.}

\institute{Kapteyn Astronomical Institute, University of Groningen, PO Box 800, 9700 AV Groningen, the Netherlands
  \and ASTRON - the Netherlands Institute for Radio Astronomy, PO Box 2, 7990 AA Dwingeloo, the Netherlands
  \and Ru{\dj}er Bo\v{s}kovi\'{c} Institute, Bijeni\v{c}ka cesta 54, 10000 Zagreb, Croatia
  \and Department of Astrophysics / IMAPP, Radboud University Nijmegen, PO Box 9010, 6500 GL Nijmegen, the Netherlands
  \and Leiden Observatory, Leiden University, PO Box 9513, 2300 RA Leiden, the Netherlands
  \and Department of Physics \& Astronomy, University College London, Gower Street, London WC1E 6BT, UK
  \and SKA SA, 3rd Floor, The Park, Park Road, Pinelands, 7405, South Africa
  \and Max-Planck Institute for Astrophysics, Karl-Schwarzschild-Strasse 1, D-85748 Garching bei M\"unchen, Germany
  \and Astronomy Centre, Department of Physics \& Astronomy, Pevensey II Building, University of Sussex, Brighton BN1 9QH, UK
  \and Department of Astronomy and Oskar Klein Centre, Stockholm University, AlbaNova, 10691 Stockholm, Sweden
  \and ASTRON~\&~IBM Center for Exascale technology, Oude Hoogeveensedijk 4, 7991 PD Dwingeloo, the Netherlands
}

\date{Received xxx / Accepted xxx}

\abstract{}{This study aims to characterize linear polarization structures in LOFAR observations of the interstellar medium (ISM) in the 3C196 field, one of the primary fields of the LOFAR-Epoch of Reionization  key science project.}
{We have used the high band antennas (HBA) of LOFAR to image this region and Rotation Measure (RM) synthesis to unravel the distribution of polarized structures in Faraday depth.} {The brightness temperature of the detected Galactic emission is $5-15~{\rm K}$ in polarized intensity and covers the range from -3 to +8 ${\rm rad~m^{-2}}$ in Faraday depth. The most interesting morphological feature is a strikingly straight 
filament at a Faraday depth of $+0.5~{\rm rad~m^{-2}}$ running from north to south, right through the centre of the field and parallel to the Galactic plane. There is also an interesting system of linear depolarization canals conspicuous in an image showing the peaks of Faraday spectra. We  used the Westerbork Synthesis Radio Telescope (WSRT) at 350 MHz to image the same region. For the first time, we see some common morphology in the RM cubes made at 150 and 350~{\rm MHz}. There is no indication of diffuse emission in 
total intensity in the interferometric data, in line with results at higher frequencies and previous LOFAR observations. Based on our results, 
we determined physical parameters of the ISM and  proposed a simple model that may explain the observed distribution of the intervening magneto-ionic medium.}{The mean line-of-sight magnetic field component, $B_\parallel$,  is determined to be $0.3\pm0.1~{\rm \mu G}$ and its spatial variation across the 3C196 field is $0.1~{\rm \mu G}$.  The filamentary structure is probably an ionized filament in the ISM, located somewhere within the Local Bubble. This filamentary structure shows an excess in thermal electron density ($n_e B_\parallel>6.2~{\rm cm^{-3}\mu G}$) compared to its surroundings.}
 

\maketitle 

\section{Introduction}\label{sec:intro}
The Galactic interstellar medium (ISM) is filled with cold, warm, and hot thermal gas \citep[the multi-phase medium; e.g.][]{mckee77} in a mixture of ionized, neutral, atomic, and molecular components. In addition, the ISM is permeated with non-thermal plasma (mostly relativistic protons and electrons) pervaded by a large-scale magnetic field. In general, the energy density in the various components is  approximately equal, but large deviations in energy balance occur in various situations. Observations of diffuse polarized emission are valuable in exploring both the distribution and the properties of the ISM. Magnetic fields, relativistic electrons, and thermal electrons can be constrained through the action of Faraday rotation, which  changes the morphology of the  observed emission.

Studies of Galactic polarized emission are mostly done at high radio frequencies \citep[$>1~{\rm GHz}$; for an overview see][and references therein]{reich06}.  At lower radio frequencies, these studies were performed mainly at 325 MHz with the Westerbork Synthesis Radio Telescope (WSRT). These studies revealed a large number of unusually-shaped polarized, small-scale structures, which have no counterpart in the total intensity \citep[e.g.][]{wieringa93, haverkorn03a, haverkorn03b, schnitzeler09}. They are attributed to a combination of Faraday rotation and depolarization effects due to variable Faraday depth in the ISM. In many areas, linear structures dominate the morphology and have so far defied explanation.  This is especially the case at higher Galactic latitudes where line-of-sight superposition effects would not inherently
 randomize anisotropic structures.   

Faraday rotation is proportional to the square of the wavelength. Therefore, low radio frequency observations are very sensitive to small column densities of magnetized plasma in the ISM that are difficult to detect at higher radio frequencies. The wide frequency coverage and good angular resolution of the Low-Frequency Array \citep[LOFAR;][]{haarlem13}, make LOFAR an excellent instrument for studying Galactic polarized emission \citep[][]{iacobelli13, jelic14}. In combination with the rotation measure (RM) synthesis technique \citep{brentjens05}, LOFAR observations allow us to  study the relative distribution of synchrotron-emitting and Faraday-rotating regions at an exquisite resolution of $\sim1~{\rm rad~m^{-2}}$ in Faraday depth. In addition, the high angular resolution available in LOFAR allows us to study the medium at a resolution hardly affected by beam depolarization. 

Recent observations with LOFAR  revealed diffuse polarization in several fields at high Galactic latitudes with surprisingly high brightness temperature. For example, polarized emission in the ELAIS-N1 field is $\sim4~{\rm K}$ at $150~{\rm MHz}$ \citep{jelic14}. This is much greater than  was anticipated on the basis of earlier WSRT \citep[e.g.][]{bernardi09, bernardi10, pizzo11} and Giant Metrewave Radio Telescope \citep[GMRT; e.g.][]{pen09} observations in the same frequency band. Early results from the Murchison Widefield Array  (MWA)  also indicated much higher levels of observed polarization \citep{bernardi13}. In this paper, we present LOFAR observations of diffuse polarized emission detected in the 3C196 field, one of the primary fields of the LOFAR-Epoch of Reionization key science project.

The 3C196 field is centred on the very bright  quasar 3C196, in a cold region towards the Galactic anticentre (l=171$^\circ$, b=+33$^\circ$). It has been observed  previously with the Low Frequency Front Ends (LFFE; $138-156~{\rm MHz}$) on the WSRT radio telescope \citep{bernardi10}. These data revealed only faint patchy polarized emission, restricted to Faraday depths smaller than $4~{\rm rad~m^{-2}}$. The surface brightness of this emission was close to the thermal noise on angular scales smaller than $10'$. On scales greater than $30'$, emission has an rms value of $0.7~{\rm K}$. The resolution in Faraday depth space of the WSRT data  was three times poorer than that of the LOFAR observations presented in this paper, which have three times wider frequency coverage. 

The paper is organized as follows. In Section 2 we give an overview of the observational set-up and the data reduction. The widefield images of the 3C196 field in total intensity and polarization are presented in Section 3. In Section 4 we discuss the fidelity of the data. In Section 5 we discuss properties of the detected diffuse polarized structures and give a possible theoretical interpretation, while analysis of the observed discrete polarized sources will be presented in a separate paper. We summarize and conclude in Section 6.

\section{Observations and data reduction}
The 3C196 field was observed multiple times during winter 2012/13 and 2013/14 with the LOFAR High Band Antennas (HBA). During this period we accumulated $\sim450$ hours of data. For this paper we have fully processed five syntheses with durations of 8h (2012) and 6h (2013),  as a representative sample of the observations. An overview of these five observations is given in Table~\ref{tab:obs}.

\begin{table}
\caption{An overview of observational parameters of the five LOFAR-HBA observations used in this study. We are using in our analysis only 6h, symmetric around transit, of each 8h observation.}
\centering                                      
\begin{tabular}{l l}          
\hline\hline                        
Observation ID  & Start Time [UTC] \\
\hline                                   
    L79324 & 06-Dec-2012 22:41:05  \\  
    L80273 & 12-Dec-2012 22:17:30  \\
    L80508 & 16-Dec-2012 22:01:46  \\
    L80897 & 21-Dec-2012 22:42:46  \\
    L192832 & 15-Dec-2013 23:06:40  \\
    \hline 
    \multicolumn{1}{l}{Phase centre (J2000.0)} & \multicolumn{1}{l}{$08^{\rm h}13^{\rm m}36.07^{\rm s}$,  $+48^\circ13'02.58''$}\\    
    \multicolumn{1}{l}{Frequency range} & \multicolumn{1}{l}{115 -- 189 MHz}\\ 
    \multicolumn{1}{l}{Spectral resolution} & \multicolumn{1}{l}{3.2~kHz}\\ 
    \multicolumn{1}{l}{Integration time} & \multicolumn{1}{l}{2 s}\\  
    \multicolumn{1}{l}{Observing time} & \multicolumn{1}{l}{8(6) hours  in 2012 (2013)}\\  
    \hline 
\end{tabular}\label{tab:obs}
\end{table}

The array was used in the so-called HBA DUAL INNER configuration \citep{haarlem13}, consisting of 48 core stations (CS) and 14 remote stations (RS), which were tapered to have the same size and shape as the CS. Data were taken in the frequency range from $115~{\rm MHz}$ to $189~{\rm MHz}$, divided into 380 sub-bands of width $195.3125~{\rm kHz}$. Each sub-band was further divided in 64 channels. The correlator integration time was $2~{\rm s}$. All four correlation products between pairs of orthogonal dipoles were recorded. The total time per observation was $8~{\rm h}$ in 2012 and $6~{\rm h}$ in 2013.  All observations were symmetric around transit and were taken during the nighttime. The \emph{uv} coverage was fully sampled up to baselines of about 800 wavelengths.  To compare observations taken in different observing cycles, we used 6h symmetric around transit of each 8h observation. 

The processing of the data was done on a CPU/GPU\footnote{CPU: Central Processing Unit; GPU: Graphics Processing Unit} cluster dedicated to the LOFAR-EoR project, located at the University of Groningen, the Netherlands. During the processing we used the LOFAR-EoR Diagnostic Database (LEDDB), which is used for the storage and management of the LOFAR-EoR observations  \citep{martinezrubi13}.

\subsection{Initial processing and calibration}
The first step in our initial processing is the automatic flagging of radio frequency interference (RFI) using \texttt{AOFlagger} \citep{offringa10,offringa12}.  In the frequency range from 115--177~MHz on average only  $\sim3\%$ of our data is flagged. This percentage, however, is much higher (>$40\%$) for frequencies above $177~{\rm MHz}$ where signals from  Digital Audio Broadcasting (DAB) corrupt the data in a few 1.75~MHz wide bands.  
For an overview of the LOFAR RFI-environment we refer to \citet{offringa13}. 

\begin{figure}
\centering \includegraphics[trim=0 0 0 0.075\textwidth, width=.5\textwidth]{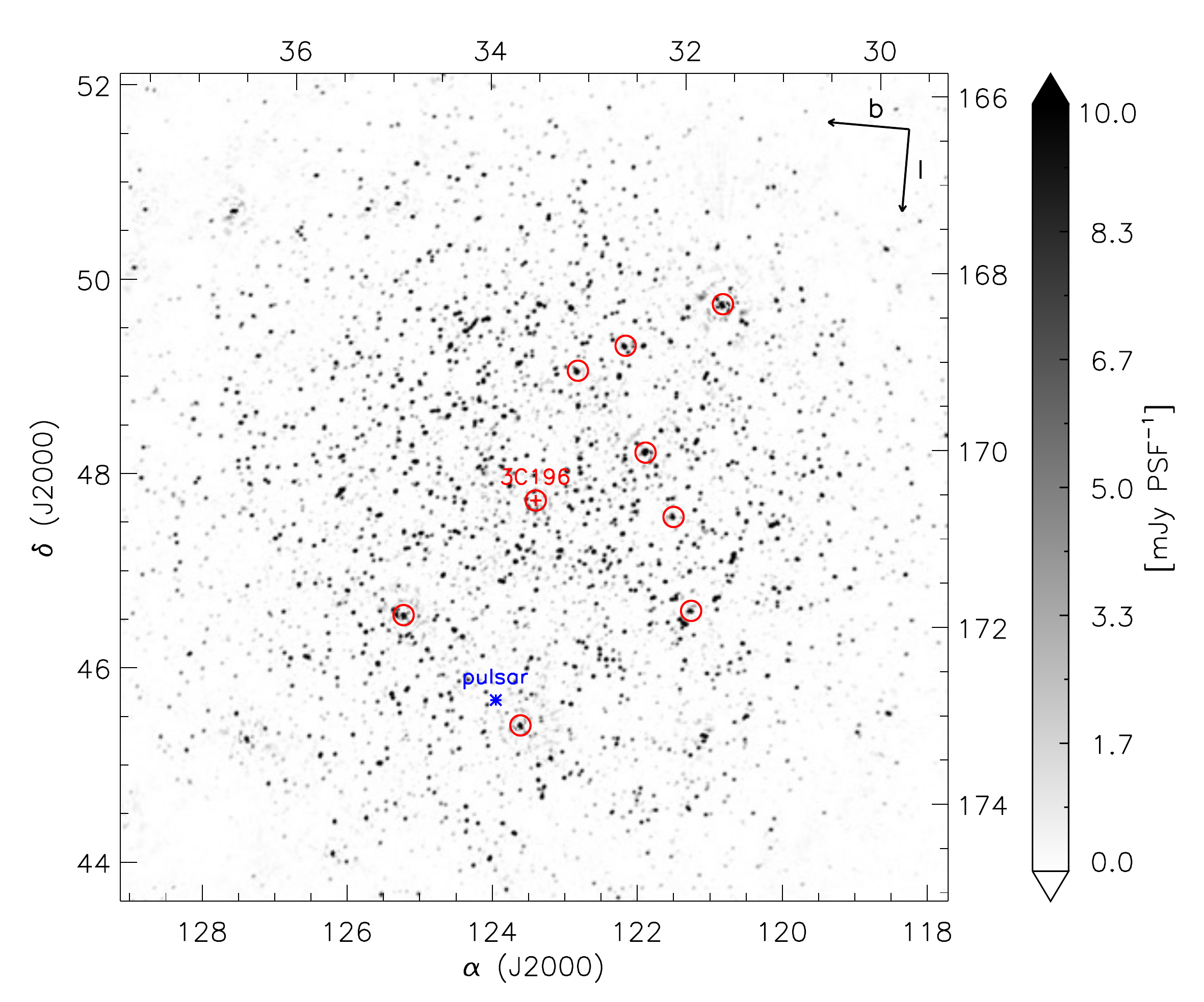}
\caption{Frequency-averaged Stokes I image of the 3C196 region, based on the four nights of data taken in December 2012.  The image is $8.3^\circ\times8.3^\circ$ in size, with a PSF of $45''$, and the noise level is $170~{\rm \mu Jy~PSF^{-1}}$. The nine bright sources towards which we performed  direction-dependent calibration are indicated with red circles. The 3C196 source is subtracted from direction-independent calibrated visibilities, while eight other sources are not. A location of the discovered pulsar J081558+461155 (see Sec.~\ref{sec:polsor}) is shown with  a blue asterisk.The image is de-convolved using the CLEAN algorithm implemented in CASA.}
\label{fig:stokesI}
\end{figure}

After flagging, the data are averaged to reduce the data volume for further processing. The bandwidth and time smearing resulting from this averaging only affects the longer baselines, which are not used in this study. To remove edge effects from the polyphase filter the 4 edge channels from the initial 64 channels are excluded in the averaging process. The resulting data set has a spectral resolution of 183~kHz (henceforth called a sub-band) and a time resolution of $10~{\rm s}$.  

The calibration proceeds in two steps. First, we perform a direction-independent calibration with the \texttt{Black Board Selfcal} (\texttt{BBS}) package \citep{pandey09}. Each sub-band is calibrated separately.  The source 3C196, which is located in the pointing and phase centre, dominates the visibilities on all baselines, making this step of the calibration straightforward.  We use a four-component model for 3C196, which accurately represents the high-resolution structure of the source (Pandey et al., in prep.). The total flux of 3C196 ($84\pm2$ Jy at 150~{\rm MHz}) is set by the broadband spectral model given by \citet{scaife12}. The BBS step removes the clock and ionospheric phase errors, and sets the frequency-dependent intensity as well as the astrometric reference frame for the field.  
The BBS solves for all four complex elements of the station Jones matrices taking the changing location of 3C196 within the dipole element beam into account .
Because the station phased array centre tracks 3C196, the varying station gain is automatically   accounted for in this  process. When applying the  
gains we also correct for the varying parallactic angle. This leads to a very stable and small polarimetric response over the inner part of station beam.  

\begin{figure}
\centering \includegraphics[trim=0 0 0 0.075\textwidth, width=.5\textwidth]{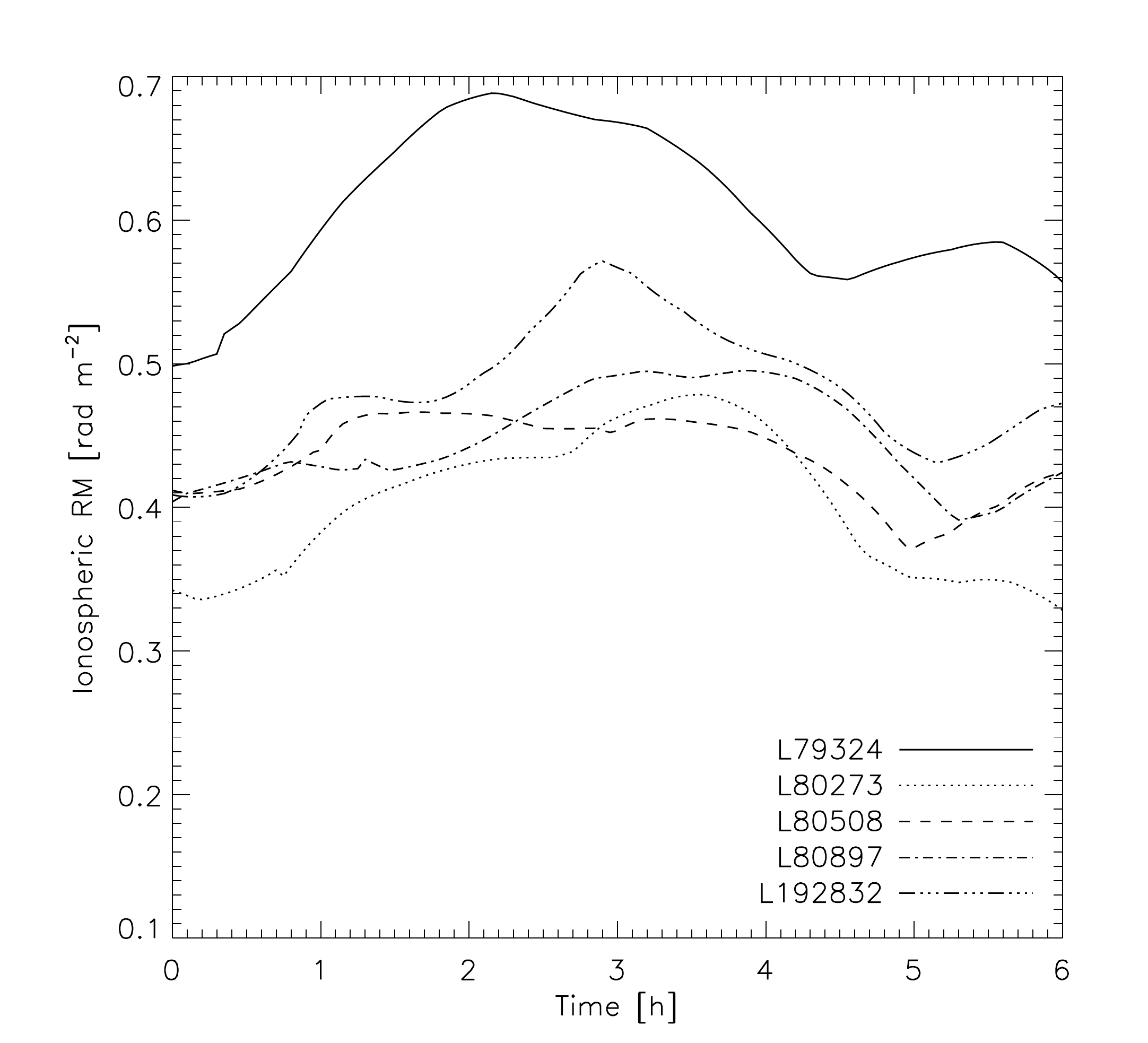}
\caption{The estimated RM variations in the ionosphere using the Global Ionospheric Maps of CODE. The RM are given as a function
of observing time and for each night.}
\label{fig:ionRM}
\end{figure}

The second step is more involved. There are at least eight additional sources with integrated apparent flux densities in excess of 2 Jy (at 150~MHz) within 2.5$^\circ$ of 3C196. To remove their time-variable polarized response, together with one of the 3C196, requires  direction-dependent calibration. This is done using \texttt{SAGECal} \citep{yatawatta08, kazemi11, kazemi13a, kazemi13b, kazemi13c} in nine directions. This corrects for time- and frequency-dependent errors due to the station beam and the slight position-dependent  ionospheric delays. Each direction is associated with one source. The solution interval is 20 minutes for the eight bright sources and 2 minutes for 3C196. This interval is sufficient to remove these sources down to the confusion noise on the short core baselines.  During \texttt{SAGECal} we are only using baselines $> 800$ wavelengths to minimize suppression of the ubiquitous extended diffuse (polarized) emission, which  is not included in the sky model and mostly appears at baselines $< 800$ wavelengths. All nine sources are unpolarized down to a fraction of a percent. These nine sources (3C196, and eight bright, discrete sources; see Fig.~\ref{fig:stokesI})  were subsequently subtracted from the residual visibilities with the directional gains derived for their directions. 

\subsection{Faraday rotation in the Earth's ionosphere}
Ionospheric Faraday rotation is a time- and direction-dependent propagation effect, whose size is proportional to the total electron content (TEC) of plasma in the ionosphere. If variability in Faraday rotation happens on a timescale that is smaller than the total integration time of an observation, the observed polarized emission is partially decorrelated. To correct for this effect we predict the RM variations in the ionosphere using the Global Ionospheric Maps of the Centre for Orbit Determination in Europe (CODE\footnote{http://aiuws.unibe.ch/ionosphere}) and the World Magnetic Model (WMM\footnote{http://www.ngdc.noaa.gov/geomag/WMM}). A full description of this procedure is given in \citet[][sec. 3.2]{jelic14}. 

Figure~\ref{fig:ionRM} shows the estimated RM variations in the ionosphere that we apply to the data. The curves are given for different nights as a function of observing time. The RM variations are the largest during the L79324 observation. This observation also shows the largest absolute RM values. We emphasize that at $150~{\rm MHz,}$ a RM variation of $0.2~{\rm rad~m^{-2}}$ rotates the plane of polarization by $45^\circ$.

\subsection{Imaging and RM synthesis}\label{sec:imgandRMsyn}
If not stated otherwise, images we present are produced using the \texttt{excon} imager \citep[][http://exconimager.sf.net/]{yatawatta14}. To analyse the polarization of diffuse emission, which mostly appears on spatial scales greater than a few arcmin, we make lower resolution images in all Stokes parameters (IQUV). These images are produced using baselines between 10 and 800 wavelengths, providing a frequency independent resolution of about 3 arcmin. We use robust (Briggs) weighting with robustness parameter equal to 0. 

\begin{figure*}[!phtb]
\centering \includegraphics[width=.45\textwidth]{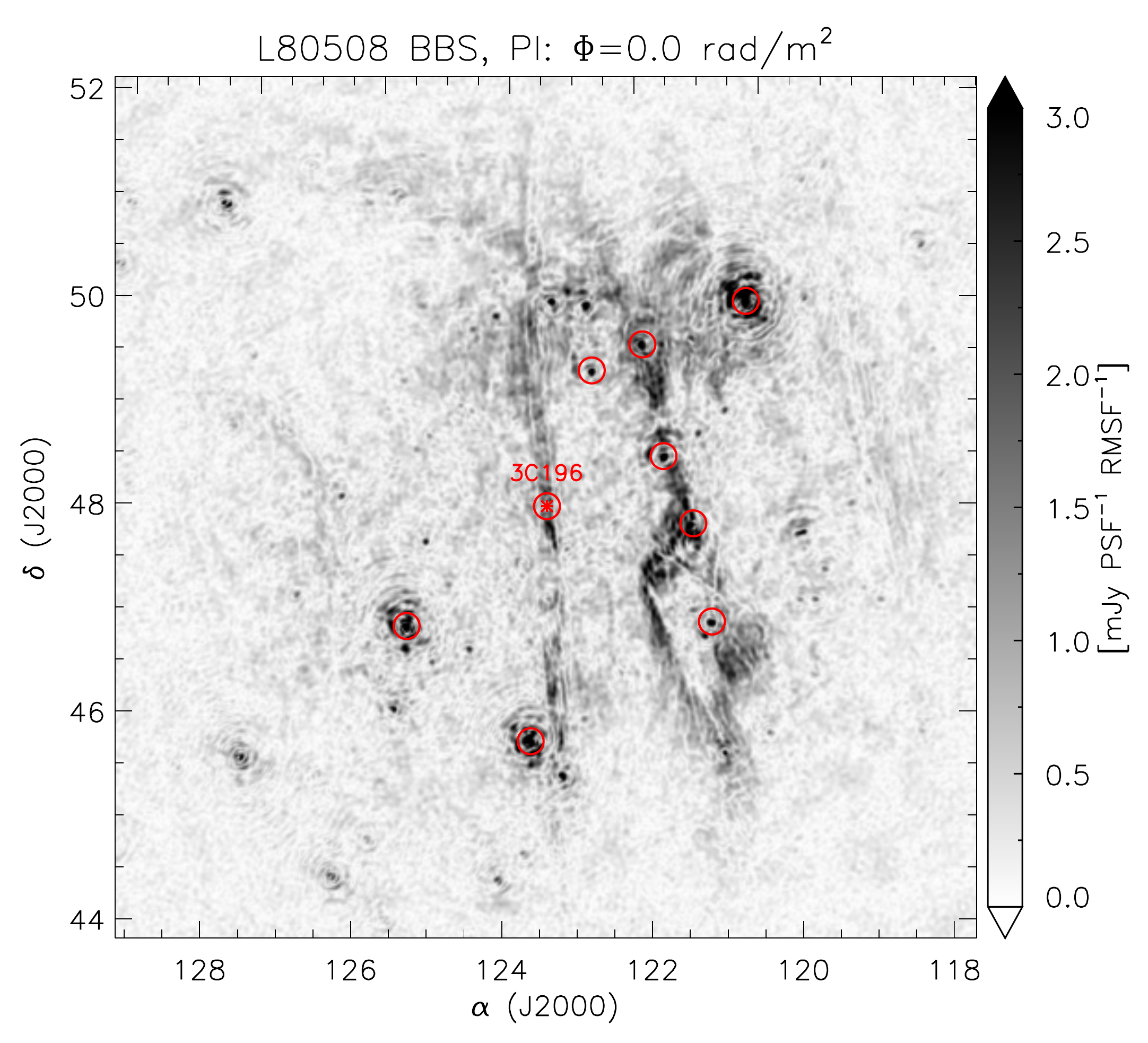}
\centering \includegraphics[width=.45\textwidth]{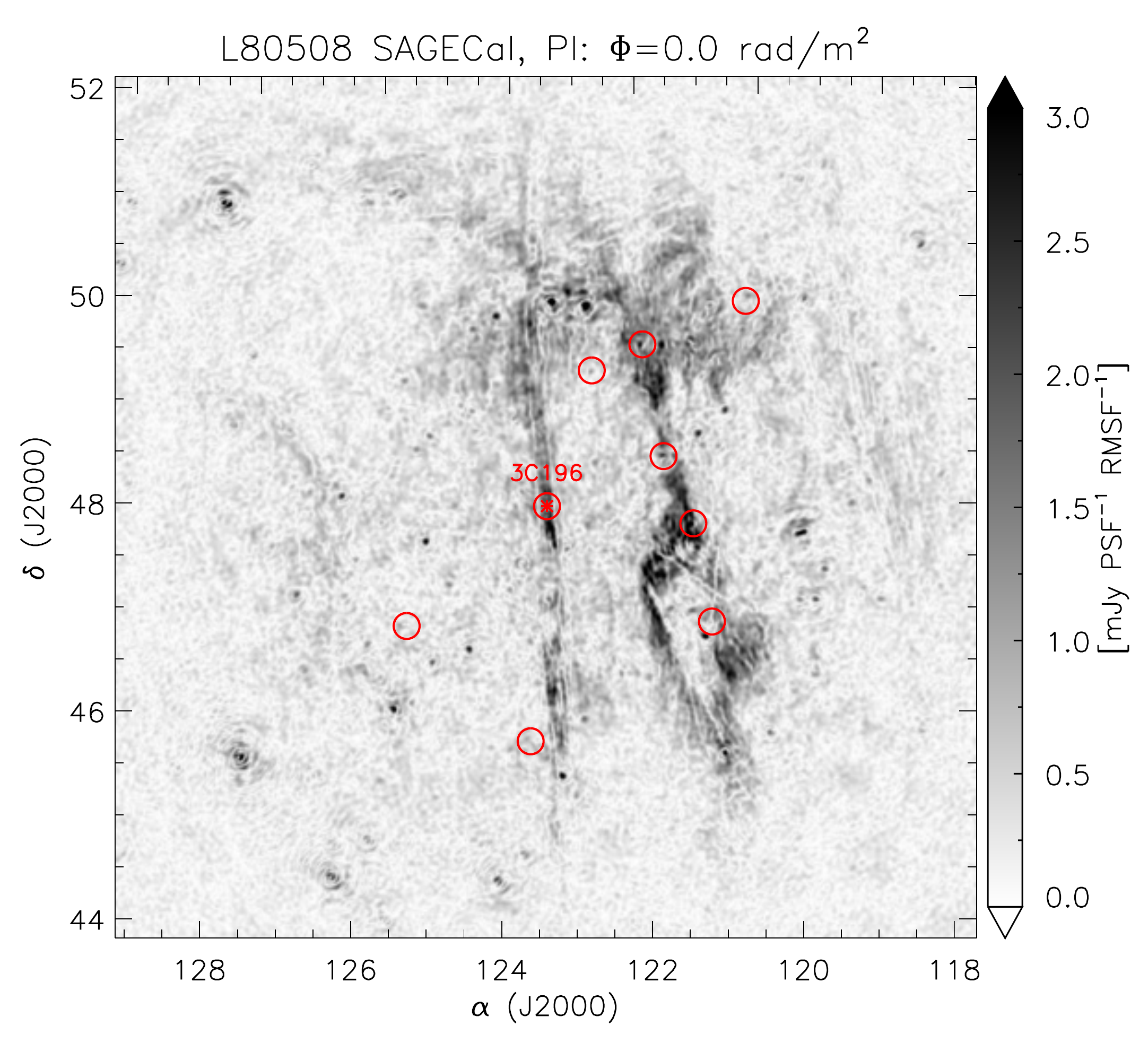}
\caption{Two images of the 3C196 region in polarized intensity at Faraday depth $0~{\rm rad~m^{-2}}$ after performing direction-independent calibration (image on the left) and direction-dependent calibration and source subtraction (image on the right). Images are $8.3^\circ\times8.3^\circ$ in size, with a PSF of $3.6'\times3.9'$. The 3C196 source is also subtracted from direction-independent calibrated visibilities. Eight bright sources, used for direction-dependent calibration and source subtraction, are indicated with red circles. The images have not been corrected for polarization noise bias nor for Faraday rotation in the Earth's ionosphere.}
\label{fig:sagecal}
\end{figure*}

To study the linearly polarized emission as a function of Faraday depth ($\Phi$), we apply RM synthesis \citep{brentjens05} to the Stokes Q,U images of 310 sub-bands, which have comparable noise levels. These sub-bands cover the frequencies between $115~{\rm MHz}$ and $175~{\rm MHz}$.  We first synthesize a cube over a wide range in Faraday depth to determine where polarized emission could be detected. The final cube covers a Faraday depth range from $-25~{\rm rad~m^{-2}}$ to $+25~{\rm rad~m^{-2}}$ in $0.25~{\rm rad~m^{-2}}$ steps. The resolution in Faraday depth space, defined by a width of a RMSF (rotation measure spread function), is $\delta\Phi=0.9~{\rm rad~m^{-2}}$, while the largest Faraday structure that can be resolved is $\Delta\Phi=1.1~{\rm rad~m^{-2}}$. Since the resolution is comparable to the maximum detectable scale, we can only detect Faraday thin structures \citep[$\lambda^2\Delta\Phi\ll 1$][]{brentjens05}.

To determine the brightness temperature of emission and its angular power spectrum, we correct the RM cube for the relative variation of the element beam pattern across the field of view and for the primary beam. These corrections are applied using the \texttt{AWimager} \citep{tasse13}.  We have not attempted to deconvolve the RM cube for the effects of the side lobes of the RMSF. The sidelobes are small and the S/N is generally so low that this would have had very little effect on the images. 
 
\section{Observational results}
\subsection{Direction-dependent calibration: instrumentally polarized sources}
The beam pattern of a LOFAR-HBA station can be described as the product of an element beam pattern and the array beam pattern of the station \citep{haarlem13}. The element beam pattern is strongly polarized. Its polarization response is related to the projection of two beam patterns of two orthogonal dipoles on the sky and the changing parallactic angle. As a result of this, spurious polarization is produced.  We correct the data for the beam pattern during calibration to first order, using BBS. The data are corrected for both the array and the element beam gain at the centre of the image. However, the relative variation of the element beam pattern across the field of view, as well as the temporal changes, are still be present in the data \cite[for more details, see][]{asad15}. The direction-dependent calibration and source subtraction can correct for these effects to some extent. 

The results of direction-dependent calibration and source subtraction are illustrated in Fig.~\ref{fig:sagecal}. We show images in polarized intensity before and after performing direction-dependent calibration and source subtraction. Images are shown at a Faraday depth of $0~{\rm rad~m^{-2}}$. The instrumental polarization leakage, after calibrating on 3C196, is to first order independent of frequency, and therefore all instrumentally polarized signals are located around a Faraday depth of $0~{\rm rad~m^{-2}}$. 

All sources included in the sky model are removed successfully. Their side lobes are significantly suppressed and they are not visible in the images. The morphology of diffuse emission is preserved.  A flux scale of diffuse emission agrees on average over all directions within $5-10\%$ with the data calibrated using only BBS (direction-independent calibration).

\subsection{Diffuse polarized emission}\label{sec:diffpe}
A series of widefield images of the 3C196 field in both polarized intensity and Stokes Q,U are presented in Fig.~\ref{fig:PI1}.
The images are $8.3^\circ\times8.3^\circ$ in size with a PSF (point spread function) of $3.6'\times3.9'$.  The noise level is $71~{\rm \mu Jy~PSF^{-1}~RMSF^{-1}}$ in polarized
intensity and $108~{\rm \mu Jy~PSF^{-1}~RMSF^{-1}}$ in Stokes Q,U. The images are given at Faraday depths of -1.5, 0, +1, +1.5, +2, +2.5,  +3.0, and $+3.5~{\rm rad~m^{-2}}$ to emphasize the various detected structures in linear polarization.  

\begin{figure*}[!phtb]
\centering \includegraphics[width=.33\textwidth]{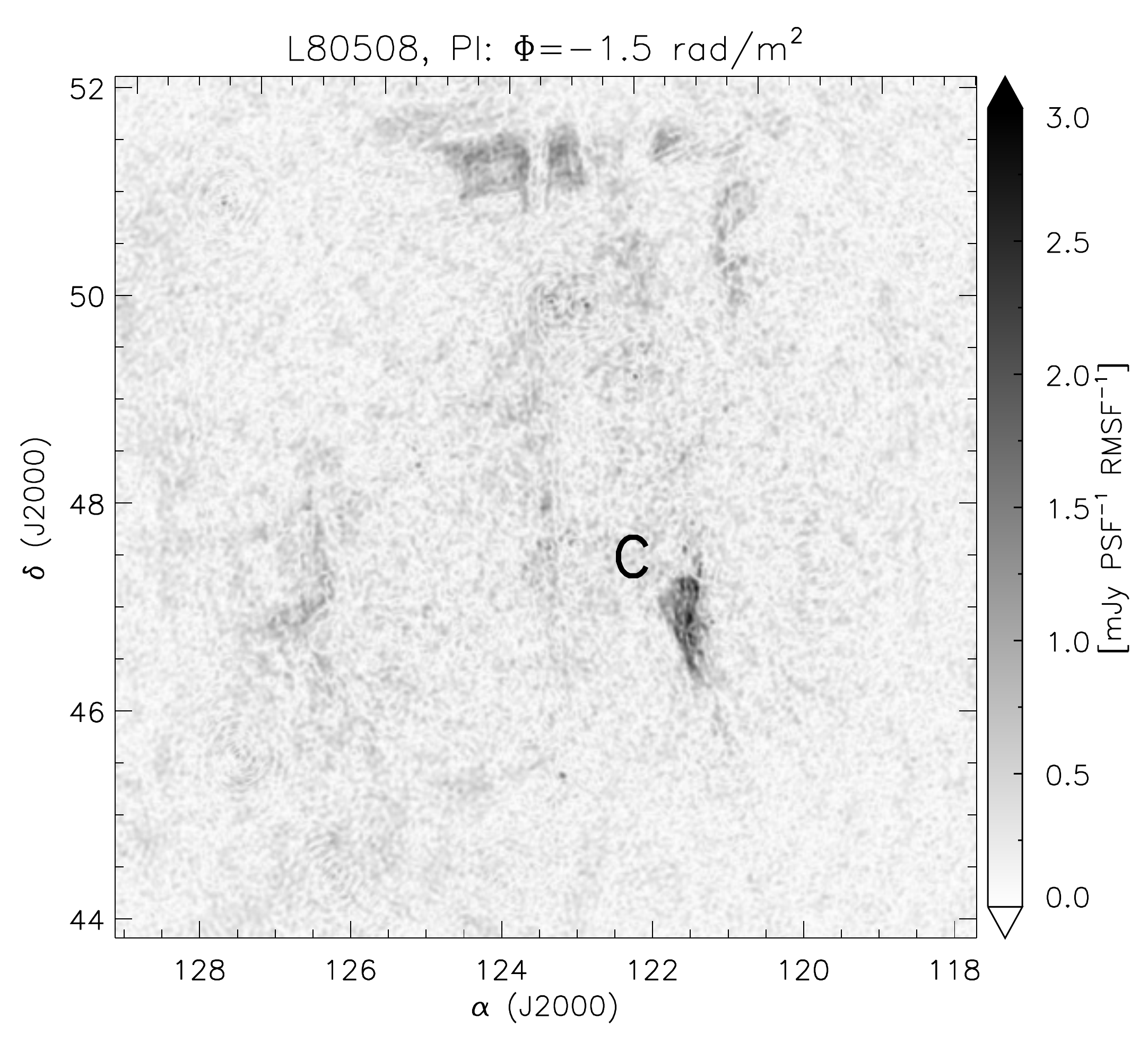}
\centering \includegraphics[width=.33\textwidth]{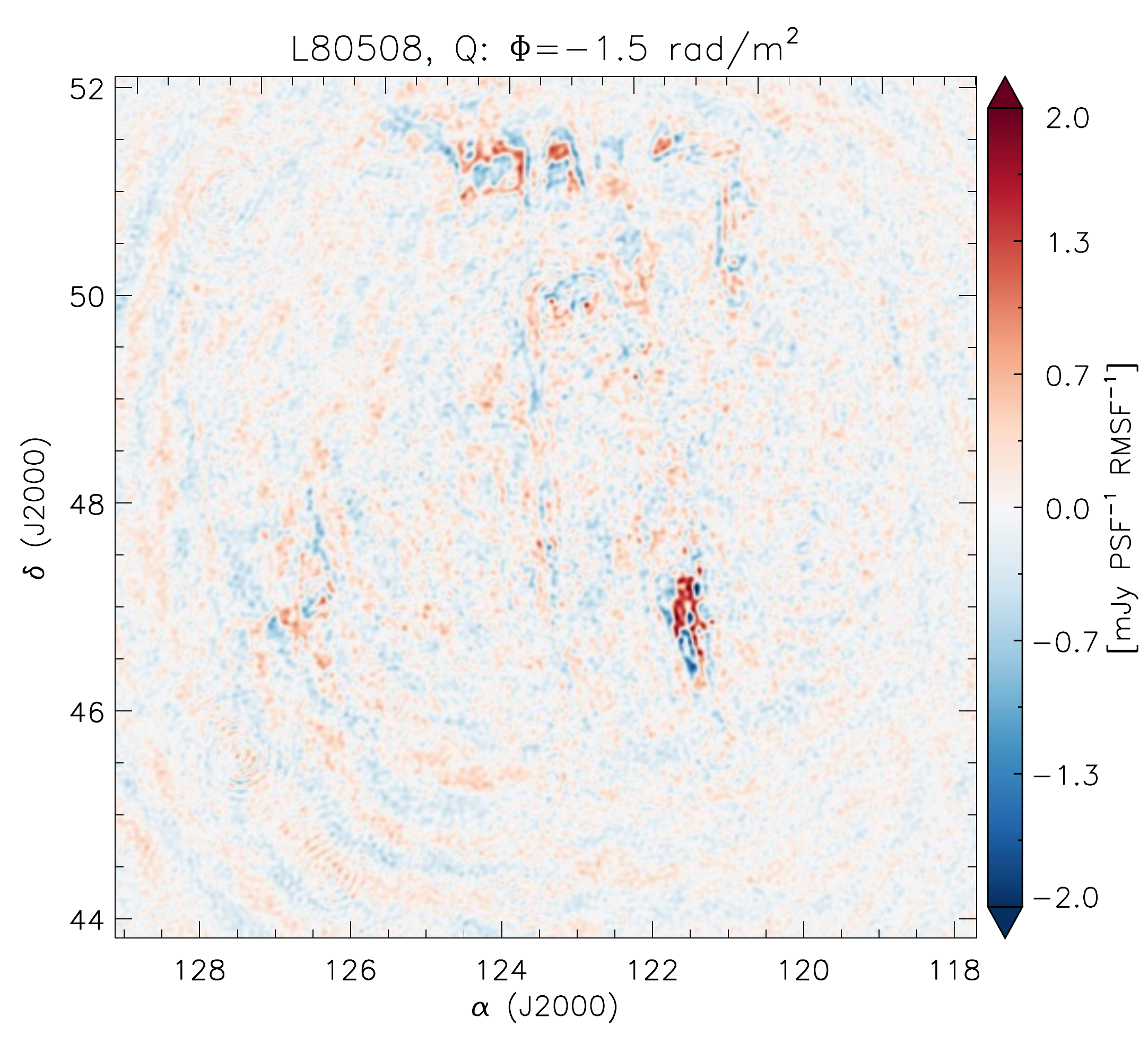}
\centering \includegraphics[width=.33\textwidth]{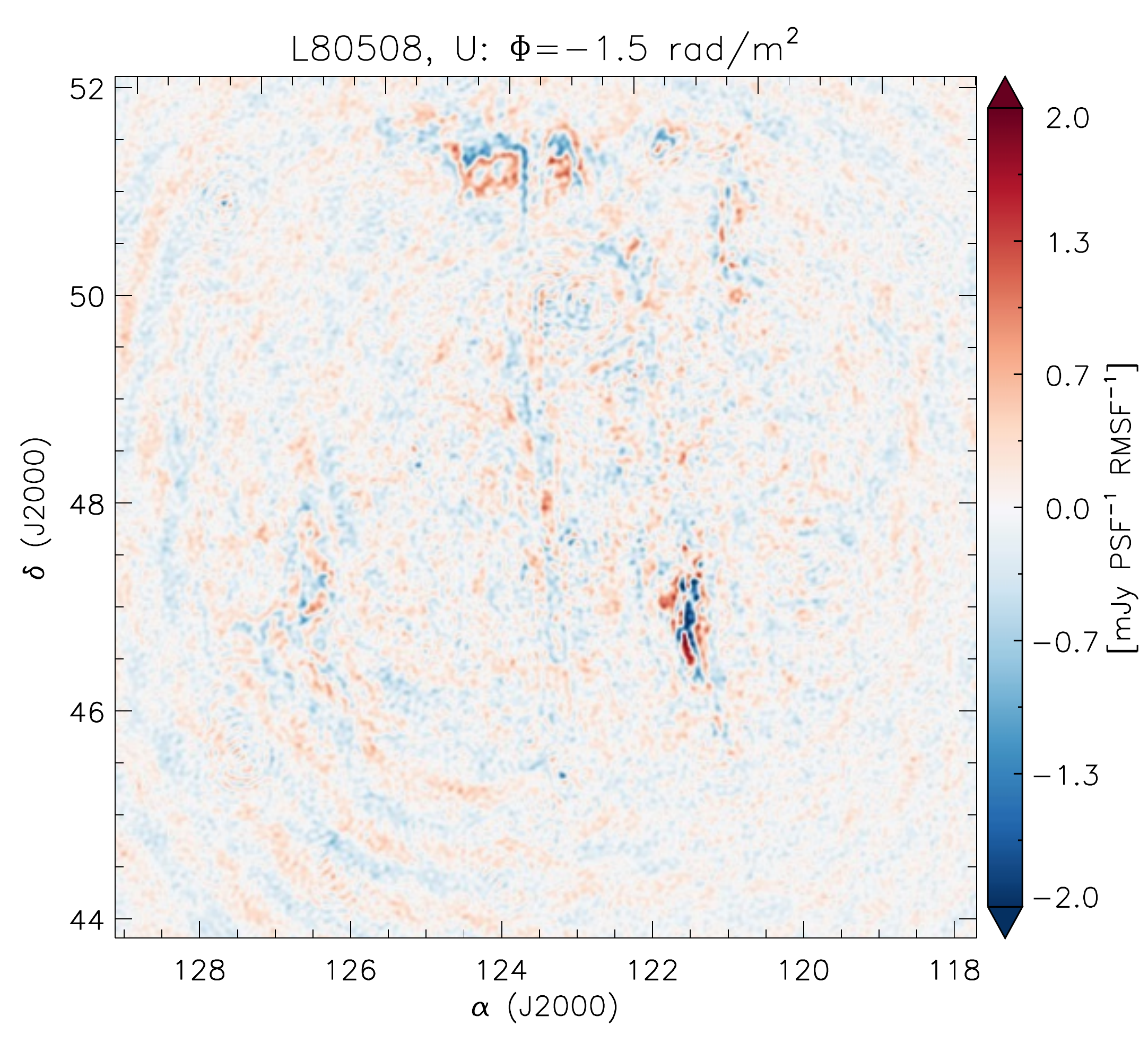}
\centering \includegraphics[width=.33\textwidth]{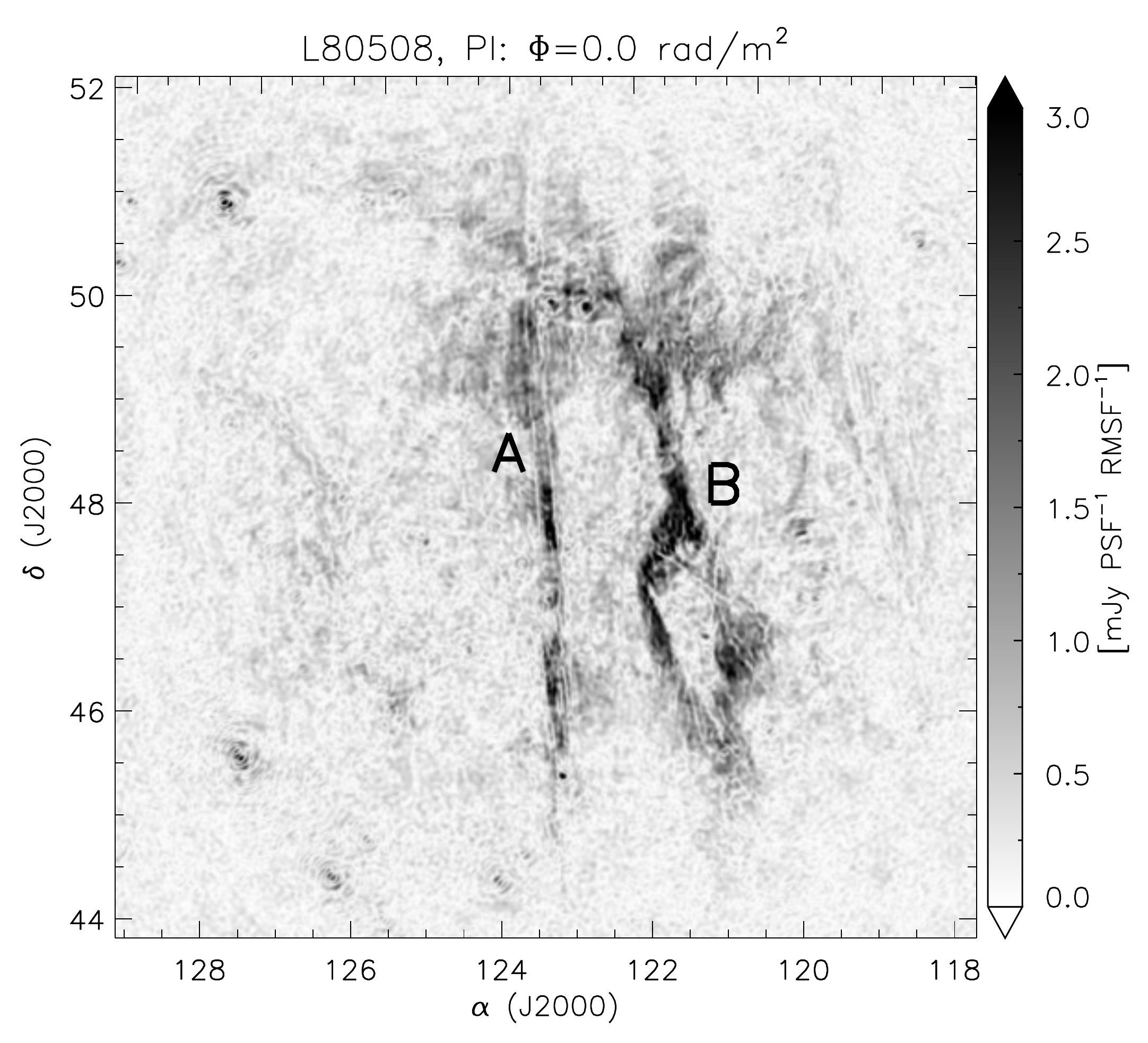}
\centering \includegraphics[width=.33\textwidth]{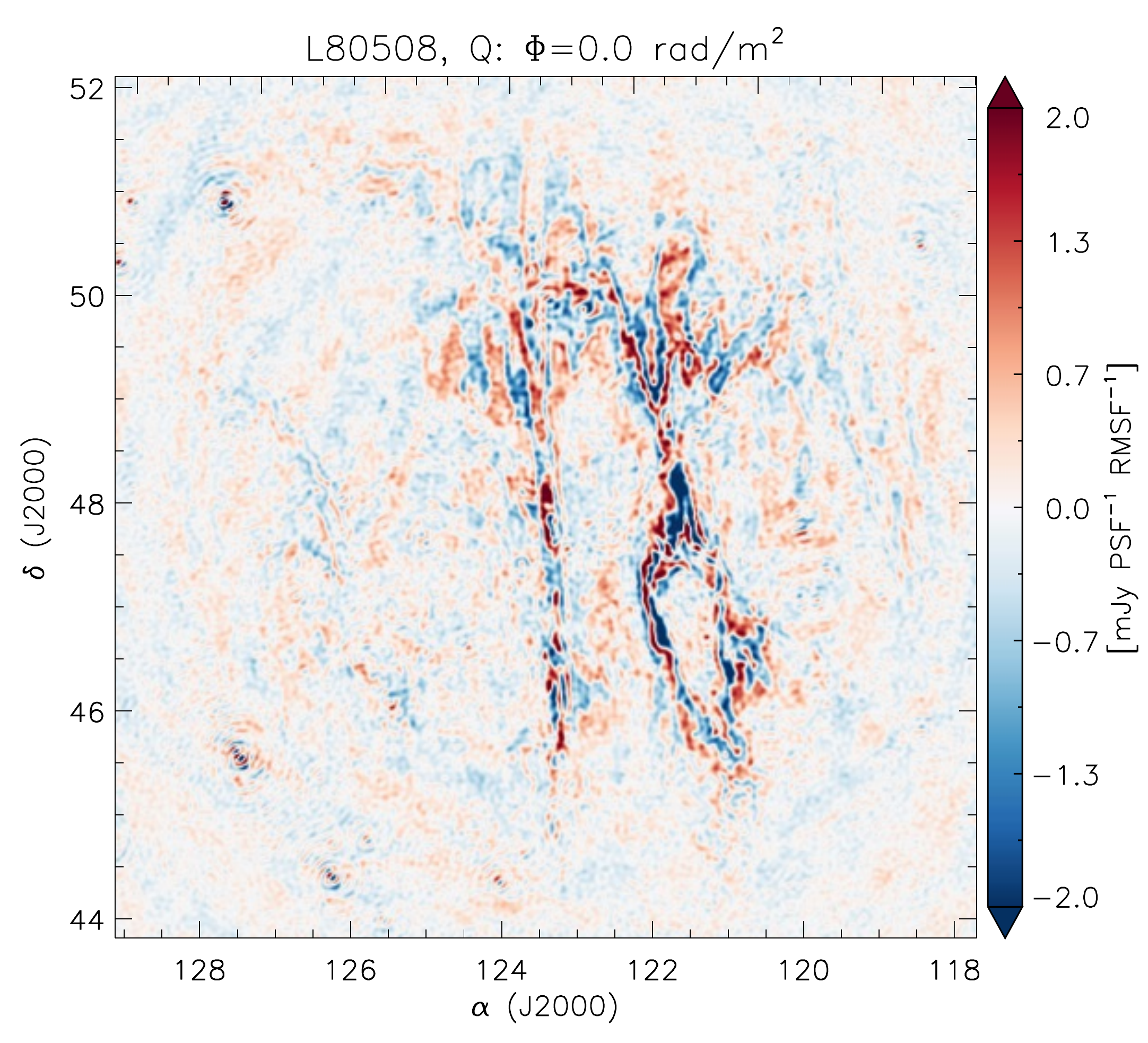}
\centering \includegraphics[width=.33\textwidth]{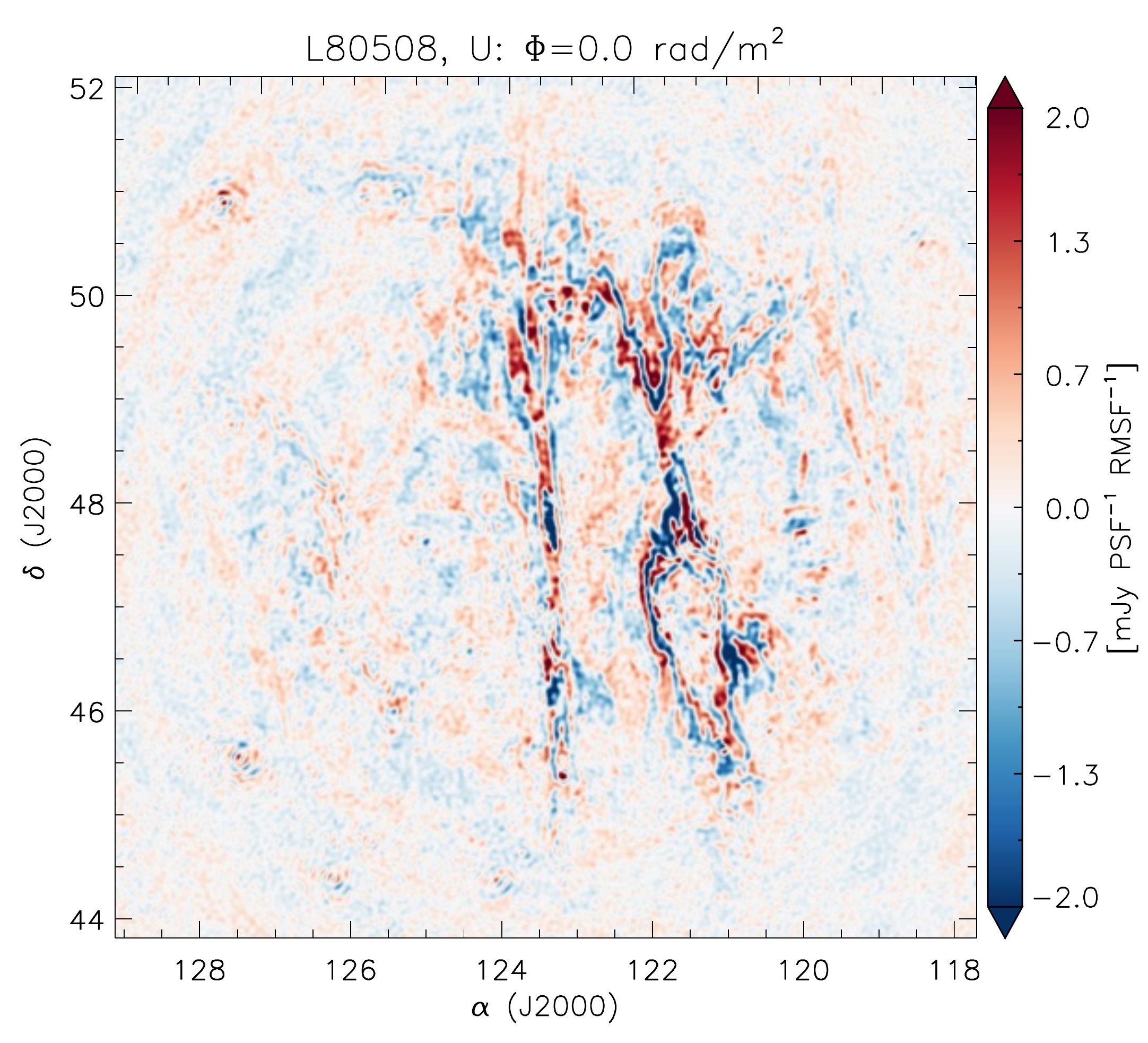}
\centering \includegraphics[width=.33\textwidth]{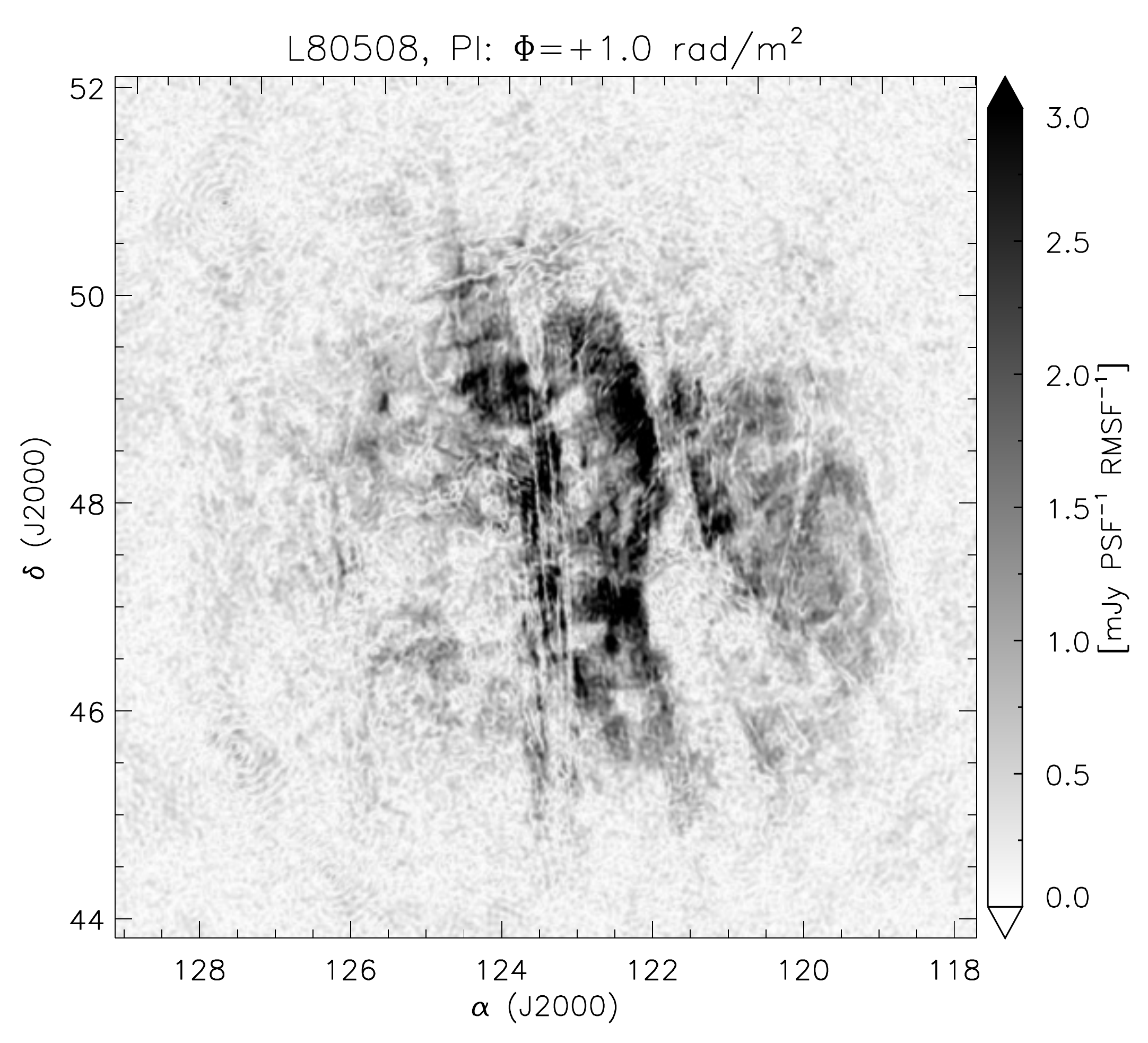}
\centering \includegraphics[width=.33\textwidth]{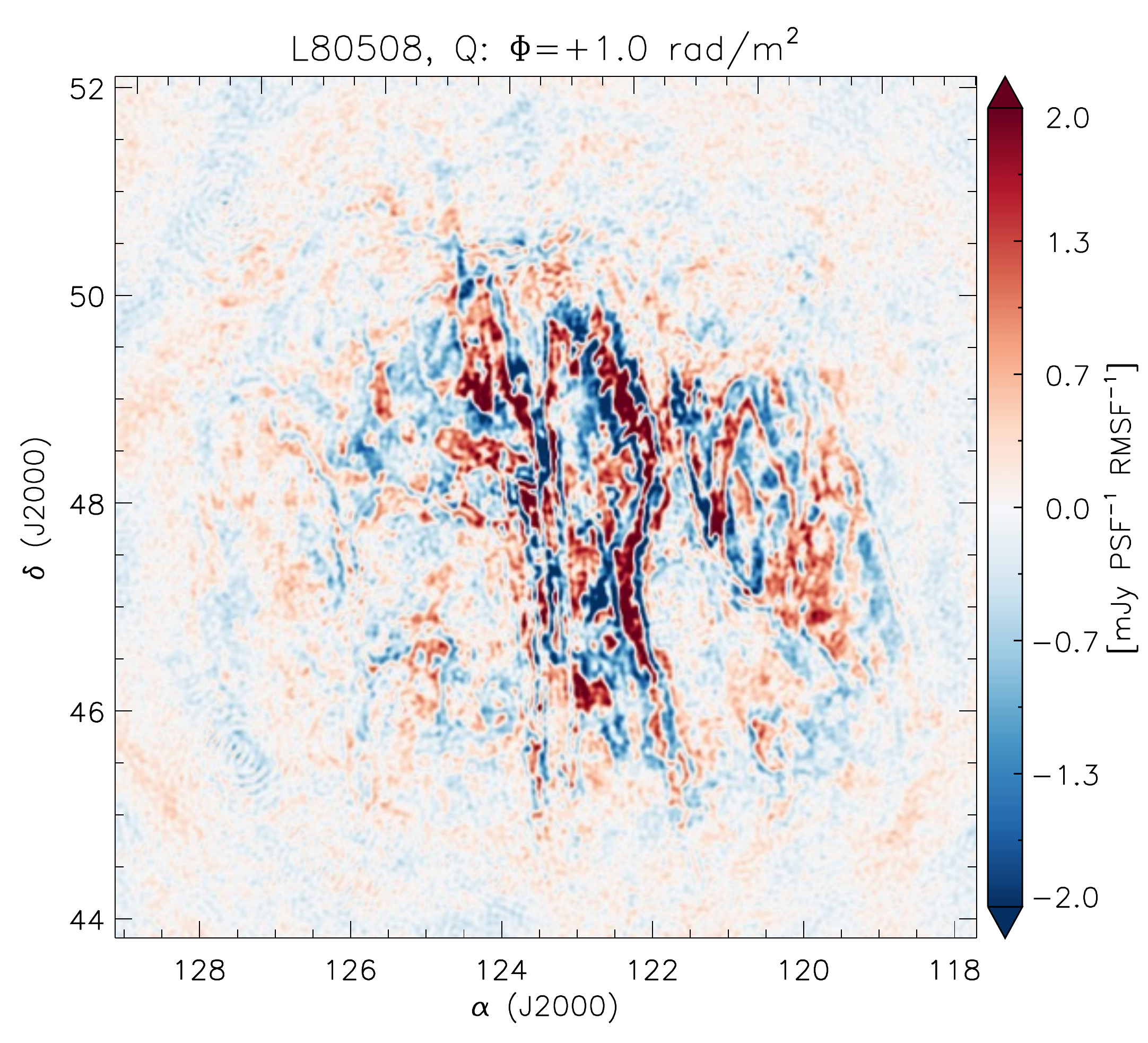}
\centering \includegraphics[width=.33\textwidth]{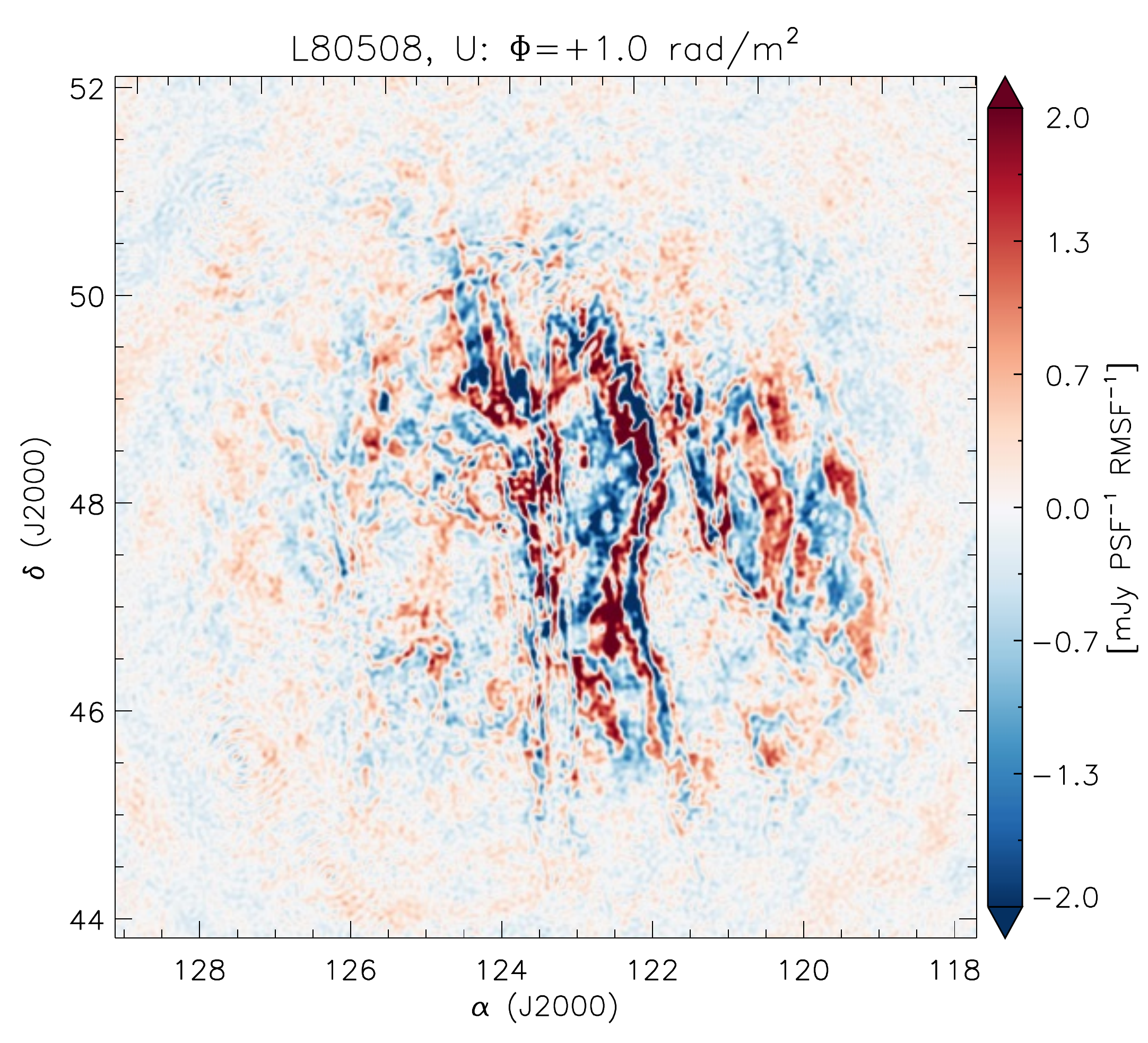}
\centering \includegraphics[width=.33\textwidth]{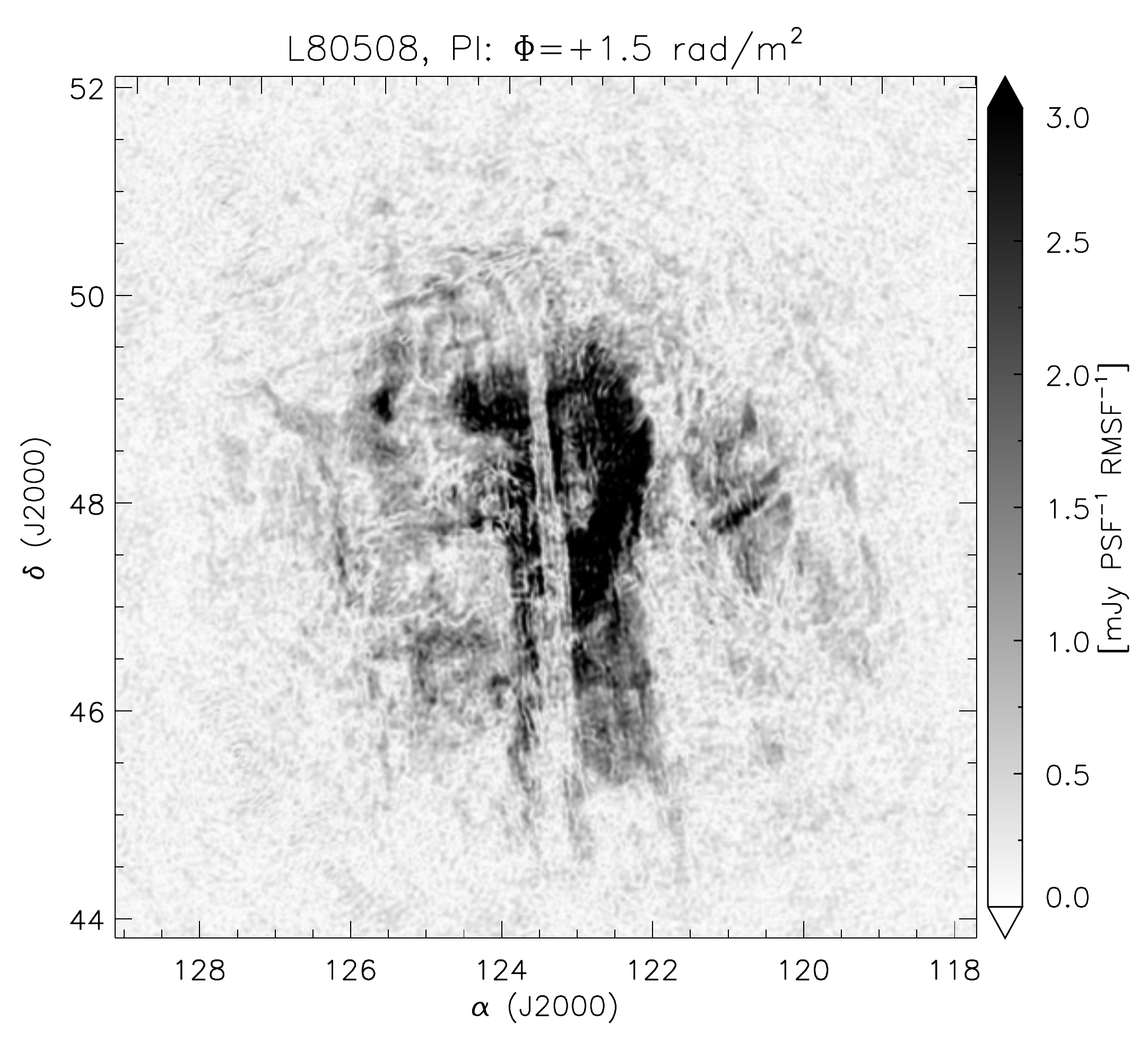}
\centering \includegraphics[width=.33\textwidth]{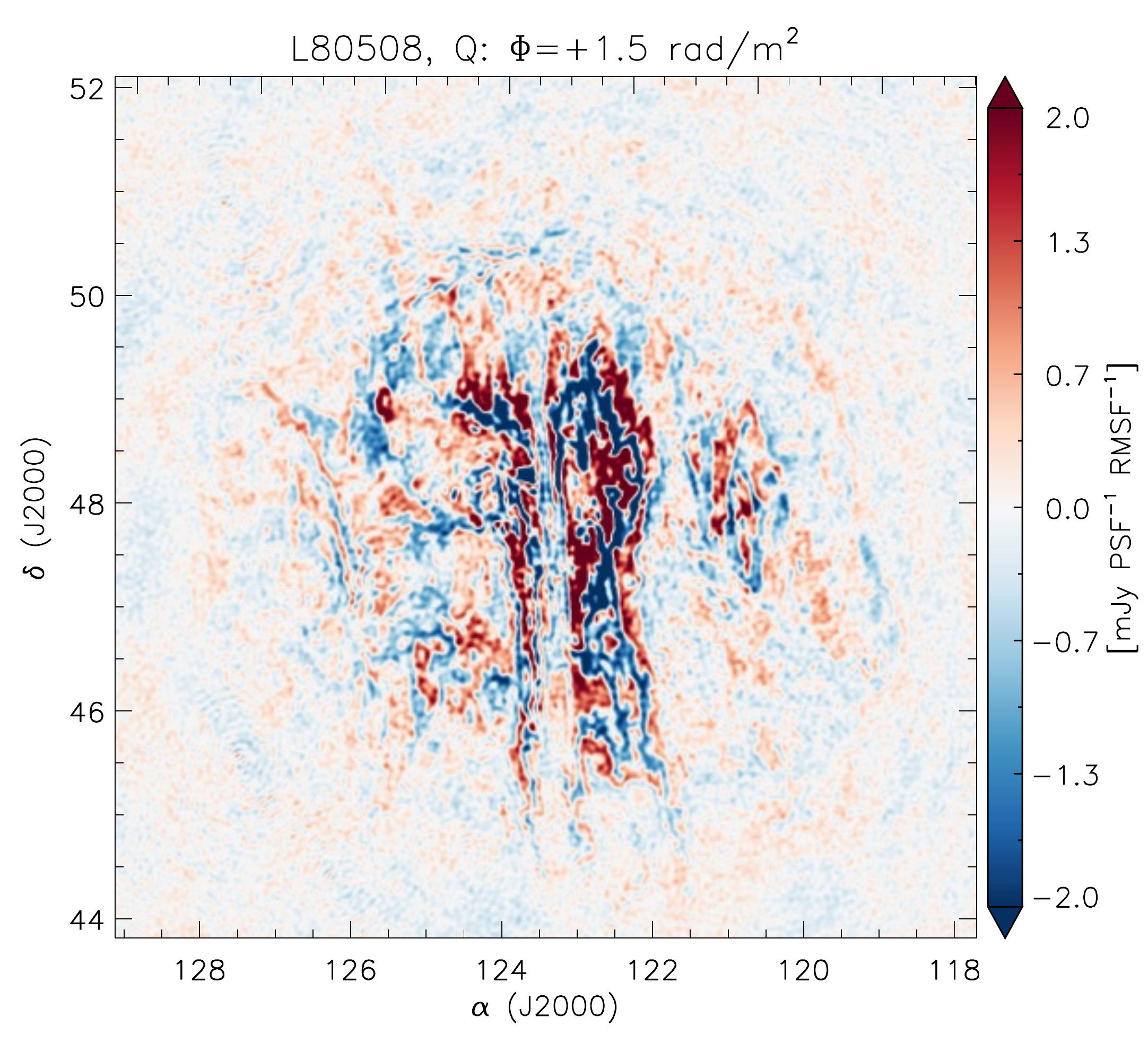}
\centering \includegraphics[width=.33\textwidth]{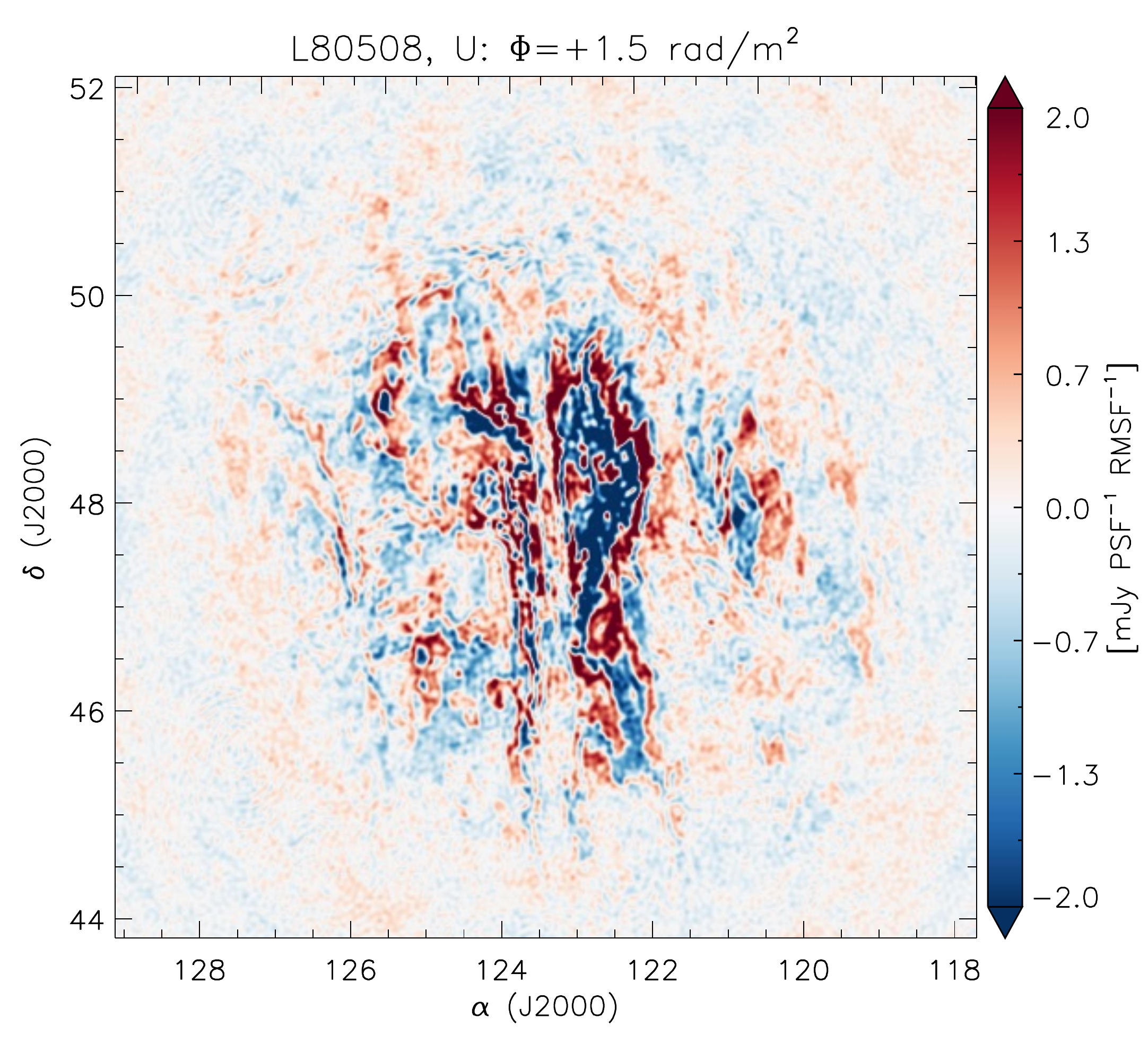}
\caption{Widefield images of the 3C196 field in polarized intensity (PI), Stokes Q, and Stokes U at Faraday depths
of -1.5, -0.0, +1.0, +1.5, +2.0, +2.5, +3.0, and +3.5~${\rm rad~m^{-2}}$. Images are $8.3^\circ\times8.3^\circ$ in size with a PSF of $3.6'\times3.9'$. 
The noise level is $71~{\rm \mu Jy~PSF^{-1}~RMSF^{-1}}$ in polarized intensity and $108~{\rm \mu Jy~PSF^{-1}~RMSF^{-1}}$ in Stokes Q,U.
The images in polarized intensity have not been corrected for polarization noise bias nor for the primary beam.}
\label{fig:PI1}
\end{figure*}

\addtocounter{figure}{-1}
\begin{figure*}[!phtb]
\centering \includegraphics[width=.33\textwidth]{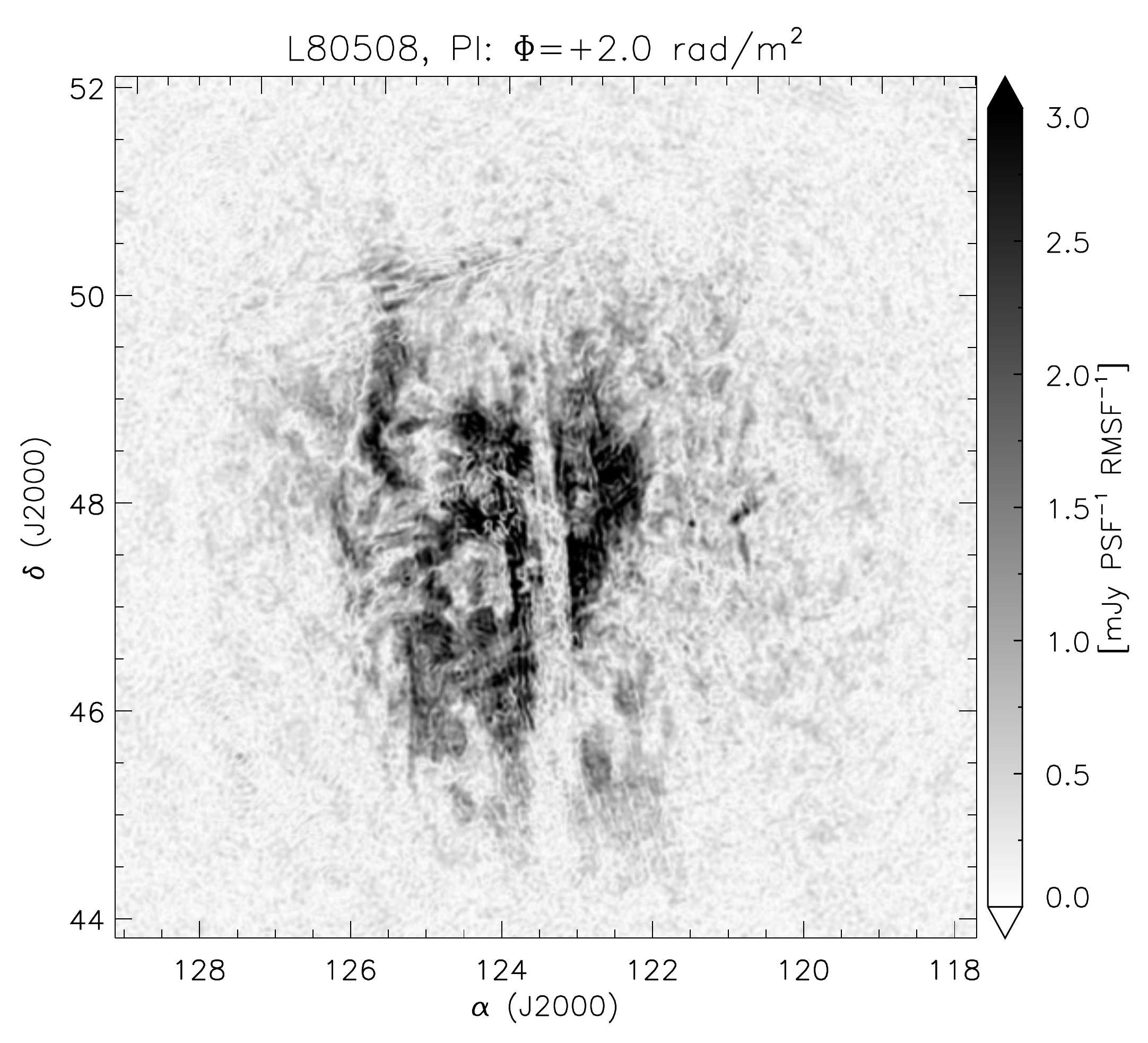}
\centering \includegraphics[width=.33\textwidth]{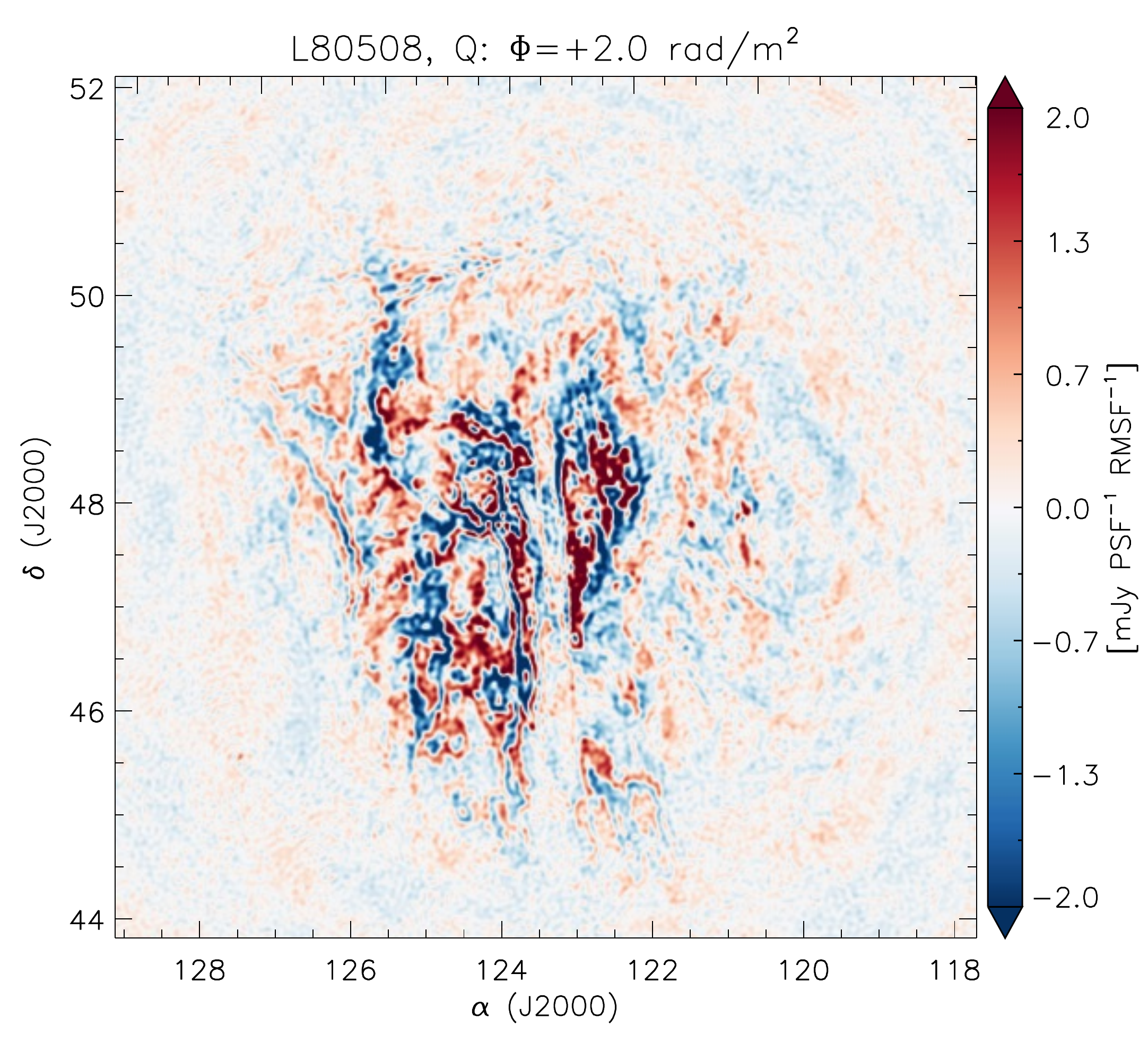}
\centering \includegraphics[width=.33\textwidth]{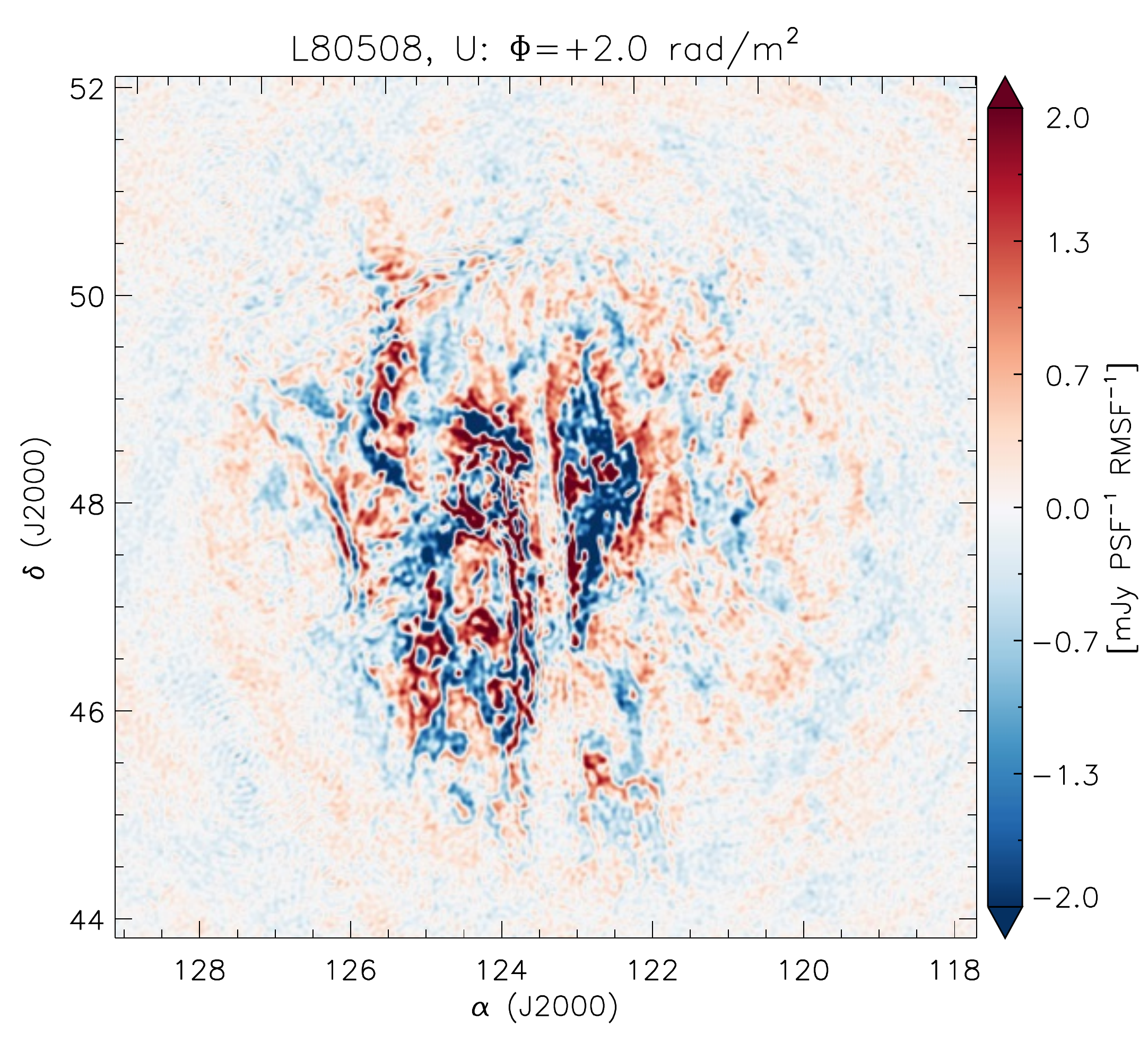}
\centering \includegraphics[width=.33\textwidth]{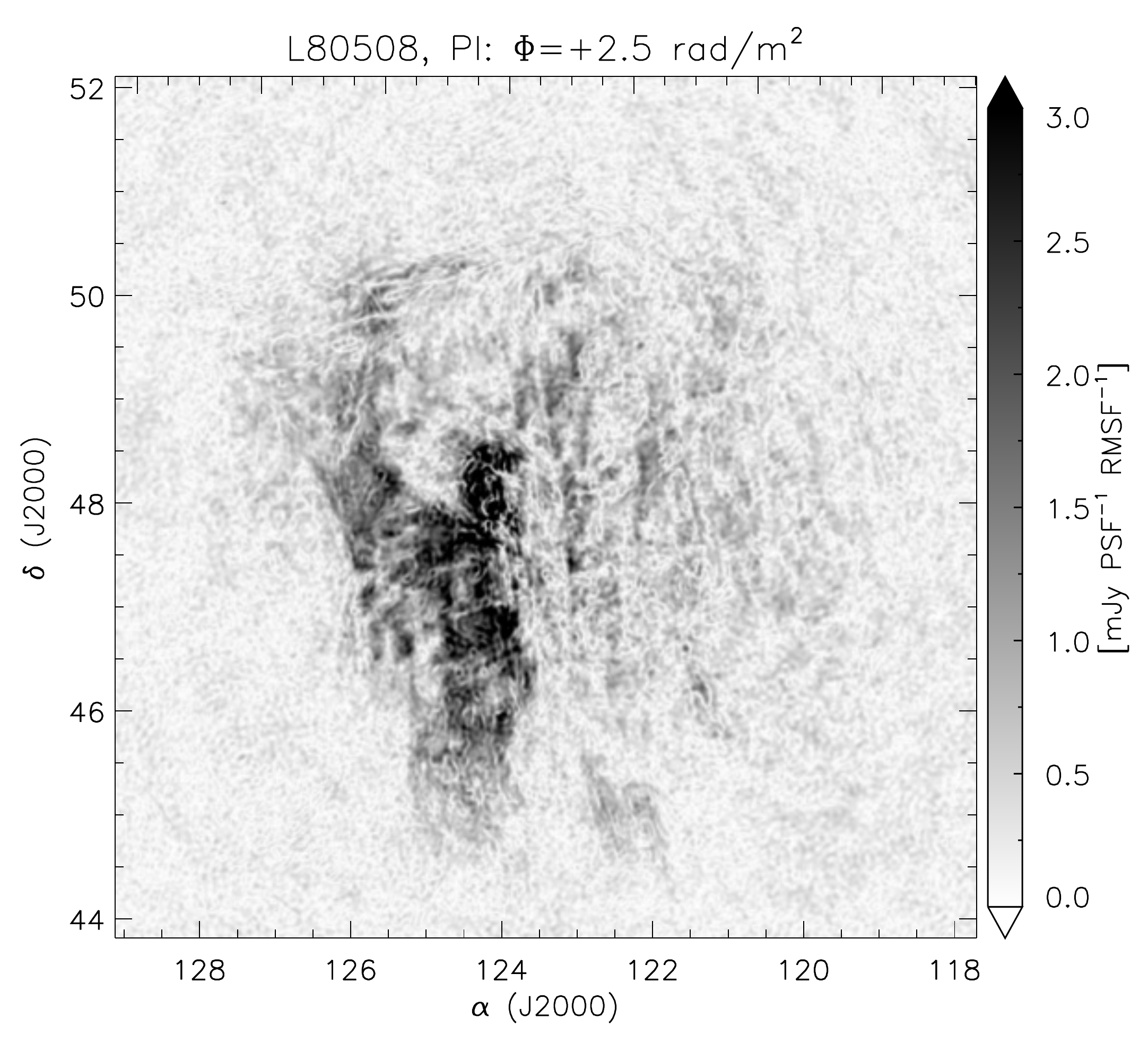}
\centering \includegraphics[width=.33\textwidth]{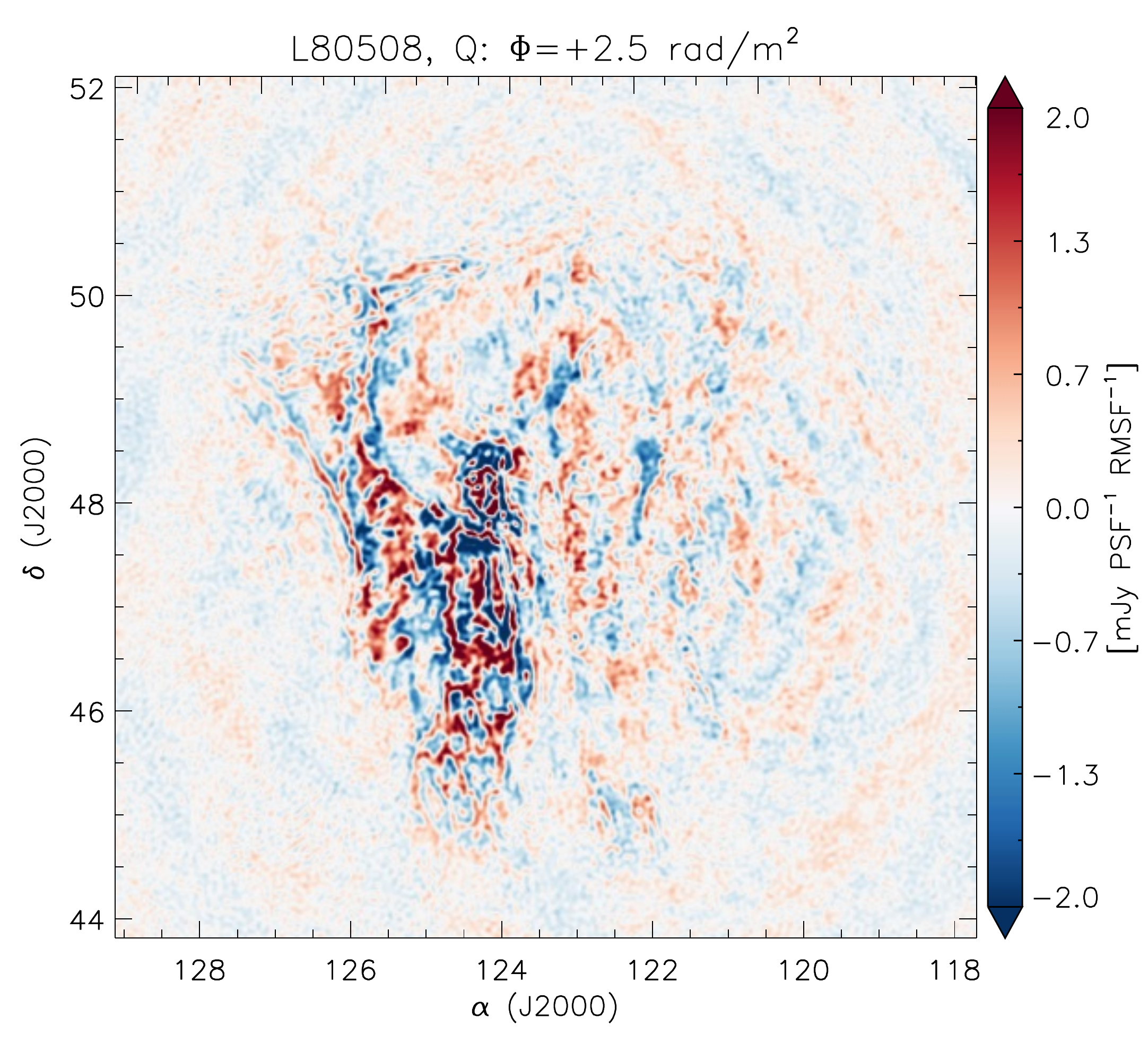}
\centering \includegraphics[width=.33\textwidth]{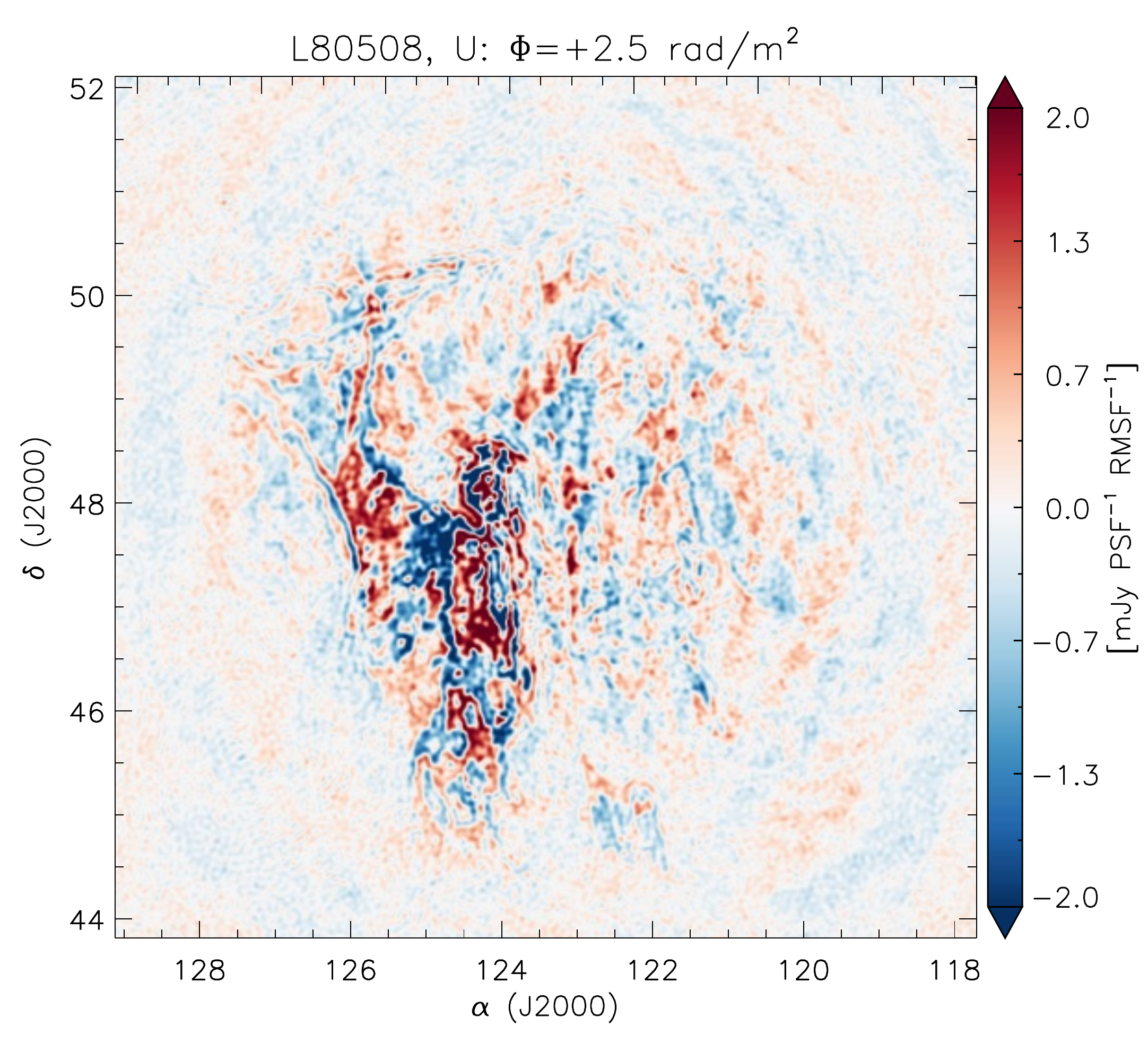}
\centering \includegraphics[width=.33\textwidth]{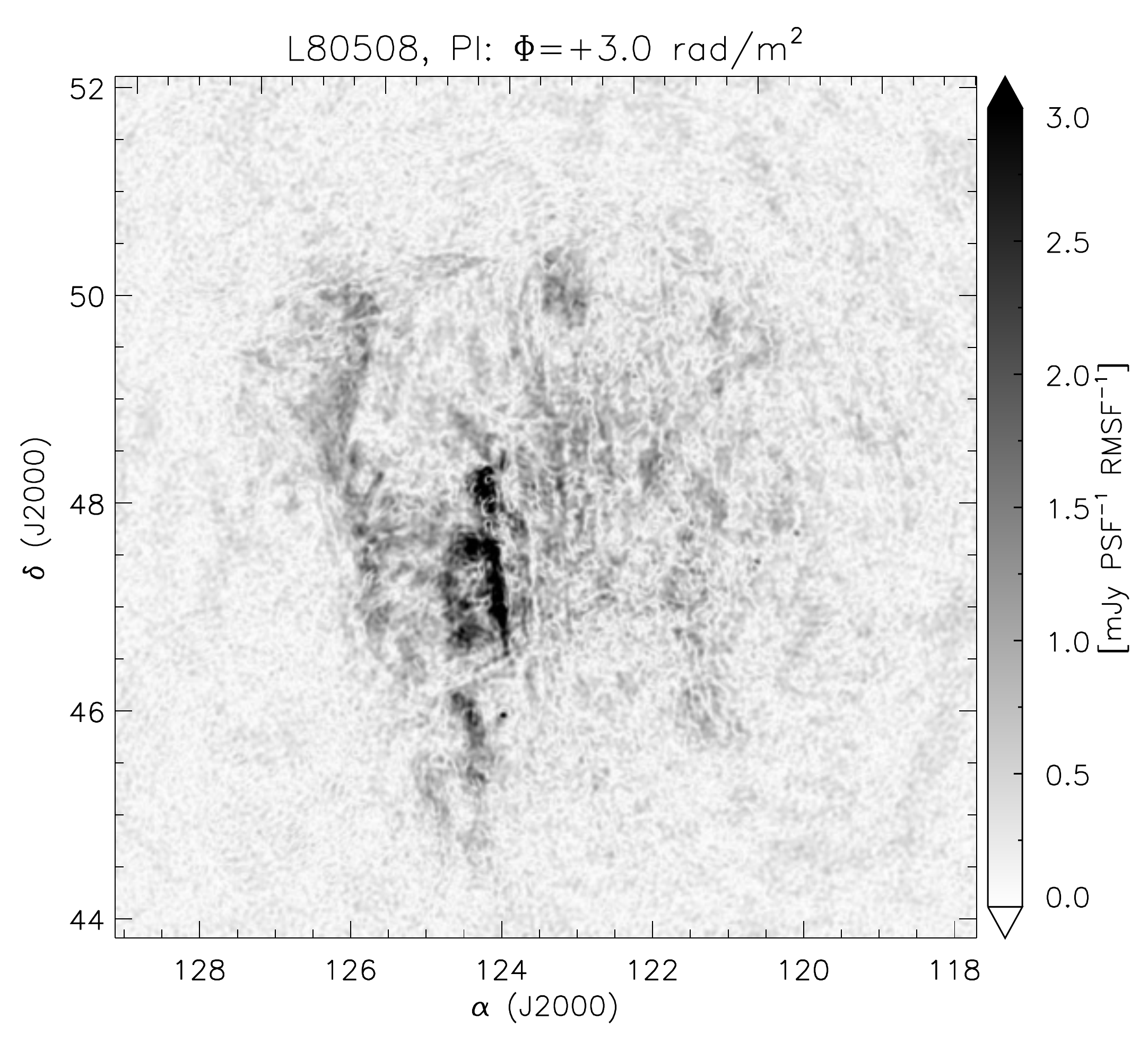}
\centering \includegraphics[width=.33\textwidth]{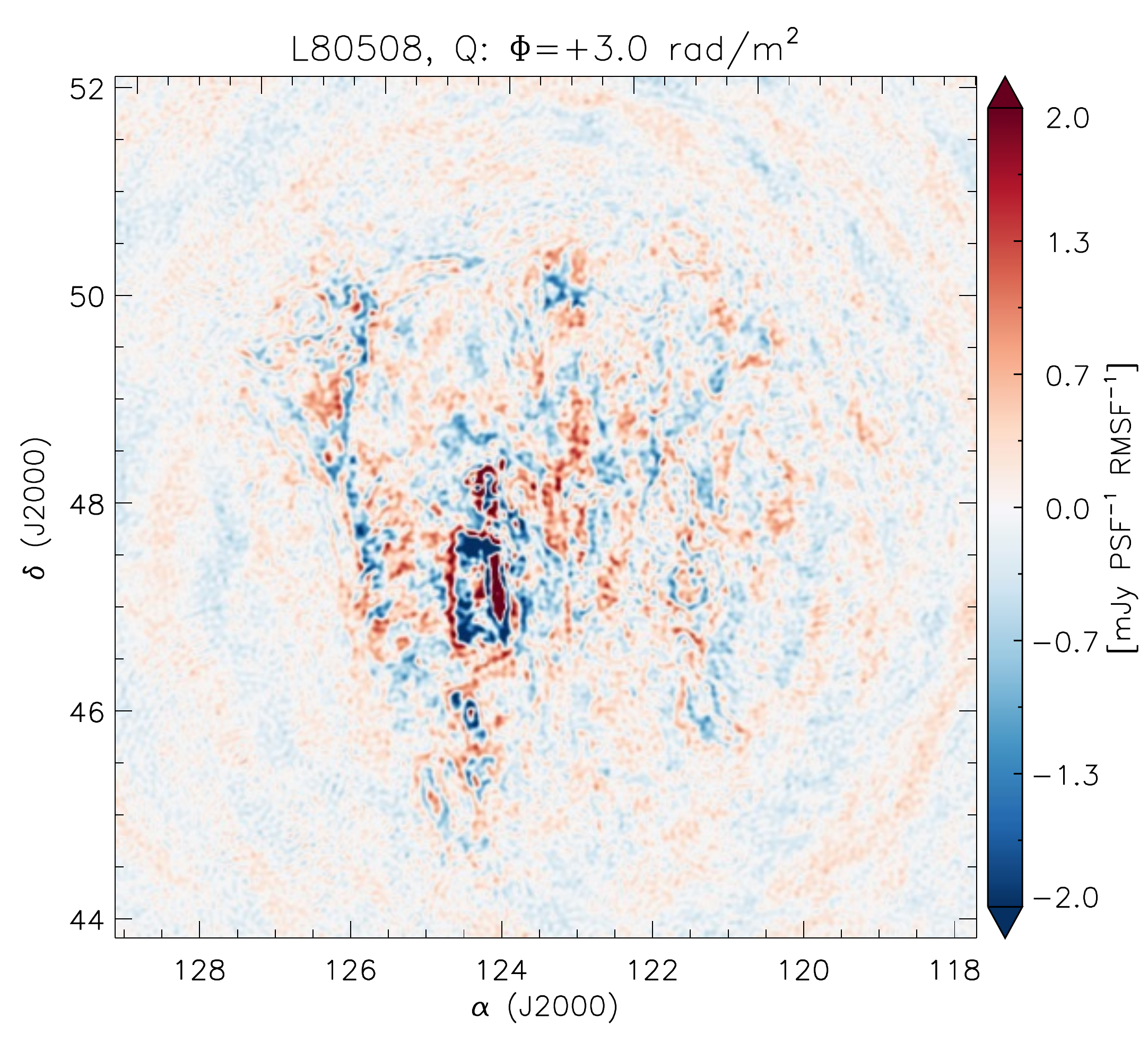}
\centering \includegraphics[width=.33\textwidth]{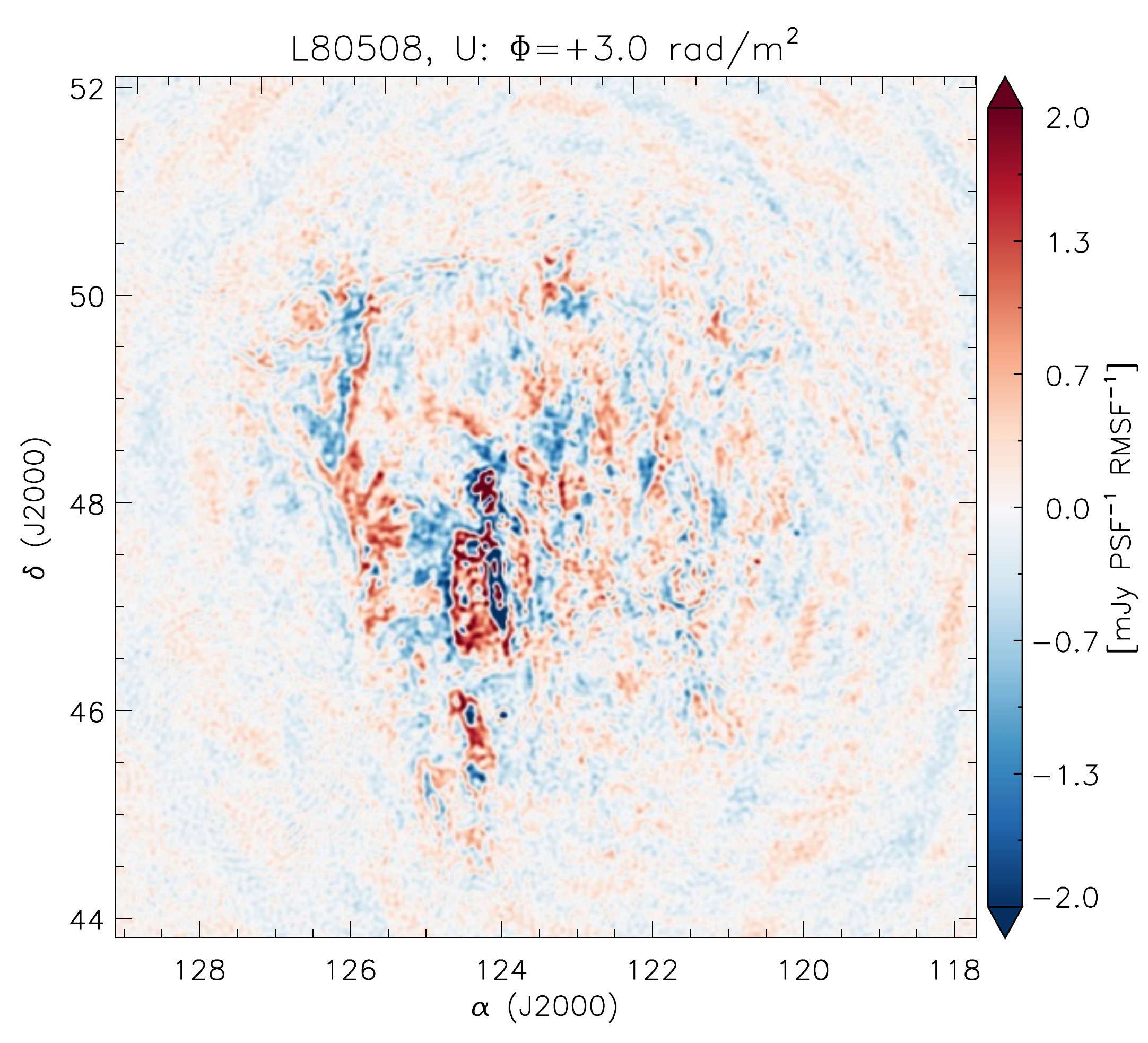}
\centering \includegraphics[width=.33\textwidth]{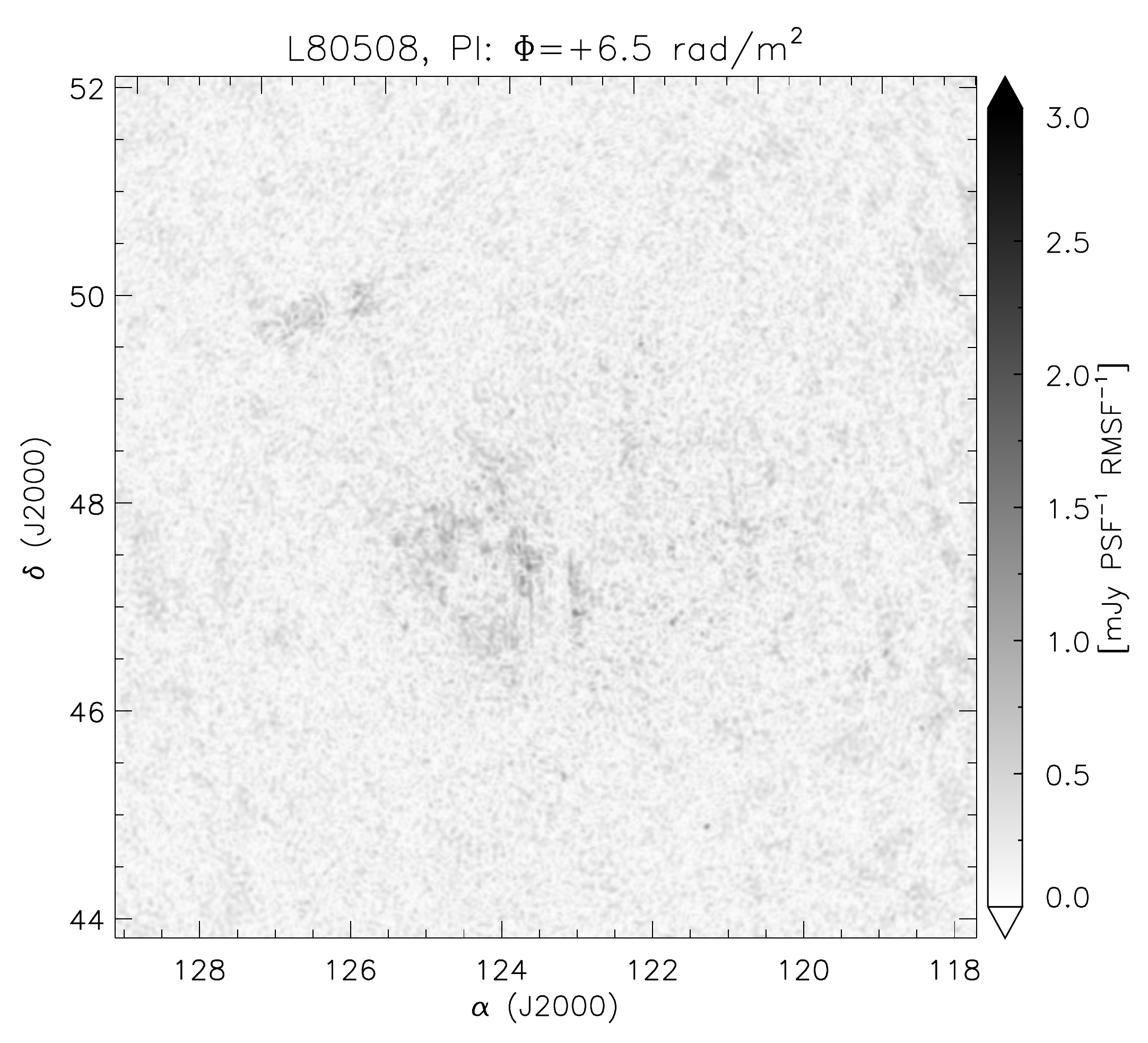}
\centering \includegraphics[width=.33\textwidth]{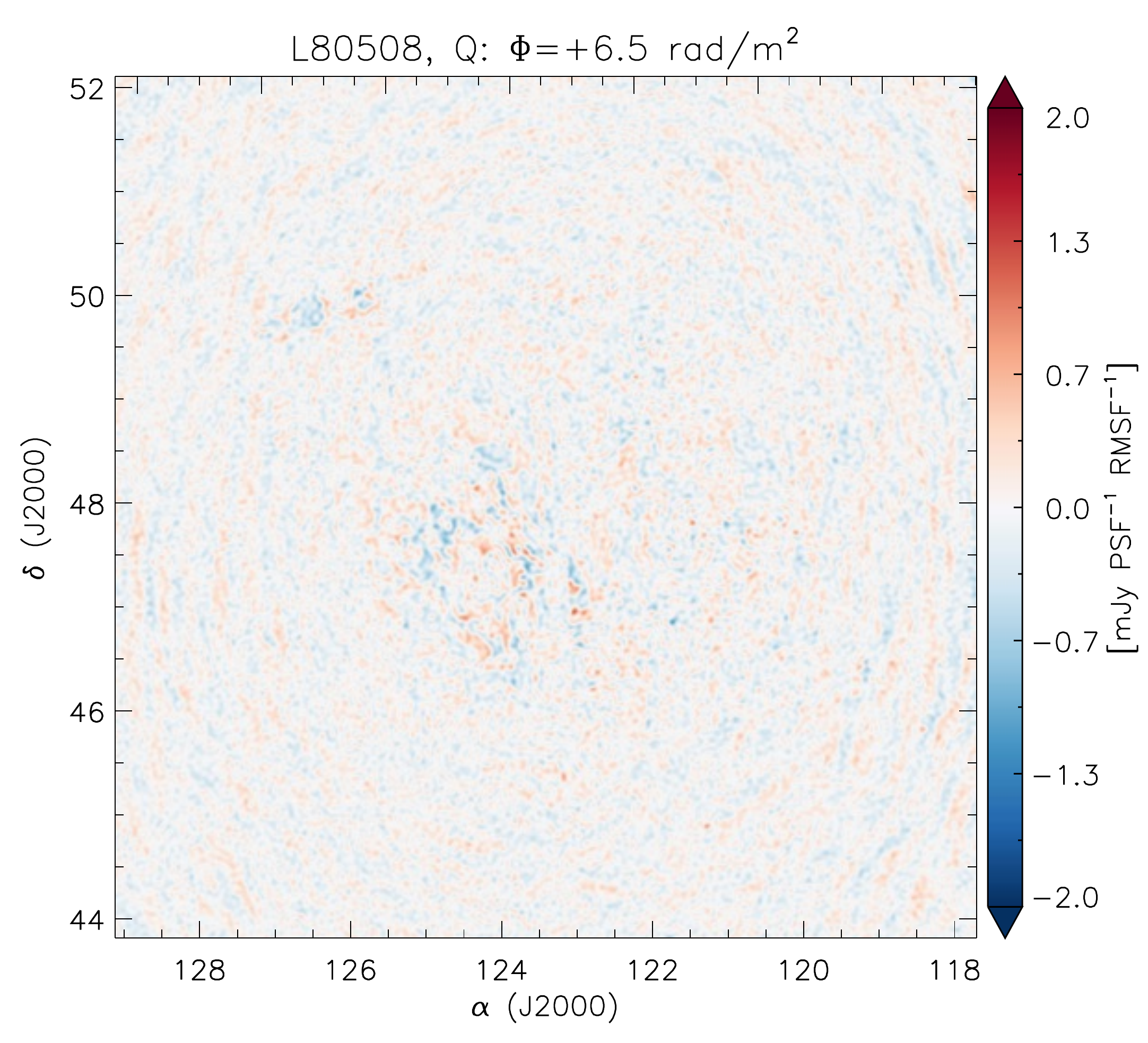}
\centering \includegraphics[width=.33\textwidth]{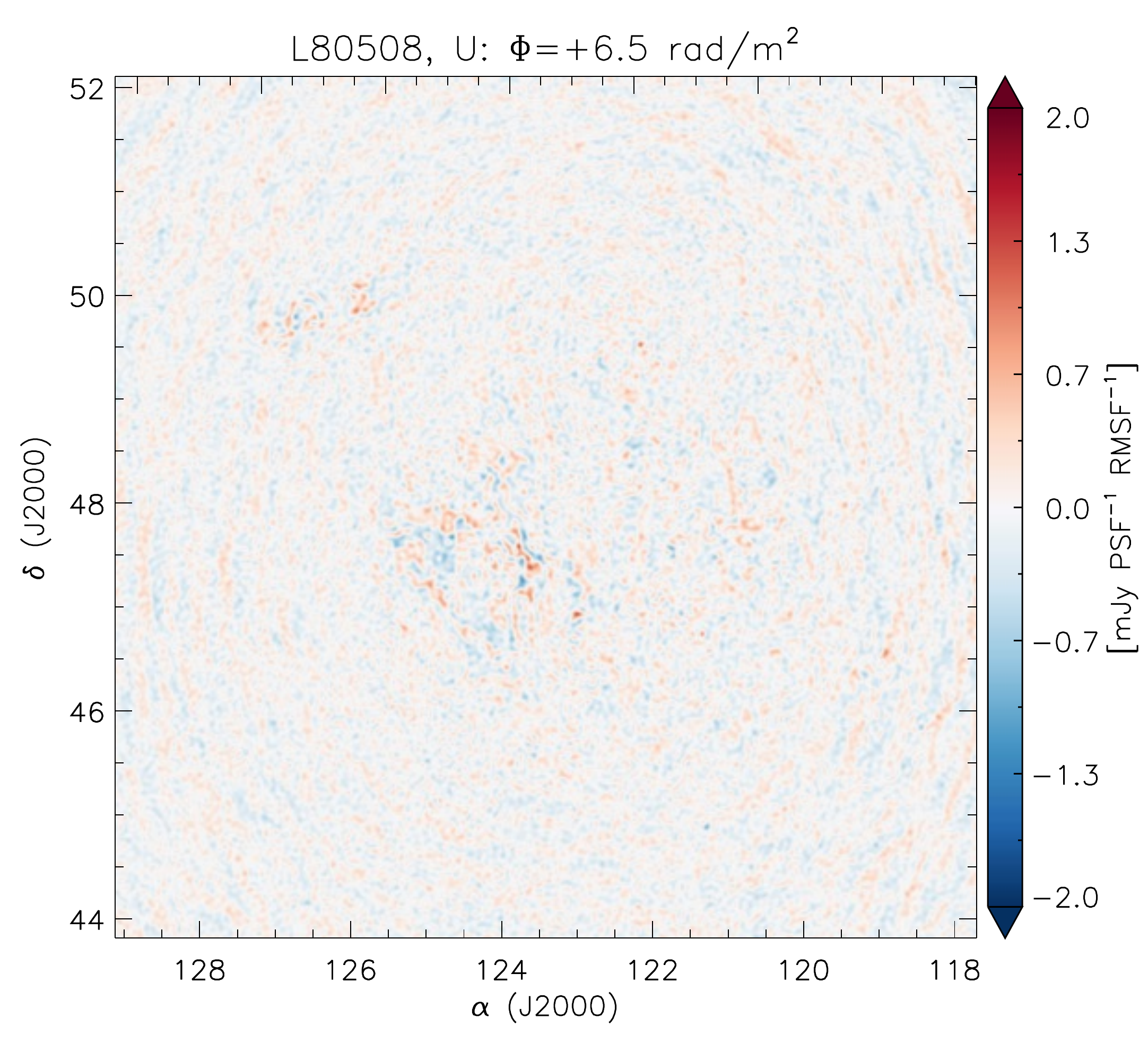}
\caption{Continued.}
\label{fig:PI2}
\end{figure*}

Diffuse emission ($\lesssim 3~{\rm mJy~PSF^{-1}~RMSF^{-1}}$) is detected over a range of Faraday depths from $-3$ to $+8~{\rm rad~m^{-2}}$. 
A triangular feature in the south-west side of the primary beam (marked with C in Fig.~\ref{fig:PI1}) dominates at negative Faraday depths.  At positive Faraday depths, emission first appears as two extended north-south structures (denoted with A and B in Fig.~\ref{fig:PI1}). A striking filamentary structure A, running right across 3C196, is at least $4^\circ$ long and $7.5'$ wide. Its mean surface brightness is $3.2~{\rm mJy~PSF^{-1}~RMSF^{-1}}$ and it peaks at Faraday depth $\sim0.5~{\rm rad~m^{-2}}$. The two-legged puppet structure B has the mean surface brightness of $2.5~{\rm mJy~PSF^{-1}~RMSF^{-1}}$, peaking at Faraday depth $\sim0.25~{\rm rad~m^{-2}}$. Towards the higher Faraday depths, emission fills up the east side of the primary beam. The mean surface brightness of this emission is $1.6~{\rm mJy~PSF^{-1}~RMSF^{-1}}$.

\begin{figure}[tb]
\centering \includegraphics[width=.5\textwidth]{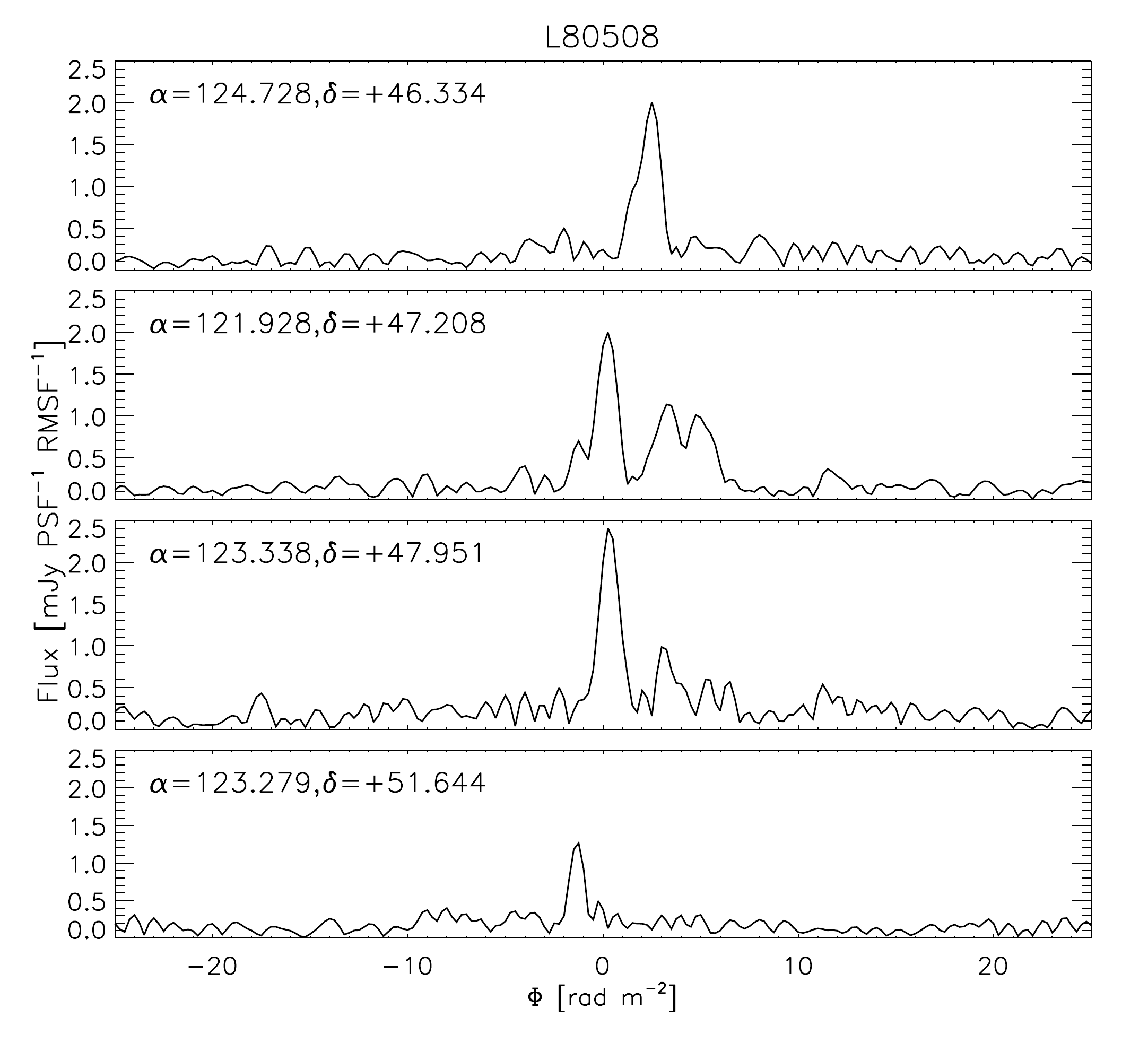}
\caption{Faraday spectra of a few representative lines-of-sight through the RM cube of polarized intensity.   We have not attempted 
to deconvolve the spectra for the effects of the side lobes of the RMSF.}
\label{fig:Fspec}
\end{figure}

A few representative Faraday spectra of detected features are shown in Fig.~\ref{fig:Fspec}. The largest Faraday structure that can be resolved is $1.1~ {\rm rad~m^{-2}}$ wide. Information contained in the Faraday spectra is summarized in Fig.~\ref{fig:RMmax}. We follow \citet{schnitzeler09} and show a map of the highest peak of the Faraday depth spectrum at each spatial pixel and a map of the RM value of each peak. The highest peaks in the Faraday spectra have on average a surface brightness of $\sim2.5~{\rm mJy~PSF^{-1}~RMSF^{-1}}$, located mostly at positive Faraday depths.

While the polarized intensity shows only the amplitude of polarized emission, images in Stokes Q,U also reflect the morphology of polarization angle.
There are a number of striking  patterns and features in Stokes Q,U (see Fig.~\ref{fig:PI1}). Most of them are outlined by regions that show emission 
in Stokes Q,U of opposite sign. They have a typical width of $~5'-10'$.
    
\subsection{Polarized sources}\label{sec:polsor}
\begin{figure*}[!phtb]
\centering \includegraphics[width=.75\textwidth]{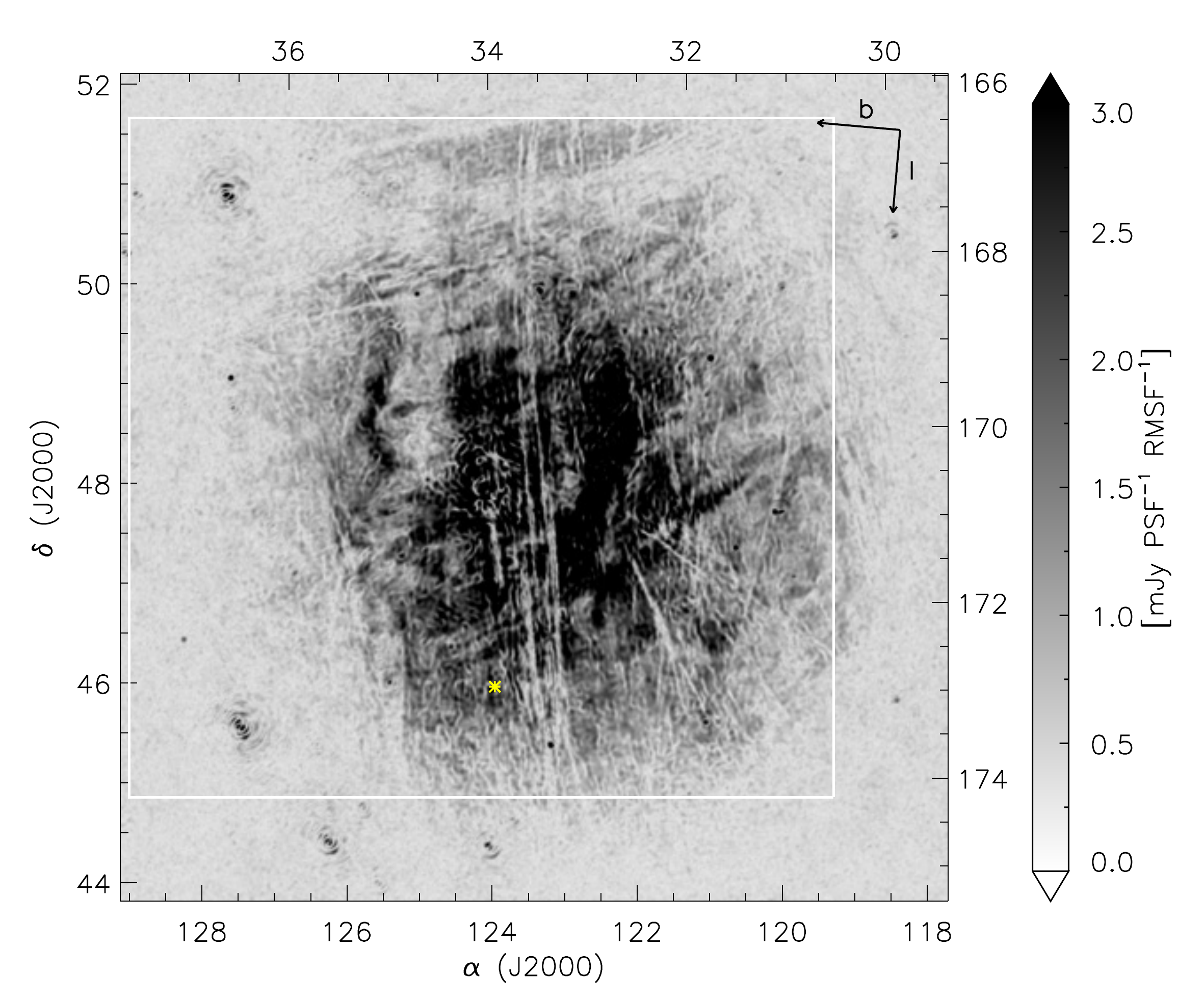}
\centering \includegraphics[width=.75\textwidth]{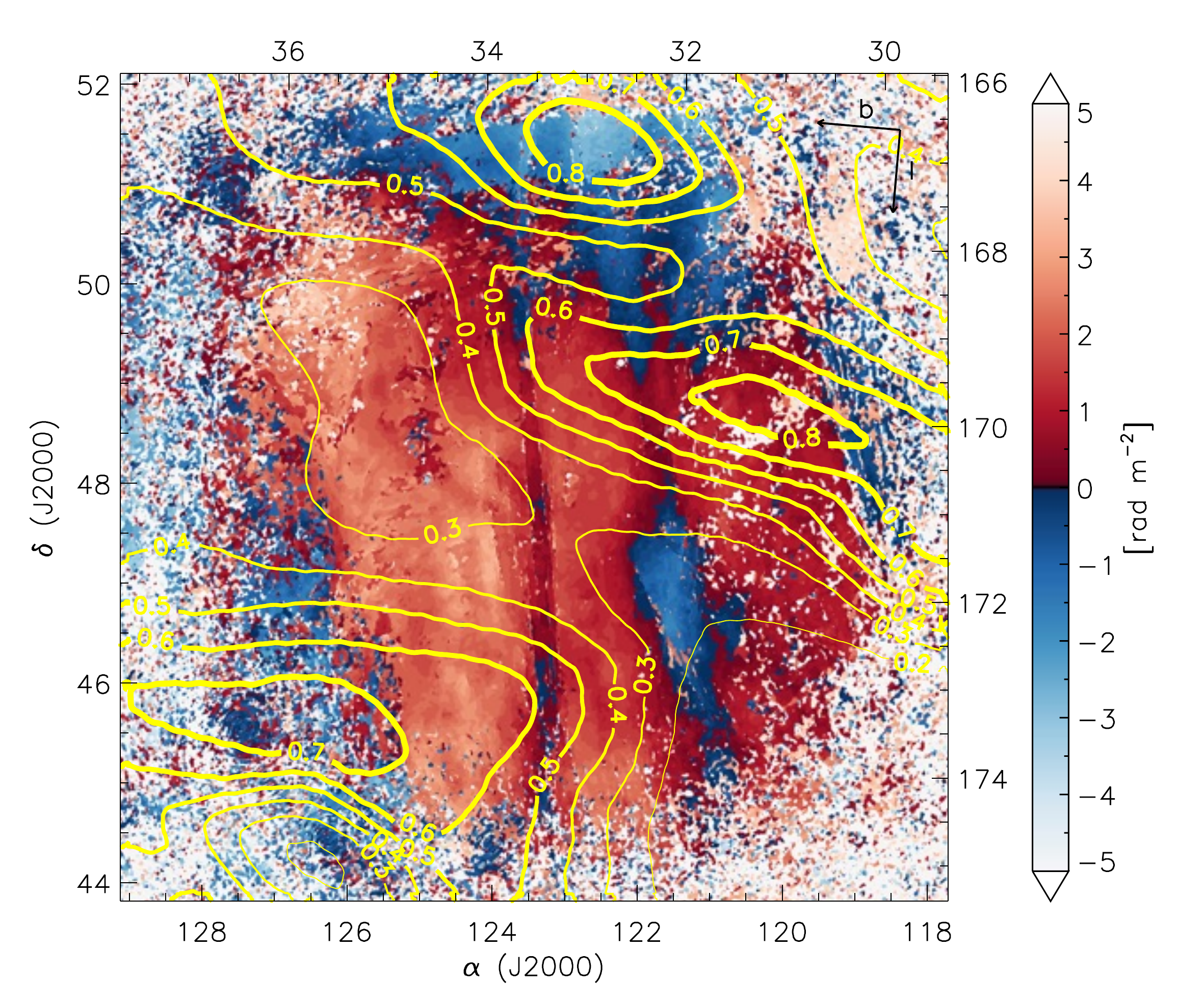}
\caption{The upper image shows the highest peak of the Faraday depth spectrum in polarization at each spatial pixel, while a map of the RM value of each peak, shown in red-blue colour scale, is on the bottom. Yellow contours are showing H$\alpha$ intensity on a one degree scale \citep{finkbeiner03, haffner03}, given in units of Rayleighs (see Sec.~\ref{sec:dis}). A white box indicates the boundaries of the WSRT 350MHz image in Fig.~\ref{fig:WSRT}. A location of the discovered pulsar J081558+461155 (see Sec.~\ref{sec:polsor}) is shown with a yellow asterisk.}
\label{fig:RMmax}
\end{figure*}

In addition to the widespread diffuse emission,  we  identified $\sim15$ discrete polarized sources in the 3C196 field. 
All but one of these sources are (giant) radio galaxies with typical polarization fractions of a few percent, and a detailed analysis of their properties will be presented in de Bruyn et al. (in prep). 
One source, J081558+461155 (see Fig.~\ref{fig:stokesI}\&\ref{fig:RMmax}), has a steep spectrum, is spatially unresolved and shows about 50\% polarization. These properties are reminiscent of those of pulsars. 

A follow-up observation of this field, in Director's Discretionary Time, 
with the LOFAR Full HBA Core in tied-array mode \citep{haarlem13} indeed revealed a pulsar with a period of $434~{\rm ms}$ and
a dispersion measure (DM) of $11.28~{\rm pc~cm^{-3}}$ (private communication, J.\ W.\ T.\ Hessels and V.\ Kondratiev; 
see the LOFAR Tied-Array All-Sky Survey Discoveries\footnote{http://www.astron.nl/lotaas/}).  

The RM of the pulsar is $+2.7\pm0.1~{\rm rad~m^{-2}}$, as determined from 
our RM cubes.  In the direction of the pulsar the \textsc{NE2001} model of the Galactic distribution of thermal electrons \citep{cordes02, cordes03} predicts a thermal electron density of $\left<n_e\right>=0.028~{\rm cm^{-3}}$. Given its DM, we estimate its distance to be $\sim400~{\rm pc}$. Given the sparsity of sources in this direction, however, both the density and the distance derived from the model must be considered uncertain. 

\section{Fidelity of the data}
In this section, we make qualitative and quantitative comparisons between five LOFAR observations. The analysis is performed
on RM cubes in polarized intensity, Stokes Q and U.  Table~\ref{tab:stats} gives an overview of the noise in the RM cubes
for different nights. Variation of the noise from night to night is small. The L79324 observation shows the largest variations in RM and has the highest noise. L80273 has the lowest noise.  The ratio of the \emph{rms} values between the polarized intensity and Stokes Q,U is $\sim0.65$. This implies consistent
normally distributed noise in Stokes Q,U.

We then compare maps of the highest peak of the Faraday depth spectrum at each spatial pixel 
for different nights. An example of this kind of  map is given 
in Fig.~\ref{fig:RMmax}. For each pair of maps, we calculate the Pearson correlation coefficient 
using only the inner $5^\circ\times5^\circ$ of the images.
We find that the peak fluxes agree well, with a correlation of $96\pm1\%$ from night to night. The L80273 and L80508
observations show the highest correlation, of 98\%. This shows that LOFAR performance from night to night
is very stable and that we are able to correct for Faraday rotation in the ionosphere to first order. The small
difference in the peak fluxes between the observations is mainly caused by the low time resolution of the 
CODE data ($2~{\rm h}$). 

To check the alignment of the RM cubes in Faraday depth, we calculate the cross-correlation between two observations
as a function of the lag in Faraday depth.  In this comparison we take only spectra that have a peak flux in 
polarized intensity $>1.5~{\rm mJy~PSF^{-1}~RMSF^{-1}}$ (at least two times above the noise). We then average the cross-correlation functions
 to determine their common peak by fitting a parabola. The position of this peak gives the misalignment of 
 the two RM cubes. On average we find that the five RM cubes are misaligned by $0.10\pm0.08~{\rm rad~m^{-2}}$.
The best alignment is between L79324 and L80508 ($0.02\pm0.01~{\rm rad~m^{-2}}$) and
the worst is between L80508 and L192832 ($0.14\pm0.01~{\rm rad~m^{-2}}$). A misalignment
between different observations can be attributed to systematic errors in the CODE data, resulting in a bias in TEC maps from day to day. This bias leads to uncertenties of $\sim1~{\rm TEC}$ unit, corresponding to 
an error in RM around $0.1~{\rm rad~m^{-2}}$.

\begin{table}
\caption{The noise in the RM cubes of different observations. The values are given for polarized intensity (PI) and Stokes Q,U.}
\centering                                      
\begin{tabular}{lccccc}          
\hline\hline        
Observation & & PI & Q & U & \\
ID &  \multicolumn{5}{c}{${\rm [\mu Jy~PSF^{-1}~RMSF^{-1}]}$}\\                              
\hline                                   
    L79324 & & 83 & 123 & 127 & \\  
    L80273 & & 67 &  101 & 102 & \\
    L80508 & & 70 &  111 & 105 & \\
    L80897 &  & 75 &  118 & 114 & \\
    L192832 & & 73 & 111 & 115 & \\
 \hline
  \end{tabular}\label{tab:stats}
\end{table}

To get a better signal-to-noise ratio in the RM cubes, one can average the cubes from different nights.
However, the RM cubes should be aligned in Faraday depth to high accuracy. Even a small 
misalignment  decorrelates the signal to some extent. The detected diffuse emission is
$\sim25\times$ above the noise, so an additional improvement of $\sqrt{5}$ does not make an
impact on our results and our current conclusions.  

Finally, we compare our LOFAR observations with observations of the same field by \citet{bernardi10} in the same frequency range, obtained using the LFFE on the WSRT radio telescope. Emission 
detected with the WSRT-LFFE is fainter and its morphology appears much simpler \citep[see Figs.~9, 10, \& 11 in][]{bernardi10} than the morphology of
emission detected with LOFAR.  There is just faint patchy emission, without any more distinctive 
morphological features.

The difference between the WSRT-LFFE and LOFAR observations can be attributed to the three times better resolution 
of LOFAR in Faraday depth. Multiple Faraday-thin structures ($\Delta\Phi<\delta\phi_{\rm WSRT-LFFE}$) detected 
in the LOFAR observations  decorrelate when we observe 
them with a broader RMSF if their emission shows different polarization angle. 
Moreover, WSRT-LFFE observations were showing some instrumental artefacts  that 
contaminated RM cubes \citep{bernardi10}. Given these 
restrictions a more quantitative comparison between the LOFAR and WSRT-LFFE observations is not very meaningful and we  
restrict our analysis to the superior quality LOFAR data.  

\section{Discussion}\label{sec:dis}
\subsection{General properties of detected diffuse polarized emission and the underlying magnetic field component}\label{sec:genprop}
\begin{figure}[tb]
\centering \includegraphics[width=.45\textwidth]{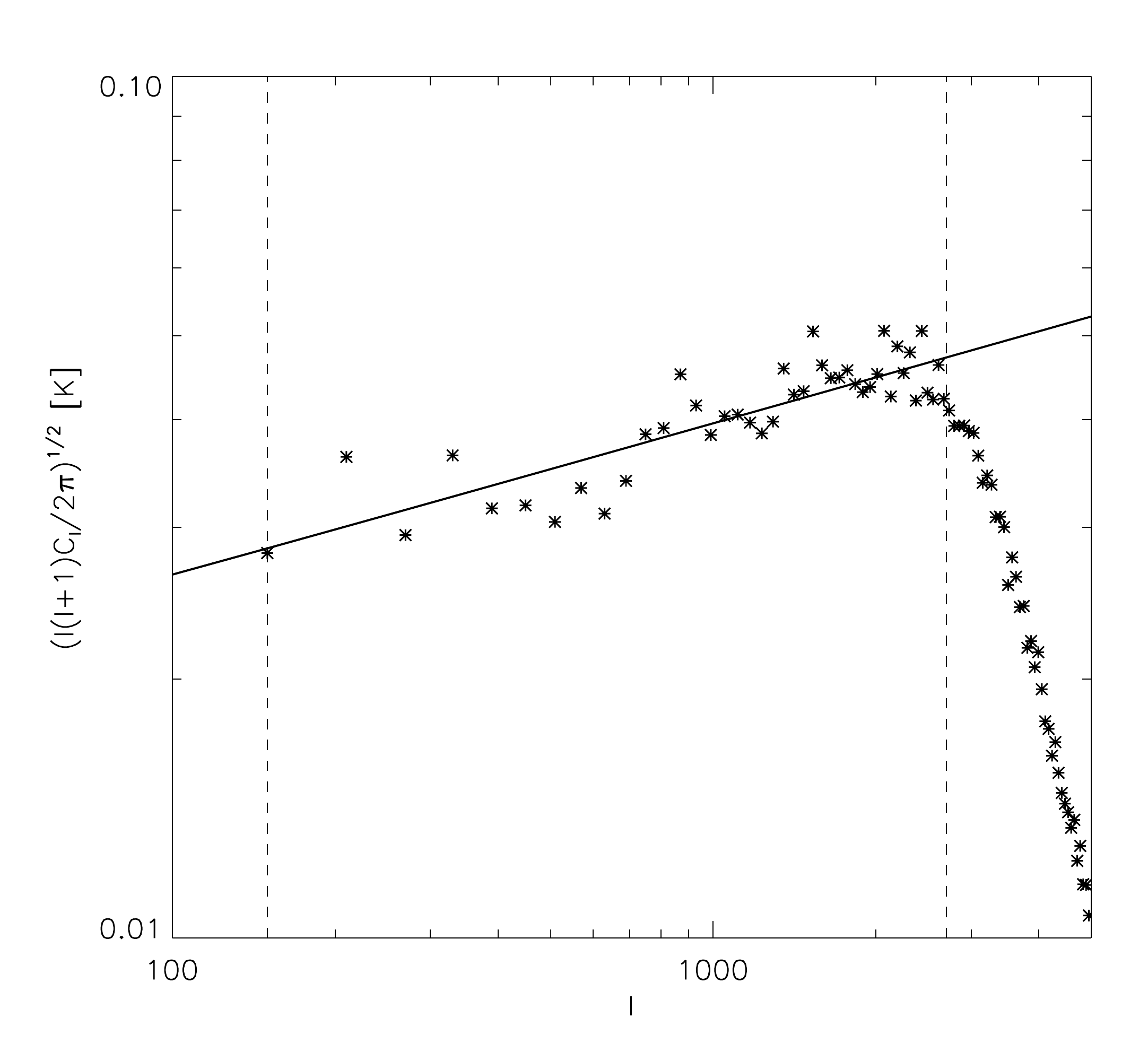}
\caption{Angular power spectrum of integrated polarized emission. The best power-law fit (solid line) to the power spectrum
(``$C_l-l$'') between $l=150$ and $l=2700$ gives a slope of $\beta^P_l=-1.65\pm0.03$. Dashed lines show a range of data points used for the power-law fit.}
\label{fig:angPS}
\end{figure}

The diffuse emission presented in Sec.~\ref{sec:diffpe} is of Galactic origin. To analyse its properties, we follow 
our previous work \citep{jelic14} and compute the total polarized intensity at each pixel by integrating
the polarized intensity RM cube along Faraday depth. The brightness temperature of the resulting integrated emission varies from $5$ to $15~{\rm K}$. 
This is comparable to the brightness temperature of emission detected in the 
ELAIS-N1 field \citep{jelic14}, also a field at high Galactic latitude. Using only the inner $3^\circ\times3^\circ$ of the field, we 
estimate the angular power spectrum of the integrated polarized emission, $C_l$. In the calculation we assume 
the ``flat-sky'' approximation \citep[e.g.][]{white99}, i.e. 
$k^2P_k\simeq\frac{l(l+1)}{(2\pi)^2}C_l$  for $l=2\pi k$, 
where $P_k$ is the 2D Fourier transform of an image and 
$l=180^\circ/\theta[^\circ]$ is a spherical harmonic multipole. 
Figure~\ref{fig:angPS} shows the resulting averaged dimensionless angular power spectrum. We then fit a power law to 
``$C_l - l$'' between $l=150$ and $l=2700$. The best fit (solid line in Fig.~\ref{fig:angPS}) gives a slope of $\beta_l=-1.64\pm0.05$. 

In total intensity a slope of the synchrotron emission is usually steeper  \citep[$-3.0\lesssim\beta_l\lesssim-2.5$ e.g.][]{baccigalupi01,tucci02,bernardi09,ghosh12}. A difference between the two slopes can be explained by redistribution of power between different angular scales of polarized emission, mostly from large scales to small scales, as a result of small-scale structure of intervening Faraday-rotating medium.  At LOFAR frequencies even a very small RM of the medium  leaves a signature in the power spectrum.

On the same angular scales, $100< l <3000$, and in the same frequency range as our observations, but in an area located in the Galactic plane
(Fan region), \citet{bernardi09} found a slope for polarized emission of $\beta_l=-1.65\pm0.15$, which is very similar to the 3C196 field. The morphology of emission in the two fields is different, yet integrated emission along Faraday depth of both fields shows a very similar power spectrum slope. This indicates that in both fields redistribution of power through Faraday rotation is similar. 

In the current data we have not detected diffuse emission in total intensity. This can be attributed to a lack of short baselines in our observations. We are only sensitive  to structures on scales smaller than two degrees, meaning that synchrotron emission in total intensity appears mostly on scales larger than two degrees. This is in line with the discussed steeper slope of the observed synchrotron emission in total intensity, and with previous observations with LOFAR \citep{jelic14} and WSRT \citep[e.g.][]{bernardi10} at high Galactic latitudes. Unfortunately, a lack of detected emission in total intensity prevents us from making a more detailed study of Galactic interstellar turbulence through fluctuations in synchrotron emission, as was done by \citet{iacobelli13} for a low Galactic latitude field. 

In near future the AARTFAAC\footnote{http://www.aartfaac.org} (Amsterdam - ASTRON Radio Transients Facility and Analysis Centre) project might help to mitigate a lack of a very short baselines in our current LOFAR observations. Six stations at the heart of LOFAR will be used as a 288-element array. This will then provide almost full uv coverage at short baselines and will enable us to map large-scale Galactic emission that is  currently not accessible.

To estimate the fraction of polarization, we divide the integrated polarized intensity by the $408~{\rm MHz}$ total intensity map \citep{haslam81,haslam82} scaled to $160~{\rm MHz}$. The spectral index between 45 and $408~{\rm MHz}$ of Galactic synchrotron emission in this region is $\beta=-2.5$ \citep{guzman11}. If we scale the brightness temperature of $\sim23~{\rm K}$ from $408~{\rm MHz}$ to $160~{\rm MHz}$, we deduce a brightness temperature of $236~{\rm K}$ in total emission. The observed polarization levels therefore imply a polarization of $\approx4\%$. 

A high degree of polarization indicates a regular large-scale magnetic field topology, coming from either uniform or anisotropically random magnetic fields.
The uniform magnetic field component, following the spiral arms of our Galaxy, is almost perpendicular to the line of sight at the direction of 3C196 \cite[for a description of Galactic magnetic field models, see][and references therein]{sun08}.
Its contribution to Faraday rotation towards the 3C196 field is therefore expected to be small. 

This assumption is supported by the estimated mean line-of-sight magnetic field component in the direction of pulsar 
J081558+461155 (see Sec.~\ref{sec:polsor}). Given the known relation between the RM and DM of a pulsar \citep[e.g.][]{manchester72}, i.e.
\begin{equation}
\frac{\left<B_{||}\right>}{[{\rm \mu G}]}=\frac{\rm RM~[{\rm rad~m^{-2}}]}{\rm 0.812~DM~[{\rm pc~cm^{-3}}]},
\end{equation}
we get $\left<B_{||}\right>=0.3\pm0.1~{\rm \mu G}$. A typical strength of magnetic field in the Galactic halo (thick disk) is $\sim3~{\rm \mu G}$,
with an expected random component of $\sim1~{\rm \mu G}$ \citep[][and references therein]{mao10,haverkorn15}.

Variations of the mean line-of-sight magnetic field component across the 3C196 field, $\sigma_{\left<B_{||}\right>}$, can be estimated from variations 
in the RM values of detected diffuse structures and in thermal electrons.  We obtain the RM variations directly from the image 
in Fig.~\ref{fig:RMmax}, $\sigma_{\rm RM}\simeq4~{\rm rad~m^{-2}}$. A good tracer of warm ionized gas is the H$\alpha$ line. 
From the composite all-sky H$\alpha$ map \citep[][v1.1]{finkbeiner03}\footnote{This map is a composite of three separate surveys. The 3C196 field is covered only by the Wisconsin H-Alpha Mapper \citep[WHAM;][]{haffner03} survey, which means that given H$\alpha$ emission is at $1^\circ$ resolution. The WHAM survey provides velocity resolved spectra, but for this work we use only integrated emission.}, we find the H$\alpha$ intensity across the 3C196 field to be $0.6\pm0.1~{\rm R}$ (Rayleighs; $1~{\rm R}=10^6~(4\pi)^{-1}$ photons ${\rm cm^{-2}~s^{-1}~sr^{-1}}$).  However, the given H$\alpha$ intensity reflects the integrated emission over different velocities ($\left|v\right|<100~{\rm km~s^{-1}}$) 
and over the line-of-sight through the whole Galaxy, while the polarized structures detected with LOFAR are 
probably limited to distances $\lesssim1-1.5~{\rm kpc}$ from us (see Sec.~\ref{sec:dist}). Electron densities that we derive from H$\alpha$ intensity should therefore be taken more as an upper limit. We do not know whether we see polarized emission all the way to the edge of the Galaxy.  

 We convert the observed H$\alpha$ intensity into emission measure (EM), according to \citep{reynolds91}, 
\begin{equation}\label{eq:em}
\frac{EM}{[\rm pc~cm^{-6}]}= \int_0^L n_e^2 dl = 2.75 \left(\frac{T}{[10^4~{\rm K}]}\right)^{0.9}\frac{I_{H\alpha}}{\rm [R]},
\end{equation}
where $T$ is the temperature of the ionized gas, $n_e$ is the local electron density, and $L$ is the total path length. Assuming 
$T=10^4~{\rm K}$ at high Galactic latitudes \citep[e.g.][and references therein]{sun08,akahori13}, we derive the related
emission measure of $EM=1.7\pm0.3~{\rm pc~cm^{-6}}$ across the 3C196 field. This corresponds 
to a large-scale mean electron density of $\left<n_e\right>\simeq0.03~{\rm cm^{-3}}$
over an assumed path length of $1.5~{\rm kpc}$. Its spatial variation across the 3C196 field is $\sigma_{\left<n_e\right>}\simeq0.003~{\rm cm^{-3}}$.
In our calculation, we do not take  observed vertical temperature gradient in the diffuse 
ionized gas into account \citep[][]{reynolds99,haffner09}, as the gradient is small and our results are not sensitive to it.

We can now use the equation for rotation measure,
\begin{equation}\label{eq:fardep}
\frac{\left<RM\right>}{\rm [rad~m^{-2}]}=0.81\int^L_0 \frac{\left<n_e\right>}{\rm [cm^{-3}]} \frac{\left<B_\parallel\right>}{\rm [\mu G]}d l
,\end{equation}
and estimate $\sigma_{\left<B_\parallel\right>}$ according to
\begin{equation}\label{eq:sigBp}
\sigma_{\left<B_\parallel\right>}=\sqrt{\left(\frac{\sigma_{\left<\rm RM\right>}}{0.81\left<n_e\right>L}\right)^2+\left(\frac{\left<{RM}\right>\sigma_{\left<n_e\right>}}{0.81\left<n_e\right>^2L}\right)^2}.
\end{equation}
In our case, the first term in Eq.~\ref{eq:sigBp} dominates over the second term. 
Using $L=1.5~{\rm kpc}$ and $\left<RM\right>=1~{\rm rad~m^{-2}}$, we get variations of the mean line-of-sight
magnetic field component across the 3C196 field to be $\sigma_{\left<B_\parallel\right>} \simeq 0.1~{\rm \mu G}$.
This is in agreement with an upper limit, $\sigma_{\left<B_\parallel\right>} \lesssim 0.4~{\rm \mu G}$, derived by
\citet{schnitzeler10} using the rotation measure of NVSS sources. 

If $B_\parallel$, $n_e$, or both show a large-scale gradient across the sky, this should be seen in observed RM cubes. Polarized structures  appear to move in the direction of the gradient as we go through different Faraday depths. We see this behaviour in our RM cubes from $+1$ to $+2.75~{\rm rad~m^{-2}}$. Polarized
emission moves mostly from the west to the east (see Fig.~\ref{fig:PI1}~\&~\ref{fig:RMmax}). To quantify this, we calculate the centroid of emission at each Faraday depth as
\begin{equation}
\vec{R_C}(\Phi)=\frac{\sum \vec {r_{i,j}} \cdot S_{i,j} (\Phi)}{\sum S_{i,j}(\Phi)},
\end{equation}
where sums are taken over all pixels of an image $S_{i,j}$ at Faraday depth $\Phi$, where $S_{i,j}>1.5~{\rm mJy~PSF^{-1}~RMSF^{-1}}$ (at least two times above the noise). The vector $\vec {r_{i,j}}$ contains the
coordinates of the pixels. 

We find that the centroid of emission changes systematically by $\sim0.8^\circ$ per ${\rm rad~m^{-2}}$ from west to east and by $0.3^\circ$ per ${\rm rad~m^{-2}}$ from north to south. From Faraday depth $+1$ to $+2.75~{\rm rad~m^{-2}}$ the polarized structures have moved by $1.4^\circ$ towards the east and by $0.6^\circ$ towards the south. This is indeed an indication of a large-scale gradient in RM.

Based on our previous discussion that the $\sigma_{\left<\rm RM\right>}$ term in Eq.~\ref{eq:sigBp} dominates over the $\sigma_{\left<n_e\right>}$ term, the large-scale gradient that we observe seems to be a gradient of $B_\parallel$ across the 3C196 field. However, the distribution of the thermal plasma is known to be extremely inhomogeneous with a very small filling factor for the Warm Ionized Medium \citep[WIM, e.g.][]{berkhuijsen06,gaensler08}, which is mostly responsible for the Faraday rotation \citep[e.g.][]{heiles12}. The estimated $n_e$  from the \citet{finkbeiner03} H$\alpha$ map is just an average of the true distribution along a line-of-sight. Therefore, it is possible that inhomogeneity of $n_e$ also contributes to the observed gradient in Faraday depth. 

Finally, $B_\parallel$ across the 3C196 field varies both in strength and in orientation. This is seen through the presence of observed structures both at positive and negative Faraday depths (see Fig.~\ref{fig:RMmax}). Positive RM values are associated with $B_\parallel$ orientated towards us and negative RM values with  $B_\parallel$ orientated away form us. This does not necessarily mean
that $B_\parallel$ has the same orientation along the line of sight.  The field can reverse, sometimes even multiple times. 

Direct evidence for field reversal(s) can be seen in the Faraday spectrum,
where structures are found at both positive and negative depths. In our RM cubes there are $\sim5\%$ of Faraday spectra that we directly associate with field reversal(s) (for an example see Fig.~\ref{fig:Fspec}). However, spectra with structures, which are only on one side of the Faraday spectrum, do not exclude field reversals. If two regions with opposite magnetic field orientations change the polarization angle of incoming emission, but they do not themselves emit, and there is no additional emission between the two, the polarization angle of incoming emission is first rotated in one direction and then derotated in the other. The observed polarization angle is then a net effect of Faraday 
rotations along these two regions \citep[for an illustration, see Fig. 2 in ][]{brentjens05}. In the Faraday spectrum, this only manifests as  one structure, despite the fact that in reality we have two physically separated regions.

\subsection{Physical picture and relative distribution of intervening magneto-ionic medium}\label{sec:dist}
The synchrotron emission of our Galaxy is known to have many components: 
the disk, a thick disk, and possibly a halo \citep[e.g.][and references therein]{subrahmanyan13}. At high Galactic latitudes the
bulk of the emission comes from the thick disk with a width of about $1-1.5~{\rm kpc}$. The polarized emission 
builds up over the same path length, but along the way it gets depolarized by the magneto-ionic ISM. Depolarization 
is stronger at lower frequencies.  A typical observed polarization at LOFAR frequencies is about $1-4\%$ \citep{iacobelli13,jelic14}, while intrinsically synchrotron radiation is polarized at the $70\%$ level. 
Hence only a small fraction of the synchrotron emission of our Galaxy is observed in polarization.

A possible geometric configuration and distribution of the random and uniform component of the magnetic field and magneto-ionic medium are very difficult to get directly from observations. Analysis of observations in Faraday depth helps to constrain the relative distribution of structures. However, the distribution in Faraday depth does not necessarily reflect the true physical distribution \citep[see discussion in][]{brentjens05}. Magnetic field reversals and inhomogeneous distribution of regions (i) that only emit; (ii) with only Faraday rotation; and (iii) with a combination of both, 
give a rise to a discrepancy between the two distributions. Full radiative transfer simulations are needed to disentangle
this problem. 

\begin{figure}[tb]
\centering \includegraphics[width=.48\textwidth]{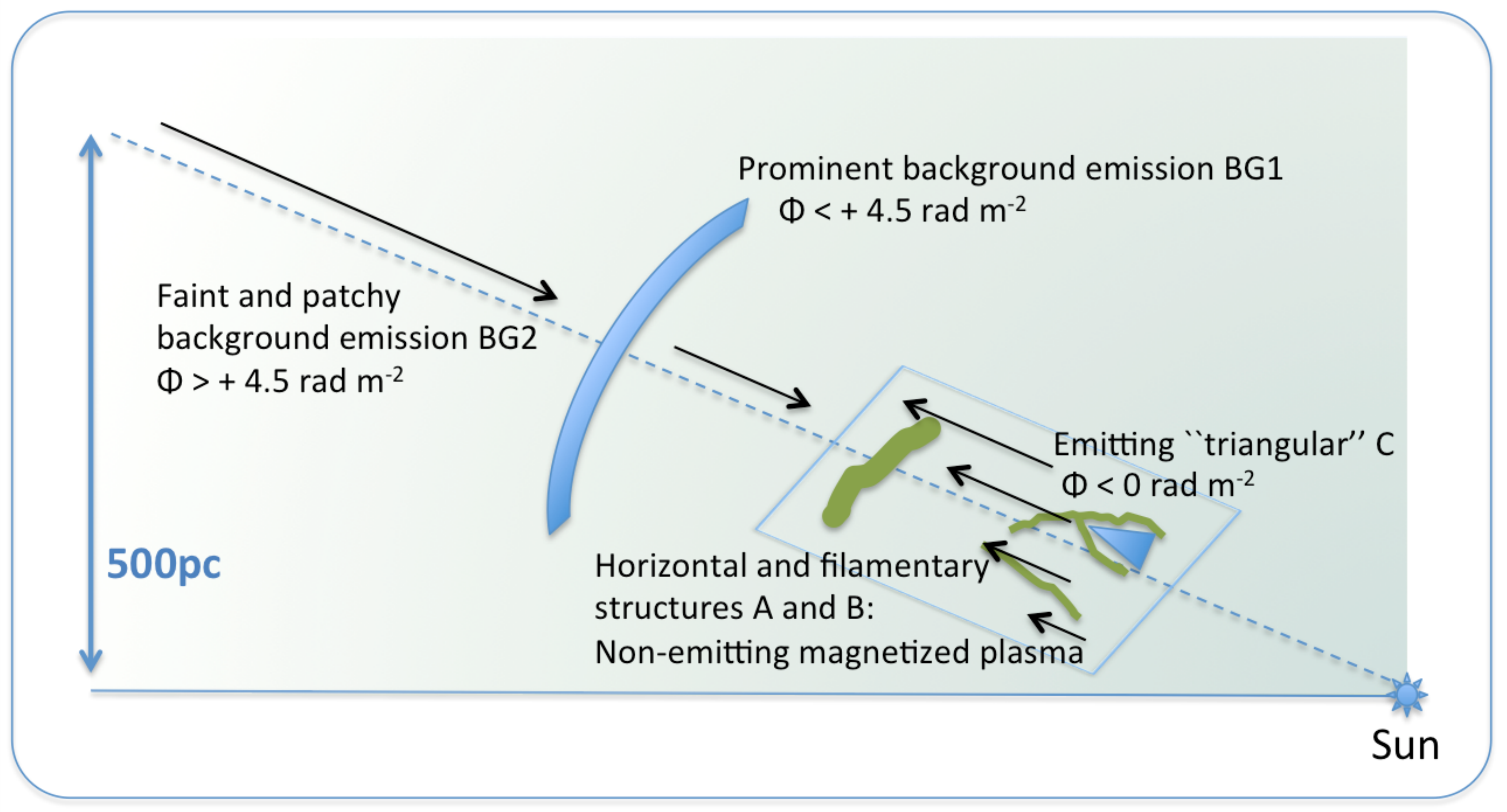}
\caption{A cartoon of the possible layout of polarized emission and magnetic fields towards the 3C196 field. 
Solid arrows indicate orientation and strength of $B_\parallel$. For display purposes, the filamentary structures are shown along the viewing direction in this sideways projection.}
\label{fig:cartoon}
\end{figure}

To visualize the complexity of the problem, we have constructed a simple physical model (see Fig.~\ref{fig:cartoon}), which we believe contains the essential ingredients to explore the conditions in the ISM that are required to explain observed structures in the 3C196 field. Observed features are organized in four distinctive groups: (i) \emph{\textup{filaments A and B}}: relatively straight filamentary structures running from north to south at Faraday depths around $+0.5~{\rm rad~m^{-2}}$ and parallel to the Galactic plane; (ii) \emph{\textup{diffuse emission BG1}}: prominent diffuse background emission at Faraday depths from +1.0 to $+4.5~{\rm rad~m^{-2}}$, showing a large-scale gradient;  (iii) \emph{\textup{diffuse emission BG2}}: very faint patchy emission at Faraday depths from +4.5 to $+7.5~{\rm rad~m^{-2}}$ and at negative Faraday depths from $-3$ to $-0.5~{\rm rad~m^{-2}}$ (iv) a \emph{\textup{horizontal structure}}
located in the upper north of the field; and (v)  a \emph{\textup{``triangular'' feature C}} in the south-west.

We  now describe and discuss our crude model. At first,  we assume that dominant orientation of $B_\parallel$ is towards us. This is consistent with the observed, mostly positive, values of Faraday depth. Based on Eq.~\ref{eq:fardep}, assuming a path length of $1.5~{\rm kpc}$, taking estimated values from Sec.~\ref{sec:genprop} of $\left<n_e\right>\simeq0.03~{\rm cm^{-3}}$ and $\left<B_\parallel\right>\simeq0.3~{\rm \mu G}$, and including a vertical gradient in electron density \footnote{The vertical electron density gradient can be described as $n_e(z)=n_{e,0}e^{-\frac{z}{h}}$, where $n_{e,0}$ is  a mid-plane electron density and $h$ is a scale height \citep[e.g.][and references therein]{haffner09,schnitzeler12}. In our calculation, we normalize it to give $\left<n_e\right>\simeq0.03~{\rm cm^{-3}}$.} , we expect to see polarized emission at Faraday depths up to $\Phi\lesssim+10~{\rm rad~m^{-2}}$. This is in an agreement with our observations, where we have detected faint patchy structures up to $+8~{\rm rad~m^{-2}}$. Simulated RM profiles towards the 3C196 field \citep[$l=+171$, $b=+33^\circ$; see Fig. 12 in][]{sun08}, based on the regular double-torus halo field and disk field models described in \citet{sun08}, also agree with our observations. These profiles predict RM values up to $\lesssim+10~{\rm rad~m^{-2}}$.

At the negative side of the Faraday spectrum, structures are only observed  up to $-3~{\rm rad~m^{-2}}$. Under the same assumptions, but with the magnetic field pointing away from us, the thickness of the negative Faraday rotating region appears to be much smaller than the positive:  Eq.~\ref{eq:fardep} gives $L\simeq250~{\rm pc}$. 

\begin{figure*}[!phtb]
\centering \includegraphics[width=.33\textwidth]{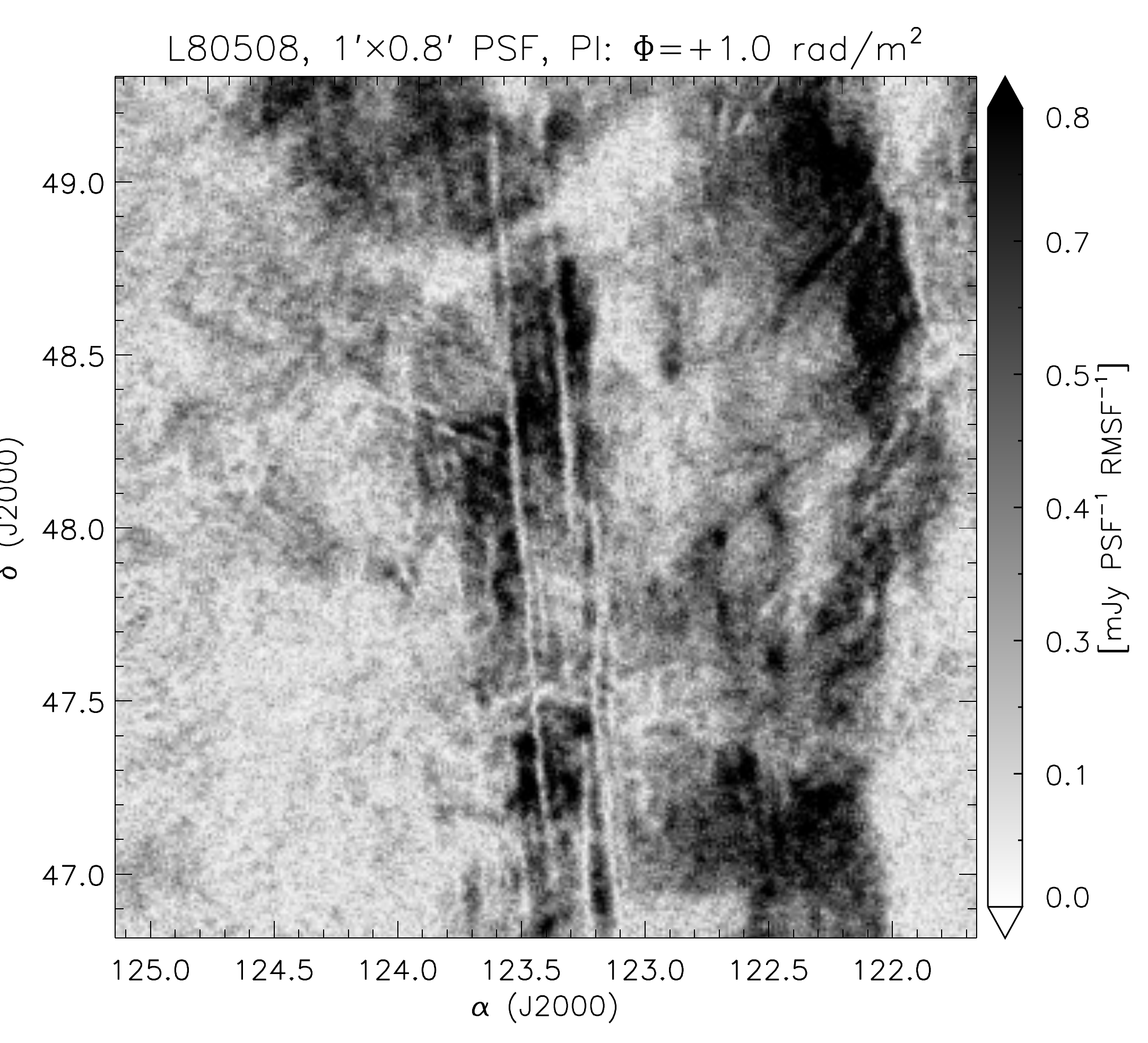}
\centering \includegraphics[width=.33\textwidth]{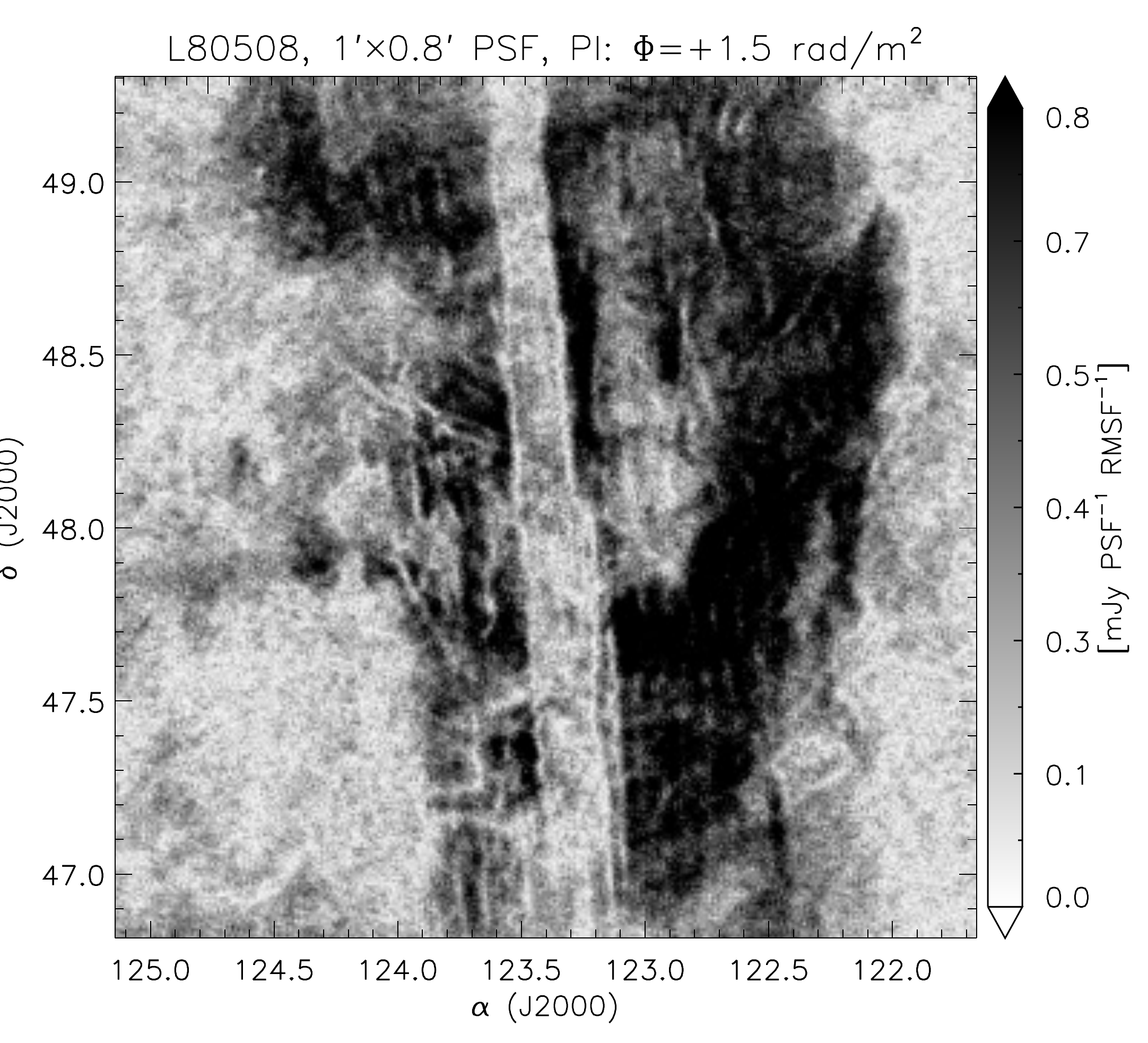}
\centering \includegraphics[width=.33\textwidth]{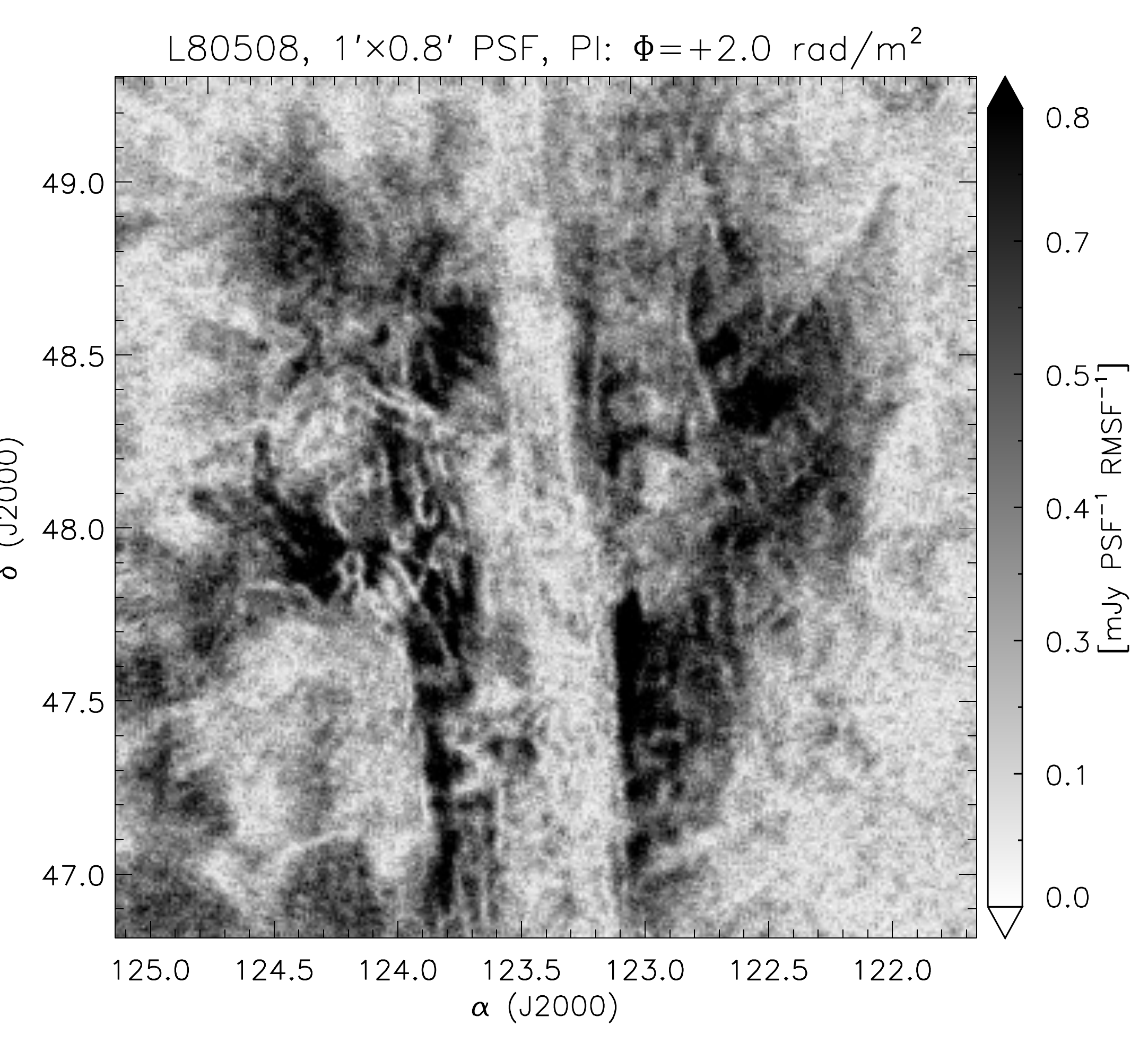}
\centering \includegraphics[width=.33\textwidth]{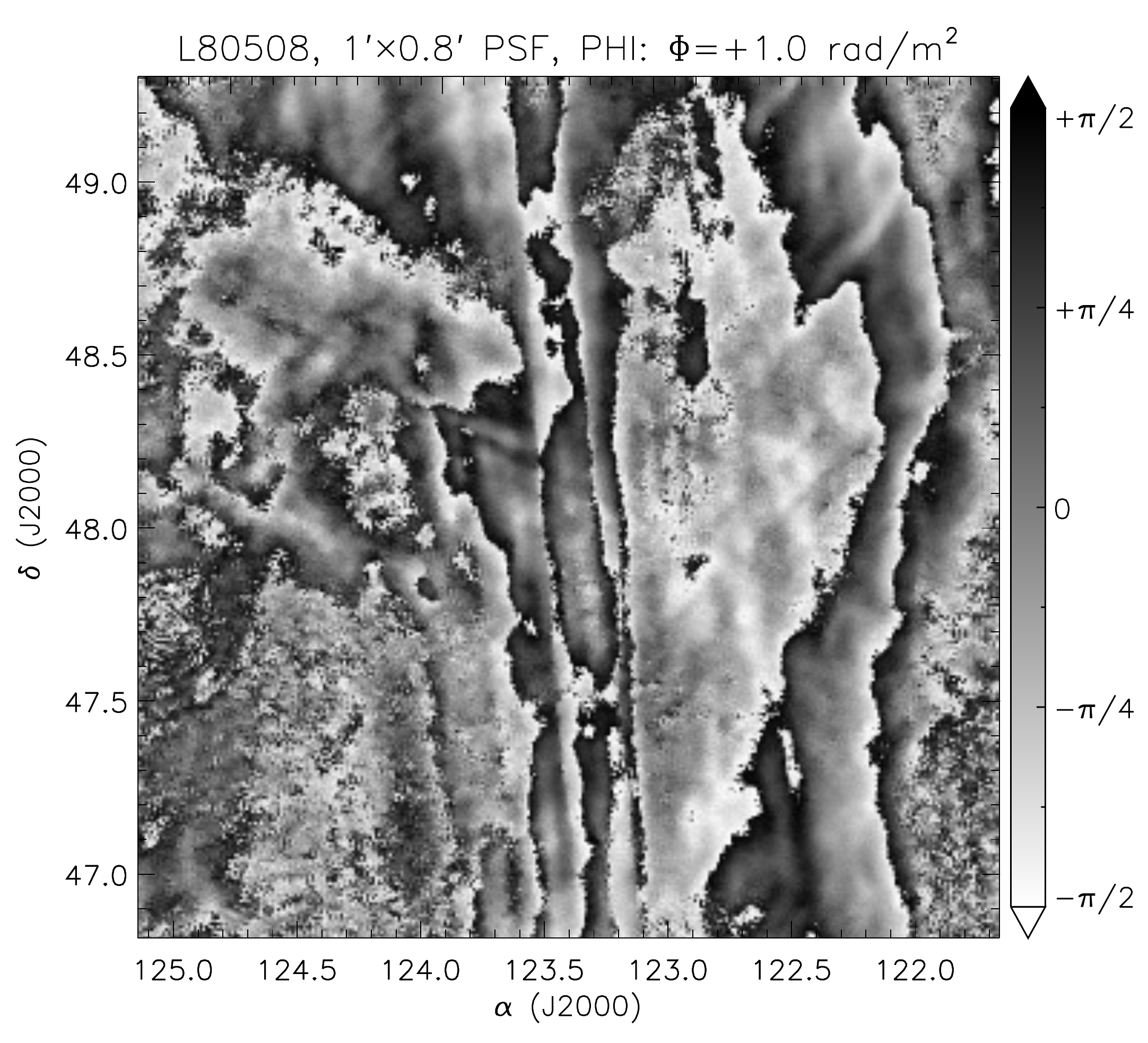}
\centering \includegraphics[width=.33\textwidth]{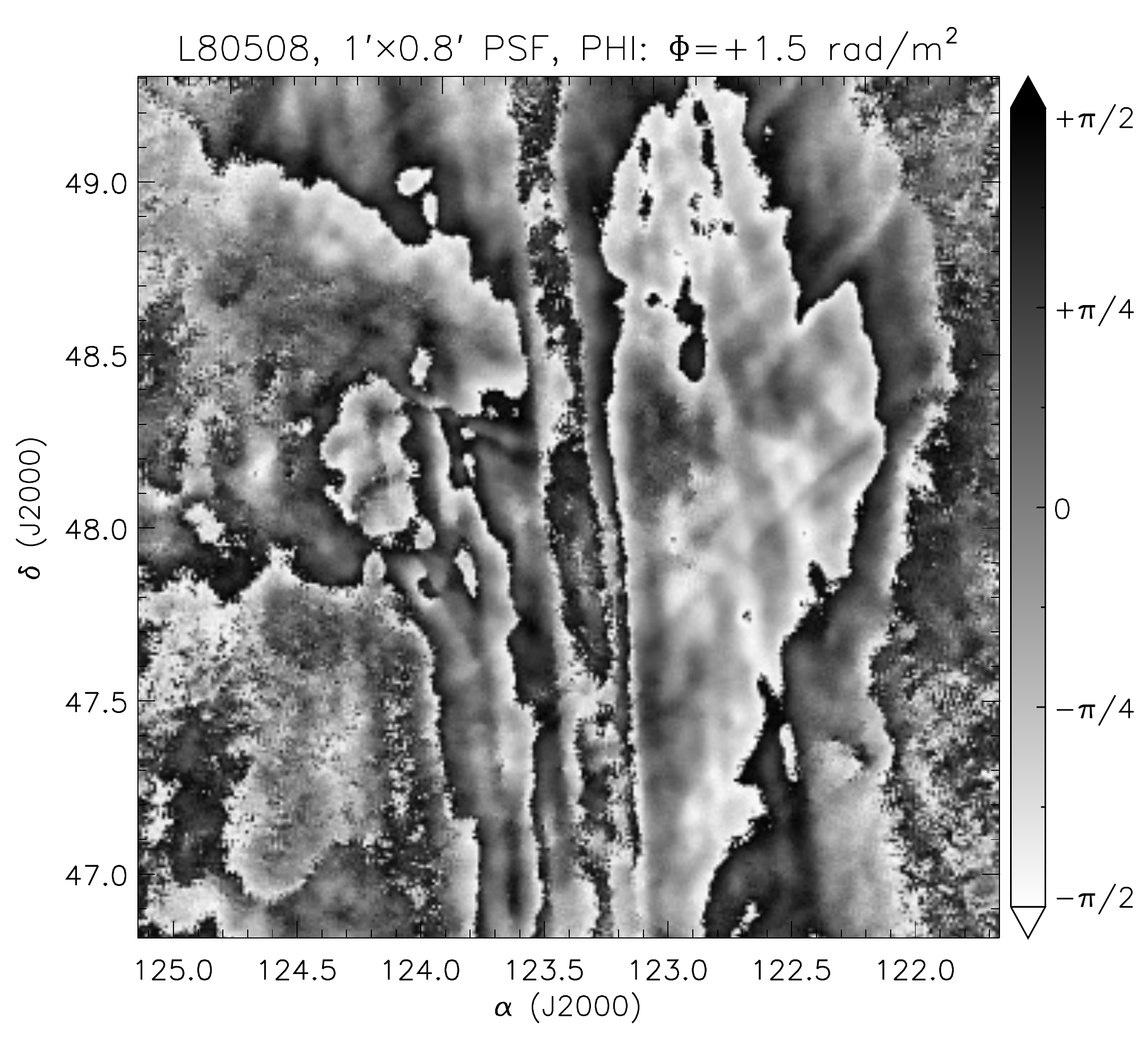}
\centering \includegraphics[width=.33\textwidth]{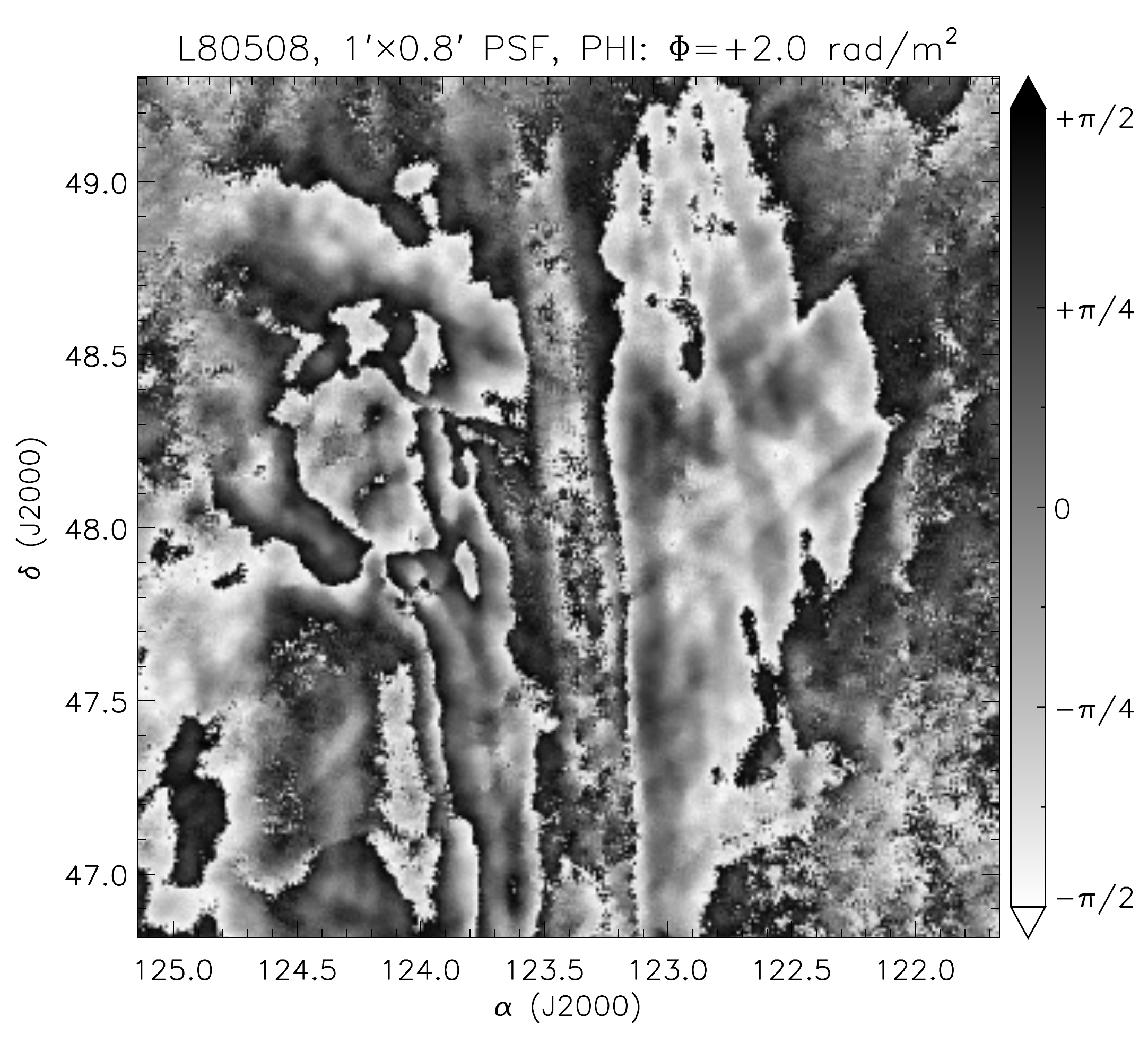}
\caption{Examples of higher resolution images of the 3C196 field in polarized intensity (PI) and polarization angle (PHI) given at Faraday depths
of +1.0, +1.5, and +2.0~${\rm rad~m^{-2}}$. Images show the central $2.5^\circ\times2.5^\circ$ of the field. During the imaging we used baselines between 10 and 4000 wavelengths, providing a frequency-independent resolution of about 1 arcmin.}
\label{fig:highres}
\end{figure*}

A relative distribution of the observed features is as follows. Filaments seem to be displaced from the background diffuse emission BG1 by only $\sim1.5~{\rm rad~m^{-2}}$. Their brightness is comparable to the background emission that shows a very uniform large-scale structure of polarization angle on both sides of the filaments (see high-resolution images in Fig.~\ref{fig:highres}). We therefore conclude that filaments are non-emitting magnetized plasma located in front of the prominent background BG1. 

The Faraday depth of the filaments is smaller than that of the background emission. This implies $B_\parallel$ with an orientation in the opposite direction to the background. Parts of the background emission are derotated within the filaments, giving a deficit in Faraday depth compared to the other parts of the background emission.

Prominent background emission BG1 is closer to us than the fainter background emission BG2, and is probably a very nearby component. BG1 emission might be associate with the synchrotron emission coming somewhere from the edge of the Local Bubble, which is estimated to have a radius of approximately $200~{\rm pc}$ \citep[see e.g.][]{sun08}. To be consistent with our observations, in front of this emission there must be a non-emitting magnetized plasma, with $B_\parallel$  orientated towards us. 
 
The same orientation of $B_\parallel$ we also expect for a region associated with the fainter emission BG2. This emission is the most distant component and probably reflects the global emission of the thick disk. Along the way, two types of depolarization occur \citep[for an overview,][ and references therein]{haverkorn04b}, resulting in the observed very faint and patchy emission. The first is wavelength-independent depolarization, due to turbulent magnetic fields in the ISM, and the second is due to differential Faraday rotation occurring in  regions where emitting and Faraday rotating areas are mixed.
 
The horizontal structure in the north, at negative Faraday depths, does not have any positive Faraday depth components along the line of sight and, therefore, it is a region with a dominant $B_\parallel$ component orientated away from us. This horizontal structure also seems to be spatially a part of the large-scale gradient observed in BG1. If this is true, in front of the synchrotron emitting region BG1, there is one layer of a non-emitting  magnetized plasma with $B_\parallel$ orientated towards us and one layer with $B_\parallel$ orientated  away from us. The second region should have a large-scale gradient in $B_\parallel$.  Since the strength of $B_\parallel$ along the gradient is different, some part of the background emission appears at positive and some at negative Faraday depths. Then, in this case, the observed filamentary structures and the horizontal structure in the north reflect the regions with an excess in thermal electron density and/or magnetic field strength. However, the filaments should be located in the front of all other components since they cast shadows both across emission observed at positive and negative Faraday depths.

Finally, Faraday spectrum at a location of the negative triangular structure C peaks both at positive and negative Faraday depths. Positive Faraday components are associated with the background emission BG1 and BG2. The negative triangular component is probably associate with an additional emitting region, located somewhere in front of the background emission BG1 and BG2. We tried to incorporate all of these  into our model.  A cartoon of this model is shown in Fig.~\ref{fig:cartoon}.
 
\subsection{Possible associations with the filamentary structure A and its proper motion}
The elongated straight filament A running in the north-to-south direction is displaced by $\sim1.5~{\rm rad~m^{-2}}$
in Faraday depth relative to the surrounding background emission, suggesting a magneto-ionic structure located in front of
the bulk of the emission. It is interesting to compare it to other linear ISM features that have been reported in the literature. \citet{mccullough01} discovered an extremely narrow, linear, ionized filament in the ISM, which resembles 
some characteristics of our filament at Faraday depth $+0.5~{\rm rad~m^{-2}}$. Their filament is $~2.5^\circ$ long,
$~20''$ wide and has an H$\alpha$ surface brightness of $I_{H_\alpha}\approx0.5~{\rm R}$. It is located at a comparable Galactic
latitude ($b=+38.5^\circ$) to the 3C196 field. \citet{mccullough01} argued  that the most probable origin of the filament 
is an ionized trail left by photoionization from a star or compact object. As an alternative, but unlikely scenario, they
 mentioned an extremely low density nearby stellar jet, or an unusually linear filament associated with a large-scale
nearby bubble, or an ionized trail left by mechanical input from a star or compact object.
They also argued that this kind of filament would show a displacement in Faraday 
depth against the background radio emission, regardless of its 
true nature of origin. 

Unfortunately, up to now there are no observations in the radio that can support this prediction. This is similarly true for our filament. Available data in H$\alpha$ of 3C196 field do not have the angular resolution needed to resolve the filament. For the purpose of further discussion, we  assume that our filament has the H$\alpha$ surface brightness of the \citet{mccullough01} filament, and
then estimate the thermal electron density of the filament and its distance.

Combining Eq.~\ref{eq:em}~\&~\ref{eq:fardep}, we find that the thermal electron density of the filament is
\begin{equation}
\frac{n_e}{[\rm cm^{-3}]}=0.81B_\parallel \frac{EM}{RM}\frac{[\rm rad~m^{-2}]}{\rm [\mu G][pc~cm^{-6}]},
\end{equation}
where to calculate $n_e$ we need to assume the strength of $B_\parallel$. Once that $n_e$ is known,
one can get $dl$ from Eq.~\ref{eq:em}~or~\ref{eq:fardep} and then estimate a distance, $d$, to the filament
based on its angular diameter, $D_\theta=9.4'$, and assuming axial symmetry. 

Taking $I_{H_\alpha}=0.5~{\rm R}$ and $RM=1.5~{\rm rad~m^{-2}}$ with the results from the previous sections, $T=10^4~{\rm K}$ and $B_\parallel=0.3~{\rm \mu G}$, we get $EM=1.37~{\rm pc~cm^{-6}}$, $n_e=0.2~{\rm cm^{-3}}$, $dl=28~{\rm pc}$, and $d=20~{\rm kpc}$. The calculated distance is obviously not realistic. The filament cannot be located behind the background emission (see discussion in Sec.~\ref{sec:dist}). This means that some of our assumptions in the calculation are not correct.

First, it is possible that our filamentary structure is not a filament but rather a sheet-like structure observed edge-on. Axial symmetry in that case is not valid, and  the distance based on this assumption would not be valid either. Second, if we keep the assumption about the filamentary nature, then either $B_\parallel$ or $n_e$  or both must be larger. If we assume that our filament is located somewhere within the Local Bubble, i.e. $d<200~{\rm pc}$, this implies that $dl<0.3~{\rm pc}$ and $n_e B_\parallel>6.2~{\rm cm^{-3}\mu G}$. A filament is a region showing an excess in $n_e$ and/or $B_\parallel$ compared to its surroundings. 

For a proper understanding of the nature of the filament, it is crucial to know its distance and proper motion. We assume that the filament is at $50~{\rm pc}$ distance and has a transverse velocity of $50~{\rm km~s^{-1}}$. It should then move $50\cdot10^{-6}~{\rm pc~year^{-1}}$, which is equivalent to $0.2''~{\rm year^{-1}}$. This causes a parallax with an amplitude of about $0.03''$ over the course of one year. Obviously it is easier to measure the filment's proper motion than its distance. We therefore try to measure the first using our 1 arcmin resolution images (see Fig.~\ref{fig:highres}), observed one year apart (L80508 \& L192832 observations).

The edges of the filament are well defined by the depolarization canals seen so clearly in the upper left image in Fig.~\ref{fig:highres}. We use these canals to create a mask and isolate the filament from its surrounding background emission. We repeat this for the two observations and then determine if the position of the filament has changed on the timescale of one year.
We find no change in position between the L80508 \& L192832 observations. The estimated error on our current measurement
is $\sim1''$. Assuming a typical transverse velocity of $50~{\rm km~s^{-1}}$, we can set a lower limit to the distance of $\gtrsim 10{\rm pc}$. 

To achieve the precision needed to measure a real proper motion, we can reduce the noise by averaging the RM cubes from many nights in each of the two seasons. However, for this to work for a very small fraction of the width of these filaments, it is essential to compare the structures at exactly the same Faraday depth. The RM cubes must therefore be aligned in RM to much better accuracy than provided by the CODE ionospheric data. Recently, Brentjens et al. (in prep.) developed a new method, which yields the required accuracy limited only by the available signal within the field of view.  The method cross-correlates the Galactic polarized emission in Faraday depth as a function of time, both within each night and between nights and season.  Initial results suggest that the ionospheric Faraday rotation can be aligned to levels on the order of $0.01~{\rm rad~m^{-2}}$, i.e. a very small fraction of the width of the RMSF. The results of this analysis will be presented in a future paper. 
 
\subsection{The system of depolarization canals}
Figure~\ref{fig:RMmax} shows many organized patterns of linear structures with no or very weak polarized signal. Many of these resemble the depolarization canals first discussed and studied by \citet{haverkorn00,haverkorn04b}. We  also refer to these structures as canals. \citet{haverkorn04b} described two characteristics of these depolarization canals. They are one PSF wide and the polarization angle changes across the canal by $90^\circ$. Mechanisms that create them are as follows. The canals are at the boundary between two regions, which each have similar polarized intensities  but a difference in polarization angle of $\Delta\Theta=(n+1/2)180^\circ$ (where $n=0,1,2,\dots$). This causes almost total depolarization through vector addition, and the resulting canal is by definition one PSF wide. The other mechanism is connected to differential Faraday rotation. A linear gradient in RM produces a long, narrow depolarization canal at a certain value $RM_c$, where $2RM_c\lambda^2=n\pi$. Across every null in the sinc-function, the polarization angle changes by $90^\circ$.
A third possibility is beam depolarization due to the presence of   turbulent magneto-ionic medium, which is extremely
small-scale, in front of a polarized background.  To explore this possibility, we  also made a RM cube with a spatial resolution of 12". At this scale the canals appear resolved, but they do not show an increased polarization signal. We therefore consider this possibility unlikely as a cause for the creation of canals.

We believe that most of the canals are probably of the first kind, that is they are due to beam depolarization due to discontinuity in the polarization angle. The best example are the canals associated with the edges of the prominent filament discussed previously and the spaghetti-like structure located to the east of it (see Fig.~\ref{fig:highres}). They are indeed one PSF wide, the intensity of the emission on either side of the canal is comparable, and there is a $90^\circ$ polarization angle change when crossing the canal. Adopting this mechanism for the creation of a canal there is, however, a clear prediction: the polarized emission cannot go to zero at all frequencies that were used in the creation of the RM cube, which by necessity spans a wide range of frequencies (in our case from 115--175~MHz). The Stokes Q and U signals at the edges of this frequency range should therefore increase again within the canals. Unfortunately the signal-to-noise ratio in our data at this stage is not yet sufficient to confirm this prediction.  We stress that the residual polarized signal within the canals also carries valuable information about the polarization signal emitted between the location of the filament and the observer at any Faraday depth.  In this picture the deepest canals must be closest to the observer. We will return to this interesting aspect  in a future paper when we have aligned and analysed, the RM cubes from the many nights we have accumulated on this field. 

There are also some canals in our observations that are broader than one PSF. One of these is located in the south-west part of the image, along the boundary between a triangular region showing emission at negative Faraday depth and surrounding emission at positive depths (see Fig.~\ref{fig:RMmax}). This type of canal must be either a region in the ISM showing no emission, all along the line of sight, or a region with an extremely turbulent magneto-ionic medium, causing wavelength-independent beam depolarization.

\subsection{Straightness of observed structures and canals}
Without regard to the origin of observed structures and canals, a real puzzle is their extraordinary straightness. Their axial ratio is larger than 100:1. It is possible that they are the results of a fortunate projection of a three-dimensional morphology of the magnetic field (e.g. the folding of magnetic sheets and loops) and the ISM.

Recent magneto-hydrodynamical simulations of the ISM,  by \citet{choi12}, showed that thermal conduction plays an important role in shaping the geometry of structures formed by the thermal instabilities by stabilizing small scales and limiting the size of the smallest condensates. If the magnetic field is uniform and weak ($\sim3~{\rm nG}$), the heat is conducted anisotropically,  primary along magnetic field lines. As a result of this, the thermal instabilities  saturate into very long thin filaments of dense gas aligned with the magnetic field. In the linear regime, this can be explained by the variation of the Field length \citep{field65} with respect to the direction of magnetic field. Isobaric perturbations with wavelength smaller than the Field length do not grow at all. Perpendicular to the magnetic field, the Field length is very small, so very short wavelength perturbations are unstable. Parallel to the magnetic field, the Field length is much longer, so only longer wavelength modes grow. As the perturbations grow non-linear, conduction tends to enforce isothermality along magnetic field lines, leading to long filaments. If the magnetic field is uniform and strong ($\sim3~{\rm \mu G}$), only motions along field lines are allowed, and, moreover, magnetic pressure provides some support in dense regions, resulting in fragmented filaments aligned with the field lines \citep[for more details we refer to][]{choi12}. Therefore, it is possible that filamentary structures 
observed in our data were shaped through thermal instabilities. To get better statistics on observed structures/filaments, and to study their distribution and orientation with regard to the regular magnetic field component, we plan to survey a larger area of the sky. This will then allow us to directly associate observed structures with the results of magneto-hydrodynamical simulations of the ISM.

Another explanation of observed linear structures can be connected to interactions of close-by moving stars with the ISM, as discussed by \citet{mccullough01}. 
Bow shocks around stars and neutron stars (pulsars) can leave long and narrow 
trails in the ISM, because of their large speeds. Or, trails can be left by an ionizing source of sufficiently low ionizing luminosity that its Str\"omgren radius is small compared with the distance it travels in a recombination time. However, to make such trails long in angular size, of a few degrees, these sources should be rather close, less than about $100~{\rm pc}$, and located somewhere within the Local Bubble. We will leave a more detailed study of this possibility for future work.

\section{WSRT observations from 312--381~MHz}
The 3C196 field was also observed with the WSRT at $350~{\rm MHz}$ in 2013 and 2014. These observations form part of a large-area mosaic survey to complement  LOFAR polarization studies. Thus far we have finished  the reduction and analysis of the part, which overlaps with the LOFAR field of view so we can make a comparison with the features seen at LOFAR frequencies.

\begin{figure}[tb]
\centering \includegraphics[width=.45\textwidth]{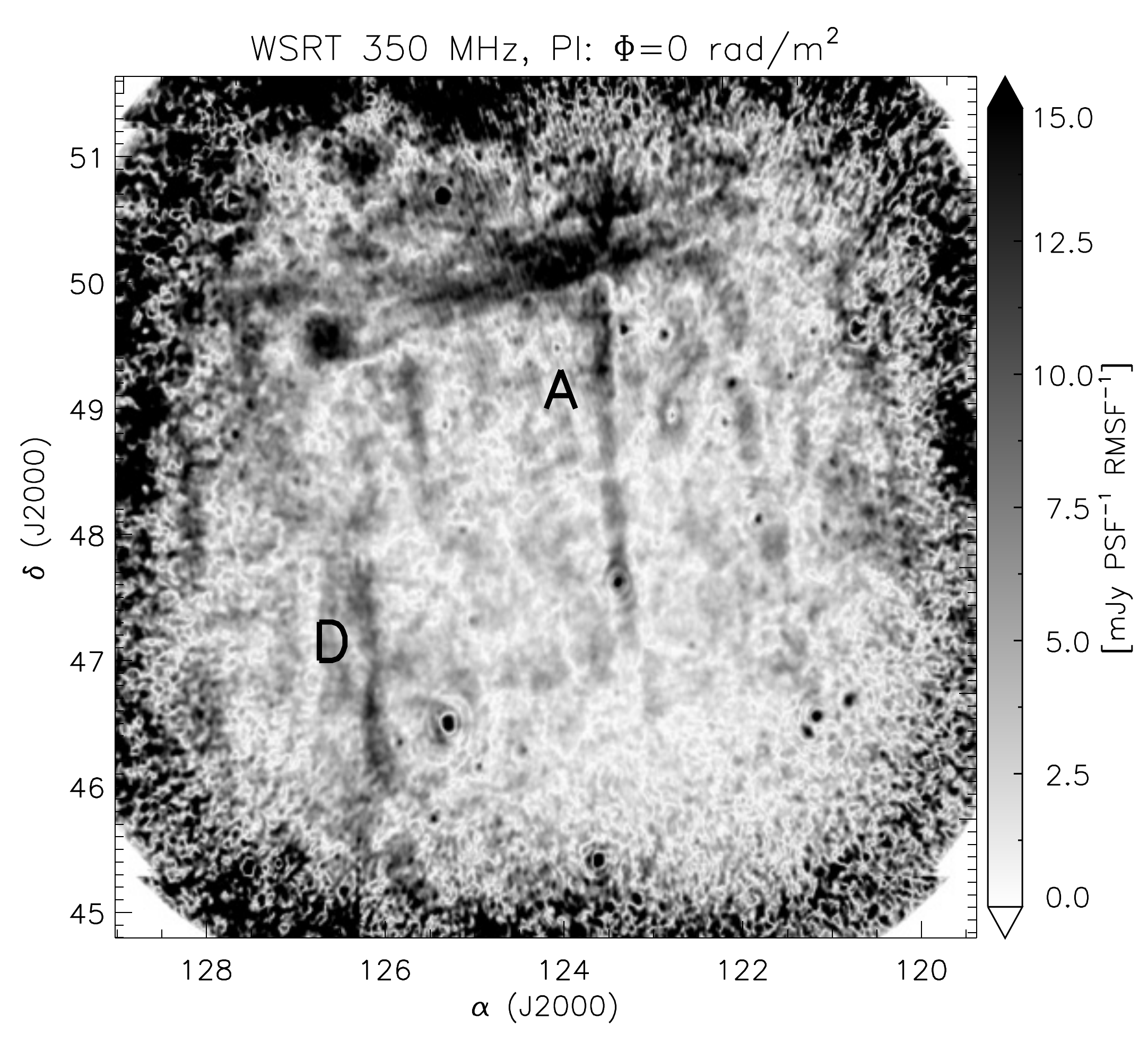}
\caption{An image of the 3C196 field in polarized intensity (PI) observed with the 28 WSRT mosaic pointings at 350 MHz (based on preliminary calibration and analysis).  In making this image, we used only the central 9 of the 28 observed pointings. The image is given at Faraday depth $0~{\rm rad~m^{-2}}$ and has not been corrected for polarization noise bias. It measures $6.9^\circ\times 6.9^\circ$ in size and the PSF is $2.8\times3.8'$.} 
\label{fig:WSRT}
\end{figure}

Figure~\ref{fig:WSRT} shows the WSRT 350 MHz image in polarized intensity at Faraday depth $0~{\rm rad~m^{-2}}$. The RMSF in the WSRT 350 MHz data ($\delta\Phi=11~{\rm rad~m^{-2}}$) is an order of magnitude less than in the LOFAR data. Hence, almost all Faraday depths that show emission in the LOFAR data are captured in this one frame in the WSRT data. The linear filament A passing over 3C196 is clearly detected. It has a surface brightness of $\sim7~{\rm mJy~PSF^{-1}~RMSF^{-1}}$. This is the first time that we see some common morphology in the RM cubes made in the LOFAR  $115-175~{\rm MHz}$ 
and WSRT $315-380~{\rm MHz}$ bands. Diffuse emission surrounding the filament is not as prominent ($\lesssim3~{\rm mJy~PSF^{-1}~RMSF^{-1}}$) as observed with LOFAR (see Fig.~\ref{fig:PI1}) and its morphology is very patchy. 

In addition to the filament A passing over 3C196, we detected yet another straight filamentary structure (marked with D in Fig.~\ref{fig:WSRT}) at the WSRT frequencies. It is located in the south-east part of the image. Its surface brightness is $\sim8~{\rm mJy~PSF^{-1}~RMSF^{-1}}$ and is probably associated with emission at $+2.5~{\rm rad~m^{-2}}$ in the LOFAR RM cubes. The two filaments in the WSRT data seem to have similar orientation.  

Bright horizontal emission ($\sim15~{\rm mJy~PSF^{-1}~RMSF^{-1}}$) located to the north of the filaments in the WSRT data coincides with the region showing weak emission and horizontal depolarization canals at LOFAR frequencies, just below horizontal emission in the north at negative Faraday depths. Some parts of this bright emission in the WSRT data are probably Faraday-thick structures at LOFAR frequencies  and not detectable with LOFAR ($\Delta\Phi>\delta\Phi_{\rm LOFAR}$). 

The different observed morphology of the diffuse emission surrounding the filament in the two different frequency bands 
may be attributed to the poor resolution in Faraday depth at 350 MHz. To test this, we  generated an RM cube at LOFAR frequencies using an RMSF that has a resolution of $\delta\Phi_{350~{\rm MHz}}\approx10~\delta\Phi_{150~{\rm MHz}}$. It is possible that multiple Faraday thin
structures ($\Delta\Phi<\delta\Phi_{\rm 350 MHz}$) detected in the LOFAR low-frequency images  decorrelate 
when we observe them with a much broader RMSF. We find that $\sim50~\%$ of the flux is depolarized, but 
the overall morphology of emission is not much different from the original LOFAR RM cubes. Thus, the lack 
of prominent background emission detected with LOFAR cannot be explained by the much broader RMSF at $350~{\rm MHz}$.

Another possible explanation for the lack of emission in the WSRT 350 MHz data may be related to the missing large-scale structures.
The shortest baseline of the WSRT is 36 m, so at 350 MHz structure on angular scales larger than about a degree is not adequately measured.
As we discussed in Sec.~\ref{sec:dis}, the synchrotron radiation is emitted on large scales, but its polarization is 
altered on smaller scales by Faraday effects. Faraday rotation is much stronger at lower radio frequencies.
We thus expect the missing large-scale emission in polarization to be more prominent at higher frequencies. This
was discussed first by \citet{haverkorn04b}. For the case of a uniformly polarized background propagating through 
a small-scale Faraday screen, the expected offset for Stokes Q and U depends on the wavelength, $\lambda$, and the width of the Faraday depth distribution, $\sigma_\Phi$, as $e^{-2\sigma^2_\Phi\lambda^4}$. In WSRT $350~{\rm MHz}$ observations, this condition is not well satisfied and one needs to take it into account during interpretation of the data \citep[e.g.][]{haverkorn04b,schnitzeler09}. 

To correct for these undetectable large-scale structures, one needs first to observe the same region at the same frequencies with a single-dish telescope with absolute intensity scaling. Then, these large-scale data can be added to the interferometer data \citep{uyaniker98,stanimirovic02}. However, for the WSRT 350 MHz observations this is not possible. We can only estimate it based on other existing absolutely calibrated observations at higher frequencies, e.g. \citet{berkhuijsen63} map of polarized emission at $408~{\rm MHz}$. This frequency is close enough to $350~{\rm MHz}$ to allow a comparison, although the polarized intensity at 408 MHz is expected to be slightly higher due to the larger polarization horizon.

The polarized brightness temperature at $408~{\rm MHz}$ at the location of the 3C196 field is 
$2.1\pm0.4~{\rm K}$ \citep{berkhuijsen63}. Using a power-law spectral index of $-2.5$, this corresponds 
to $\sim3~{\rm K}$ at $350~{\rm MHz}$. If we now smooth our $350~{\rm MHz}$ data to the $2^\circ$
FWHM of \citet{berkhuijsen63} we get $0.08~{\rm K}$. This means that indeed we have a missing 
large-scale component in polarized intensity.

\section{Summary and conclusions}
We  have presented results from LOFAR HBA observations of a field centred on the 3C196 source, taken as part
of the LOFAR-EoR project. We  applied RM synthesis to the data to reveal a very rich morphology of polarized 
emission. Detected structures are spread over a wide range of Faraday depths, ranging from $-3$
to $+8~{\rm rad~m^{-2}}$. Their average brightness temperature is $5-15~{\rm K}$, which is comparable to
the emission in the ELAIS-N1 field and also to a field observed with LOFAR at high Galactic latitudes  \citep{jelic14}.

The most interesting features detected in polarization are (i) a straight and long filamentary structure at Faraday depth $+0.5~{\rm rad~m^{-2}}$, parallel to the Galactic plane and (ii) a system of linear depolarization canals. The first is probably  an ionized structure in the ISM located very close by, somewhere within the Local Bubble. It is displaced by $1.5~{\rm rad~m^{-2}}$ from the surrounding background
emission. This implies that, within the filament, the thermal electron density and magnetic field must satisfy 
$n_e B_\parallel >6.2~{\rm cm^{-3}\mu G}$. With the current accuracy of the position measurement of $~1''$,
we can set a lower limit to its distance of $10~{\rm pc}$, assuming a transverse velocity of $50~{\rm km s^{-1}}$.

The same filament was also observed with the WSRT at $350~{\rm MHz}$. 
This is the first time that we see some common morphology in the RM cubes made in two different frequency
regimes. Interestingly, in the south-east part of the WSRT mosaic there is another filament. 
This filament is probably associated with emission at $+2.5~{\rm rad~m^{-2}}$ in the LOFAR RM cubes. 
Its orientation is the same as the filament located in the centre of the image. This is a tentative indication that these two filaments might be a part of the same large-scale structure in the ISM. 

A follow up study of the filaments discussed in this paper showed their strong correlation with the orientation of the sky 
projected magnetic field component as probed by the \textit{Planck} maps of  the dust emission in polarization \citep{zaroubi15}. 
Dust particles are usually embedded within neutral hydrogen (H{\sc i}) gas, as seen through H{\sc i}/dust correlation at high Galactic 
latitudes \citep{planckXVII}. Therefore, high angular resolution H{\sc i} data of 3C196 field can be very usefully in providing potential kinematic
information about the features studied in this paper. This is also supported by \citet{clark14} analysis of H{\sc i} data from 
the Galactic Arecibo L-Band Feed Array H{\sc i} (GALFA-H{\sc i})  Survey and the Parkes Galactic All Sky Survey (GASS), which  
revealed many slender, linear H{\sc i} features (``fibers'') that extent for many degrees at high Galactic latitude. 
These fibers are orientated along the interstellar magnetic field as probed by starlight polarization and are most likely a component 
of the Local Bubble wall.

Most of the depolarization canals detected in our LOFAR data are due to beam depolarization. However, some of them might
be associated with regions in the ISM that show a very turbulent magneto-ionic medium, causing wavelength-independent beam depolarization. 
Without regard to their origin, the straightness of these canals is a real puzzle. Are they a projection effect of complicated morphology
in the magnetic field and the ISM, or are they associated with fast-moving close-by stars interacting with the ISM? This needs to
be investigated further. Planned observations covering a much larger area of this region, along with observations at other wavelengths,
will help towards this goal. 

Results presented in this paper are also of concern for epoch of reionization experiments. The instrumentally polarized response 
of an interferometer needs to be calibrated to a small fraction of a percent to limit leakage of polarization signals to
levels of a few mK. Otherwise, the leakage can contaminate the cosmological 21-cm signal in total intensity  \citep[e.g.][]{jelic10}.
The contamination depends strongly on the brightness temperature of polarized emission, its morphology, and its spread in Faraday depth. 
In the case of the polarized emission detected in the 3C196 field, which is limited to small Faraday depths 
($|\Phi|\lesssim10~{\rm rad~m^{-2}}$), this is not a big concern. \citet{asad15} showed that in the spherical power spectrum the leakage 
power of emission presented in this paper is lower than the cosmological signal at $k<0.3~{\rm Mpc^{-1}}$. Moreover,
there is a window in the cylindrical power spectrum where the cosmological signal dominates over the leakage. 

\begin{acknowledgements} 
We thank an anonymous referee for useful comments that improved the manuscript. VJ would like to thank the Netherlands Foundation for Scientific Research (NWO) for financial support through VENI grant 639.041.336.  AGdB, ARO, MM, SY and VNP acknowledge support by the ERC for project 339743 (LOFARCORE). LVEK, AG, KMBA and HKV acknowledge the financial support from the European Research Council under ERC-Starting Grant FIRSTLIGHT - 258942. FBA acknowledges the support of the Royal Society for a University Research Fellowship. GH acknowledges funding from the People Programme (Marie Curie Actions) of the European Union's Seventh Framework Programme (FP7/2007--2013) under REA grant agreement no.\ 327999. ITI was supported by the Science and Technology Facilities
Council [grant number ST/L000652/1]. The Low-Frequency Array (LOFAR) was designed and constructed by ASTRON, the Netherlands Institute for Radio Astronomy, and has facilities in several countries, which are owned by various parties (each with their own funding sources) and are collectively operated by the International LOFAR Telescope (ILT) foundation under a joint scientific policy. The Westerbork Synthesis Radio Telescope (WSRT) is operated by the ASTRON  with support from the Netherlands Foundation for Scientific Research (NWO).  The Wisconsin H-Alpha Mapper is funded by the National Science Foundation.
\end{acknowledgements} 

\bibliographystyle{aa}
\bibliography{reflist_polar}

\begin{thebibliography}{66}
\expandafter\ifx\csname natexlab\endcsname\relax\def\natexlab#1{#1}\fi

\bibitem[{{Akahori} {et~al.}(2013){Akahori}, {Ryu}, {Kim}, \&
  {Gaensler}}]{akahori13}
{Akahori}, T., {Ryu}, D., {Kim}, J., \& {Gaensler}, B.~M. 2013, \apj, 767, 150

\bibitem[{{Asad} {et~al.}(2015){Asad}, {Koopmans}, {Jeli{\'c}}, {Pandey},
  {Ghosh}, {Abdalla}, {Bernardi}, {Brentjens}, {de Bruyn}, {Bus}, {Ciardi},
  {Chapman}, {Daiboo}, {Fernandez}, {Harker}, {Iliev}, {Jensen},
  {Martinez-Rubi}, {Mellema}, {Mevius}, {Offringa}, {Patil}, {Schaye},
  {Thomas}, {van der Tol}, {Vedantham}, {Yatawatta}, \& {Zaroubi}}]{asad15}
{Asad}, K.~M.~B., {Koopmans}, L.~V.~E., {Jeli{\'c}}, V., {et~al.} 2015, \mnras,
  451, 3709

\bibitem[{{Baccigalupi} {et~al.}(2001){Baccigalupi}, {Burigana}, {Perrotta},
  {De Zotti}, {La Porta}, {Maino}, {Maris}, \& {Paladini}}]{baccigalupi01}
{Baccigalupi}, C., {Burigana}, C., {Perrotta}, F., {et~al.} 2001, \aap, 372, 8

\bibitem[{{Berkhuijsen} \& {Brouw}(1963)}]{berkhuijsen63}
{Berkhuijsen}, E.~M. \& {Brouw}, W.~N. 1963, \bain, 17, 185

\bibitem[{{Berkhuijsen} {et~al.}(2006){Berkhuijsen}, {Mitra}, \&
  {Mueller}}]{berkhuijsen06}
{Berkhuijsen}, E.~M., {Mitra}, D., \& {Mueller}, P. 2006, Astronomische
  Nachrichten, 327, 82

\bibitem[{{Bernardi} {et~al.}(2009){Bernardi}, {de Bruyn}, {Brentjens},
  {Ciardi}, {Harker}, {Jeli{\'c}}, {Koopmans}, {Labropoulos}, {Offringa},
  {Pandey}, {Schaye}, {Thomas}, {Yatawatta}, \& {Zaroubi}}]{bernardi09}
{Bernardi}, G., {de Bruyn}, A.~G., {Brentjens}, M.~A., {et~al.} 2009, \aap,
  500, 965

\bibitem[{{Bernardi} {et~al.}(2010){Bernardi}, {de Bruyn}, {Harker},
  {Brentjens}, {Ciardi}, {Jeli{\'c}}, {Koopmans}, {Labropoulos}, {Offringa},
  {Pandey}, {Schaye}, {Thomas}, {Yatawatta}, \& {Zaroubi}}]{bernardi10}
{Bernardi}, G., {de Bruyn}, A.~G., {Harker}, G., {et~al.} 2010, \aap, 522, A67

\bibitem[{{Bernardi} {et~al.}(2013){Bernardi}, {Greenhill}, {Mitchell}, {Ord},
  {Hazelton}, {Gaensler}, {de Oliveira-Costa}, {Morales}, {Udaya Shankar},
  {Subrahmanyan}, {Wayth}, {Lenc}, {Williams}, {Arcus}, {Arora}, {Barnes},
  {Bowman}, {Briggs}, {Bunton}, {Cappallo}, {Corey}, {Deshpande}, {deSouza},
  {Emrich}, {Goeke}, {Herne}, {Hewitt}, {Johnston-Hollitt}, {Kaplan}, {Kasper},
  {Kincaid}, {Koenig}, {Kratzenberg}, {Lonsdale}, {Lynch}, {McWhirter},
  {Morgan}, {Oberoi}, {Pathikulangara}, {Prabu}, {Remillard}, {Rogers},
  {Roshi}, {Salah}, {Sault}, {Srivani}, {Stevens}, {Tingay}, {Waterson},
  {Webster}, {Whitney}, {Williams}, \& {Wyithe}}]{bernardi13}
{Bernardi}, G., {Greenhill}, L.~J., {Mitchell}, D.~A., {et~al.} 2013, \apj,
  771, 105

\bibitem[{{Brentjens} \& {de Bruyn}(2005)}]{brentjens05}
{Brentjens}, M.~A. \& {de Bruyn}, A.~G. 2005, \aap, 441, 1217

\bibitem[{{Choi} \& {Stone}(2012)}]{choi12}
{Choi}, E. \& {Stone}, J.~M. 2012, \apj, 747, 86

\bibitem[{{Clark} {et~al.}(2014){Clark}, {Peek}, \& {Putman}}]{clark14}
{Clark}, S.~E., {Peek}, J.~E.~G., \& {Putman}, M.~E. 2014, \apj, 789, 82

\bibitem[{{Cordes} \& {Lazio}(2002)}]{cordes02}
{Cordes}, J.~M. \& {Lazio}, T.~J.~W. 2002, ArXiv Astrophysics e-prints
  [\eprint{astro-ph/0207156}]

\bibitem[{{Cordes} \& {Lazio}(2003)}]{cordes03}
{Cordes}, J.~M. \& {Lazio}, T.~J.~W. 2003, ArXiv Astrophysics e-prints
  [\eprint{astro-ph/0301598}]

\bibitem[{{Field}(1965)}]{field65}
{Field}, G.~B. 1965, \apj, 142, 531

\bibitem[{{Finkbeiner}(2003)}]{finkbeiner03}
{Finkbeiner}, D.~P. 2003, \apjs, 146, 407

\bibitem[{{Gaensler} {et~al.}(2008){Gaensler}, {Madsen}, {Chatterjee}, \&
  {Mao}}]{gaensler08}
{Gaensler}, B.~M., {Madsen}, G.~J., {Chatterjee}, S., \& {Mao}, S.~A. 2008,
  \pasa, 25, 184

\bibitem[{{Ghosh} {et~al.}(2012){Ghosh}, {Prasad}, {Bharadwaj}, {Ali}, \&
  {Chengalur}}]{ghosh12}
{Ghosh}, A., {Prasad}, J., {Bharadwaj}, S., {Ali}, S.~S., \& {Chengalur}, J.~N.
  2012, \mnras, 426, 3295

\bibitem[{{Guzm{\'a}n} {et~al.}(2011){Guzm{\'a}n}, {May}, {Alvarez}, \&
  {Maeda}}]{guzman11}
{Guzm{\'a}n}, A.~E., {May}, J., {Alvarez}, H., \& {Maeda}, K. 2011, \aap, 525,
  A138

\bibitem[{{Haffner} {et~al.}(2009){Haffner}, {Dettmar}, {Beckman}, {Wood},
  {Slavin}, {Giammanco}, {Madsen}, {Zurita}, \& {Reynolds}}]{haffner09}
{Haffner}, L.~M., {Dettmar}, R.-J., {Beckman}, J.~E., {et~al.} 2009, Reviews of
  Modern Physics, 81, 969

\bibitem[{{Haffner} {et~al.}(2003){Haffner}, {Reynolds}, {Tufte}, {Madsen},
  {Jaehnig}, \& {Percival}}]{haffner03}
{Haffner}, L.~M., {Reynolds}, R.~J., {Tufte}, S.~L., {et~al.} 2003, \apjs, 149,
  405

\bibitem[{{Haslam} {et~al.}(1981){Haslam}, {Klein}, {Salter}, {Stoffel},
  {Wilson}, {Cleary}, {Cooke}, \& {Thomasson}}]{haslam81}
{Haslam}, C.~G.~T., {Klein}, U., {Salter}, C.~J., {et~al.} 1981, \aap, 100, 209

\bibitem[{{Haslam} {et~al.}(1982){Haslam}, {Salter}, {Stoffel}, \&
  {Wilson}}]{haslam82}
{Haslam}, C.~G.~T., {Salter}, C.~J., {Stoffel}, H., \& {Wilson}, W.~E. 1982,
  \aaps, 47, 1

\bibitem[{{Haverkorn}(2015)}]{haverkorn15}
{Haverkorn}, M. 2015, in Astrophysics and Space Science Library, Vol. 407,
  Astrophysics and Space Science Library, ed. A.~{Lazarian}, E.~M. {de Gouveia
  Dal Pino}, \& C.~{Melioli}, 483

\bibitem[{{Haverkorn} {et~al.}(2000){Haverkorn}, {Katgert}, \& {de
  Bruyn}}]{haverkorn00}
{Haverkorn}, M., {Katgert}, P., \& {de Bruyn}, A.~G. 2000, \aap, 356, L13

\bibitem[{{Haverkorn} {et~al.}(2003{\natexlab{a}}){Haverkorn}, {Katgert}, \&
  {de Bruyn}}]{haverkorn03a}
{Haverkorn}, M., {Katgert}, P., \& {de Bruyn}, A.~G. 2003{\natexlab{a}}, \aap,
  403, 1031

\bibitem[{{Haverkorn} {et~al.}(2003{\natexlab{b}}){Haverkorn}, {Katgert}, \&
  {de Bruyn}}]{haverkorn03b}
{Haverkorn}, M., {Katgert}, P., \& {de Bruyn}, A.~G. 2003{\natexlab{b}}, \aap,
  404, 233

\bibitem[{{Haverkorn} {et~al.}(2004){Haverkorn}, {Katgert}, \& {de
  Bruyn}}]{haverkorn04b}
{Haverkorn}, M., {Katgert}, P., \& {de Bruyn}, A.~G. 2004, \aap, 427, 549

\bibitem[{{Heiles} \& {Haverkorn}(2012)}]{heiles12}
{Heiles}, C. \& {Haverkorn}, M. 2012, \ssr, 166, 293

\bibitem[{{Iacobelli} {et~al.}(2013){Iacobelli}, {Haverkorn}, {Orr{\'u}},
  {Pizzo}, {Anderson}, {Beck}, {Bell}, {Bonafede}, {Chyzy}, {Dettmar},
  {En{\ss}lin}, {Heald}, {Horellou}, {Horneffer}, {Jurusik}, {Junklewitz},
  {Kuniyoshi}, {Mulcahy}, {Paladino}, {Reich}, {Scaife}, {Sobey},
  {Sotomayor-Beltran}, {Alexov}, {Asgekar}, {Avruch}, {Bell}, {van Bemmel},
  {Bentum}, {Bernardi}, {Best}, {B{\i}rzan}, {Breitling}, {Broderick}, {Brouw},
  {Br{\"u}ggen}, {Butcher}, {Ciardi}, {Conway}, {de Gasperin}, {de Geus},
  {Duscha}, {Eisl{\"o}ffel}, {Engels}, {Falcke}, {Fallows}, {Ferrari},
  {Frieswijk}, {Garrett}, {Grie{\ss}meier}, {Gunst}, {Hamaker}, {Hassall},
  {Hessels}, {Hoeft}, {H{\"o}randel}, {Jelic}, {Karastergiou}, {Kondratiev},
  {Koopmans}, {Kramer}, {Kuper}, {van Leeuwen}, {Macario}, {Mann}, {McKean},
  {Munk}, {Pandey-Pommier}, {Polatidis}, {R{\"o}ttgering}, {Schwarz}, {Sluman},
  {Smirnov}, {Stappers}, {Steinmetz}, {Tagger}, {Tang}, {Tasse}, {Toribio},
  {Vermeulen}, {Vocks}, {Vogt}, {van Weeren}, {Wise}, {Wucknitz}, {Yatawatta},
  {Zarka}, \& {Zensus}}]{iacobelli13}
{Iacobelli}, M., {Haverkorn}, M., {Orr{\'u}}, E., {et~al.} 2013, \aap, 558, A72

\bibitem[{{Jeli{\'c}} {et~al.}(2014){Jeli{\'c}}, {de Bruyn}, {Mevius},
  {Abdalla}, {Asad}, {Bernardi}, {Brentjens}, {Bus}, {Chapman}, {Ciardi},
  {Daiboo}, {Fernandez}, {Ghosh}, {Harker}, {Jensen}, {Kazemi}, {Koopmans},
  {Labropoulos}, {Martinez-Rubi}, {Mellema}, {Offringa}, {Pandey}, {Patil},
  {Thomas}, {Vedantham}, {Veligatla}, {Yatawatta}, {Zaroubi}, {Alexov},
  {Anderson}, {Avruch}, {Beck}, {Bell}, {Bentum}, {Best}, {Bonafede},
  {Bregman}, {Breitling}, {Broderick}, {Brouw}, {Br{\"u}ggen}, {Butcher},
  {Conway}, {de Gasperin}, {de Geus}, {Deller}, {Dettmar}, {Duscha},
  {Eisl{\"o}ffel}, {Engels}, {Falcke}, {Fallows}, {Fender}, {Ferrari},
  {Frieswijk}, {Garrett}, {Grie{\ss}meier}, {Gunst}, {Hamaker}, {Hassall},
  {Haverkorn}, {Heald}, {Hessels}, {Hoeft}, {H{\"o}randel}, {Horneffer}, {van
  der Horst}, {Iacobelli}, {Juette}, {Karastergiou}, {Kondratiev}, {Kramer},
  {Kuniyoshi}, {Kuper}, {van Leeuwen}, {Maat}, {Mann}, {McKay-Bukowski},
  {McKean}, {Munk}, {Nelles}, {Norden}, {Paas}, {Pandey-Pommier}, {Pietka},
  {Pizzo}, {Polatidis}, {Reich}, {R{\"o}ttgering}, {Rowlinson}, {Scaife},
  {Schwarz}, {Serylak}, {Smirnov}, {Steinmetz}, {Stewart}, {Tagger}, {Tang},
  {Tasse}, {ter Veen}, {Thoudam}, {Toribio}, {Vermeulen}, {Vocks}, {van
  Weeren}, {Wijers}, {Wijnholds}, {Wucknitz}, \& {Zarka}}]{jelic14}
{Jeli{\'c}}, V., {de Bruyn}, A.~G., {Mevius}, M., {et~al.} 2014, \aap, 568,
  A101

\bibitem[{{Jeli{\'c}} {et~al.}(2010){Jeli{\'c}}, {Zaroubi}, {Labropoulos},
  {Bernardi}, {de Bruyn}, \& {Koopmans}}]{jelic10}
{Jeli{\'c}}, V., {Zaroubi}, S., {Labropoulos}, P., {et~al.} 2010, \mnras, 409,
  1647

\bibitem[{{Kazemi} \& {Yatawatta}(2013)}]{kazemi13c}
{Kazemi}, S. \& {Yatawatta}, S. 2013, \mnras, 435, 597

\bibitem[{{Kazemi} {et~al.}(2013{\natexlab{a}}){Kazemi}, {Yatawatta}, \&
  {Zaroubi}}]{kazemi13a}
{Kazemi}, S., {Yatawatta}, S., \& {Zaroubi}, S. 2013{\natexlab{a}}, \mnras,
  430, 1457

\bibitem[{{Kazemi} {et~al.}(2013{\natexlab{b}}){Kazemi}, {Yatawatta}, \&
  {Zaroubi}}]{kazemi13b}
{Kazemi}, S., {Yatawatta}, S., \& {Zaroubi}, S. 2013{\natexlab{b}}, \mnras,
  434, 3130

\bibitem[{{Kazemi} {et~al.}(2011){Kazemi}, {Yatawatta}, {Zaroubi},
  {Lampropoulos}, {de Bruyn}, {Koopmans}, \& {Noordam}}]{kazemi11}
{Kazemi}, S., {Yatawatta}, S., {Zaroubi}, S., {et~al.} 2011, \mnras, 414, 1656

\bibitem[{{Manchester}(1972)}]{manchester72}
{Manchester}, R.~N. 1972, \apj, 172, 43

\bibitem[{{Mao} {et~al.}(2010){Mao}, {Gaensler}, {Haverkorn}, {Zweibel},
  {Madsen}, {McClure-Griffiths}, {Shukurov}, \& {Kronberg}}]{mao10}
{Mao}, S.~A., {Gaensler}, B.~M., {Haverkorn}, M., {et~al.} 2010, \apj, 714,
  1170

\bibitem[{{Martinez-Rubi} {et~al.}(2013){Martinez-Rubi}, {Veligatla}, {de
  Bruyn}, {Lampropoulos}, {Offringa}, {Jelic}, {Yatawatta}, {Koopmans}, \&
  {Zaroubi}}]{martinezrubi13}
{Martinez-Rubi}, O., {Veligatla}, V.~K., {de Bruyn}, A.~G., {et~al.} 2013, in
  Astronomical Society of the Pacific Conference Series, Vol. 475, Astronomical
  Data Analysis Software and Systems XXII, ed. D.~N. {Friedel}, 377

\bibitem[{{McCullough} \& {Benjamin}(2001)}]{mccullough01}
{McCullough}, P.~R. \& {Benjamin}, R.~A. 2001, \aj, 122, 1500

\bibitem[{{McKee} \& {Ostriker}(1977)}]{mckee77}
{McKee}, C.~F. \& {Ostriker}, J.~P. 1977, \apj, 218, 148

\bibitem[{{Offringa} {et~al.}(2010){Offringa}, {de Bruyn}, {Biehl}, {Zaroubi},
  {Bernardi}, \& {Pandey}}]{offringa10}
{Offringa}, A.~R., {de Bruyn}, A.~G., {Biehl}, M., {et~al.} 2010, \mnras, 405,
  155

\bibitem[{{Offringa} {et~al.}(2013){Offringa}, {de Bruyn}, {Zaroubi}, {van
  Diepen}, {Martinez-Ruby}, {Labropoulos}, {Brentjens}, {Ciardi}, {Daiboo},
  {Harker}, {Jeli{\'c}}, {Kazemi}, {Koopmans}, {Mellema}, {Pandey}, {Pizzo},
  {Schaye}, {Vedantham}, {Veligatla}, {Wijnholds}, {Yatawatta}, {Zarka},
  {Alexov}, {Anderson}, {Asgekar}, {Avruch}, {Beck}, {Bell}, {Bell}, {Bentum},
  {Bernardi}, {Best}, {Birzan}, {Bonafede}, {Breitling}, {Broderick},
  {Br{\"u}ggen}, {Butcher}, {Conway}, {de Vos}, {Dettmar}, {Eisloeffel},
  {Falcke}, {Fender}, {Frieswijk}, {Gerbers}, {Griessmeier}, {Gunst},
  {Hassall}, {Heald}, {Hessels}, {Hoeft}, {Horneffer}, {Karastergiou},
  {Kondratiev}, {Koopman}, {Kuniyoshi}, {Kuper}, {Maat}, {Mann}, {McKean},
  {Meulman}, {Mevius}, {Mol}, {Nijboer}, {Noordam}, {Norden}, {Paas}, {Pandey},
  {Pizzo}, {Polatidis}, {Rafferty}, {Rawlings}, {Reich}, {R{\"o}ttgering},
  {Schoenmakers}, {Sluman}, {Smirnov}, {Sobey}, {Stappers}, {Steinmetz},
  {Swinbank}, {Tagger}, {Tang}, {Tasse}, {van Ardenne}, {van Cappellen}, {van
  Duin}, {van Haarlem}, {van Leeuwen}, {van Weeren}, {Vermeulen}, {Vocks},
  {Wijers}, {Wise}, \& {Wucknitz}}]{offringa13}
{Offringa}, A.~R., {de Bruyn}, A.~G., {Zaroubi}, S., {et~al.} 2013, \aap, 549,
  A11

\bibitem[{{Offringa} {et~al.}(2012){Offringa}, {van de Gronde}, \&
  {Roerdink}}]{offringa12}
{Offringa}, A.~R., {van de Gronde}, J.~J., \& {Roerdink}, J.~B.~T.~M. 2012,
  \aap, 539, A95

\bibitem[{{Pandey} {et~al.}(2009){Pandey}, {van Zwieten}, {de Bruyn}, \&
  {Nijboer}}]{pandey09}
{Pandey}, V.~N., {van Zwieten}, J.~E., {de Bruyn}, A.~G., \& {Nijboer}, R.
  2009, in Astronomical Society of the Pacific Conference Series, Vol. 407, The
  Low-Frequency Radio Universe, ed. D.~J. {Saikia}, D.~A. {Green}, Y.~{Gupta},
  \& T.~{Venturi}, 384

\bibitem[{{Pen} {et~al.}(2009){Pen}, {Chang}, {Hirata}, {Peterson}, {Roy},
  {Gupta}, {Odegova}, \& {Sigurdson}}]{pen09}
{Pen}, U.-L., {Chang}, T.-C., {Hirata}, C.~M., {et~al.} 2009, \mnras, 399, 181

\bibitem[{{Pizzo} {et~al.}(2011){Pizzo}, {de Bruyn}, {Bernardi}, \&
  {Brentjens}}]{pizzo11}
{Pizzo}, R.~F., {de Bruyn}, A.~G., {Bernardi}, G., \& {Brentjens}, M.~A. 2011,
  \aap, 525, A104

\bibitem[{{Planck Collaboration} {et~al.}(2014){Planck Collaboration},
  {Abergel}, {Ade}, {Aghanim}, {Alves}, {Aniano}, {Arnaud}, {Ashdown},
  {Aumont}, {Baccigalupi}, {Banday}, {Barreiro}, {Bartlett}, {Battaner},
  {Benabed}, {Benoit-L{\'e}vy}, {Bernard}, {Bersanelli}, {Bielewicz}, {Bobin},
  {Bonaldi}, {Bond}, {Bouchet}, {Boulanger}, {Burigana}, {Cardoso}, {Catalano},
  {Chamballu}, {Chiang}, {Christensen}, {Clements}, {Colombi}, {Colombo},
  {Couchot}, {Crill}, {Cuttaia}, {Danese}, {Davis}, {de Bernardis}, {de Rosa},
  {de Zotti}, {Delabrouille}, {D{\'e}sert}, {Dickinson}, {Diego}, {Dole},
  {Donzelli}, {Dor{\'e}}, {Douspis}, {Dupac}, {Efstathiou}, {En{\ss}lin},
  {Eriksen}, {Falgarone}, {Finelli}, {Forni}, {Frailis}, {Franceschi},
  {Galeotta}, {Ganga}, {Ghosh}, {Giard}, {Giraud-H{\'e}raud},
  {Gonz{\'a}lez-Nuevo}, {G{\'o}rski}, {Gregorio}, {Gruppuso}, {Guillet},
  {Hansen}, {Harrison}, {Helou}, {Henrot-Versill{\'e}},
  {Hern{\'a}ndez-Monteagudo}, {Herranz}, {Hildebrandt}, {Hivon}, {Hobson},
  {Holmes}, {Hornstrup}, {Hovest}, {Huffenberger}, {Jaffe}, {Jaffe}, {Joncas},
  {Jones}, {Jones}, {Juvela}, {Kalberla}, {Keih{\"a}nen}, {Kerp}, {Keskitalo},
  {Kisner}, {Kneissl}, {Knoche}, {Kunz}, {Kurki-Suonio}, {Lagache},
  {L{\"a}hteenm{\"a}ki}, {Lamarre}, {Lasenby}, {Lawrence}, {Leonardi},
  {Levrier}, {Liguori}, {Lilje}, {Linden-V{\o}rnle}, {L{\'o}pez-Caniego},
  {Lubin}, {Mac{\'{\i}}as-P{\'e}rez}, {Maffei}, {Maino}, {Mandolesi}, {Maris},
  {Marshall}, {Martin}, {Mart{\'{\i}}nez-Gonz{\'a}lez}, {Masi}, {Massardi},
  {Matarrese}, {Mazzotta}, {Melchiorri}, {Mendes}, {Mennella}, {Migliaccio},
  {Mitra}, {Miville-Desch{\^e}nes}, {Moneti}, {Montier}, {Morgante},
  {Mortlock}, {Munshi}, {Murphy}, {Naselsky}, {Nati}, {Natoli}, {Noviello},
  {Novikov}, {Novikov}, {Oxborrow}, {Pagano}, {Pajot}, {Paoletti}, {Pasian},
  {Perdereau}, {Perotto}, {Perrotta}, {Piacentini}, {Piat}, {Pierpaoli},
  {Pietrobon}, {Plaszczynski}, {Pointecouteau}, {Polenta}, {Ponthieu}, {Popa},
  {Pratt}, {Prunet}, {Puget}, {Rachen}, {Reach}, {Rebolo}, {Reinecke},
  {Remazeilles}, {Renault}, {Ricciardi}, {Riller}, {Ristorcelli}, {Rocha},
  {Rosset}, {Roudier}, {Rusholme}, {Sandri}, {Savini}, {Spencer}, {Starck},
  {Sureau}, {Sutton}, {Suur-Uski}, {Sygnet}, {Tauber}, {Terenzi}, {Toffolatti},
  {Tomasi}, {Tristram}, {Tucci}, {Umana}, {Valenziano}, {Valiviita}, {Van
  Tent}, {Verstraete}, {Vielva}, {Villa}, {Wade}, {Wandelt}, {Winkel}, {Yvon},
  {Zacchei}, \& {Zonca}}]{planckXVII}
{Planck Collaboration}, {Abergel}, A., {Ade}, P.~A.~R., {et~al.} 2014, \aap,
  566, A55

\bibitem[{{Reich}(2006)}]{reich06}
{Reich}, W. 2006, ArXiv Astrophysics e-prints [\eprint{astro-ph/0603465}]

\bibitem[{{Reynolds}(1991)}]{reynolds91}
{Reynolds}, R.~J. 1991, \apjl, 372, L17

\bibitem[{{Reynolds} {et~al.}(1999){Reynolds}, {Haffner}, \&
  {Tufte}}]{reynolds99}
{Reynolds}, R.~J., {Haffner}, L.~M., \& {Tufte}, S.~L. 1999, \apjl, 525, L21

\bibitem[{{Scaife} \& {Heald}(2012)}]{scaife12}
{Scaife}, A.~M.~M. \& {Heald}, G.~H. 2012, \mnras, 423, L30

\bibitem[{{Schnitzeler}(2010)}]{schnitzeler10}
{Schnitzeler}, D.~H.~F.~M. 2010, \mnras, 409, L99

\bibitem[{{Schnitzeler}(2012)}]{schnitzeler12}
{Schnitzeler}, D.~H.~F.~M. 2012, \mnras, 427, 664

\bibitem[{{Schnitzeler} {et~al.}(2009){Schnitzeler}, {Katgert}, \& {de
  Bruyn}}]{schnitzeler09}
{Schnitzeler}, D.~H.~F.~M., {Katgert}, P., \& {de Bruyn}, A.~G. 2009, \aap,
  494, 611

\bibitem[{{Stanimirovic}(2002)}]{stanimirovic02}
{Stanimirovic}, S. 2002, in Astronomical Society of the Pacific Conference
  Series, Vol. 278, Single-Dish Radio Astronomy: Techniques and Applications,
  ed. S.~{Stanimirovic}, D.~{Altschuler}, P.~{Goldsmith}, \& C.~{Salter},
  375--396

\bibitem[{{Subrahmanyan} \& {Cowsik}(2013)}]{subrahmanyan13}
{Subrahmanyan}, R. \& {Cowsik}, R. 2013, \apj, 776, 42

\bibitem[{{Sun} {et~al.}(2008){Sun}, {Reich}, {Waelkens}, \&
  {En{\ss}lin}}]{sun08}
{Sun}, X.~H., {Reich}, W., {Waelkens}, A., \& {En{\ss}lin}, T.~A. 2008, \aap,
  477, 573

\bibitem[{{Tasse} {et~al.}(2013){Tasse}, {van der Tol}, {van Zwieten}, {van
  Diepen}, \& {Bhatnagar}}]{tasse13}
{Tasse}, C., {van der Tol}, S., {van Zwieten}, J., {van Diepen}, G., \&
  {Bhatnagar}, S. 2013, \aap, 553, A105

\bibitem[{{Tucci} {et~al.}(2002){Tucci}, {Carretti}, {Cecchini}, {Nicastro},
  {Fabbri}, {Gaensler}, {Dickey}, \& {McClure-Griffiths}}]{tucci02}
{Tucci}, M., {Carretti}, E., {Cecchini}, S., {et~al.} 2002, \apj, 579, 607

\bibitem[{{Uyaniker} {et~al.}(1998){Uyaniker}, {Fuerst}, {Reich}, {Reich}, \&
  {Wielebinski}}]{uyaniker98}
{Uyaniker}, B., {Fuerst}, E., {Reich}, W., {Reich}, P., \& {Wielebinski}, R.
  1998, \aaps, 132, 401

\bibitem[{{van Haarlem} {et~al.}(2013){van Haarlem}, {Wise}, {Gunst}, {Heald},
  {McKean}, {Hessels}, {de Bruyn}, {Nijboer}, {Swinbank}, {Fallows},
  {Brentjens}, {Nelles}, {Beck}, {Falcke}, {Fender}, {H{\"o}randel},
  {Koopmans}, {Mann}, {Miley}, {R{\"o}ttgering}, {Stappers}, {Wijers},
  {Zaroubi}, {van den Akker}, {Alexov}, {Anderson}, {Anderson}, {van Ardenne},
  {Arts}, {Asgekar}, {Avruch}, {Batejat}, {B{\"a}hren}, {Bell}, {Bell}, {van
  Bemmel}, {Bennema}, {Bentum}, {Bernardi}, {Best}, {B{\^i}rzan}, {Bonafede},
  {Boonstra}, {Braun}, {Bregman}, {Breitling}, {van de Brink}, {Broderick},
  {Broekema}, {Brouw}, {Br{\"u}ggen}, {Butcher}, {van Cappellen}, {Ciardi},
  {Coenen}, {Conway}, {Coolen}, {Corstanje}, {Damstra}, {Davies}, {Deller},
  {Dettmar}, {van Diepen}, {Dijkstra}, {Donker}, {Doorduin}, {Dromer}, {Drost},
  {van Duin}, {Eisl{\"o}ffel}, {van Enst}, {Ferrari}, {Frieswijk}, {Gankema},
  {Garrett}, {de Gasparin}, {Gerbers}, {de Geus}, {Grie{\ss}meier}, {Grit},
  {Gruppen}, {Hamaker}, {Hassall}, {Hoeft}, {Holties}, {Horneffer}, {van der
  Horst}, {van Houwelingen}, {Huijgen}, {Iacobelli}, {Intema}, {Jackson},
  {Jelic}, {de Jong}, {Kant}, {Karastergiou}, {Koers}, {Kollen}, {Kondratiev},
  {Kooistra}, {Koopman}, {Koster}, {Kuniyoshi}, {Kramer}, {Kuper},
  {Lambropoulos}, {Law}, {van Leeuwen}, {Lemaitre}, {Loose}, {Maat}, {Macario},
  {Markoff}, {Masters}, {McFadden}, {McKay-Bukowski}, {Meijering}, {Meulman},
  {Mevius}, {Millenaar}, {Miller-Jones}, {Mohan}, {Mol}, {Morawietz},
  {Morganti}, {Mulcahy}, {Mulder}, {Munk}, {Nieuwenhuis}, {van Nieuwpoort},
  {Noordam}, {Norden}, {Noutsos}, {Offringa}, {Olofsson}, {Omar}, {Orr{\'u}},
  {Overeem}, {Paas}, {Pandey-Pommier}, {Pandey}, {Pizzo}, {Polatidis},
  {Rafferty}, {Rawlings}, {Reich}, {de Reijer}, {Reitsma}, {Renting},
  {Riemers}, {Rol}, {Romein}, {Roosjen}, {Ruiter}, {Scaife}, {van der Schaaf},
  {Scheers}, {Schellart}, {Schoenmakers}, {Schoonderbeek}, {Serylak},
  {Shulevski}, {Sluman}, {Smirnov}, {Sobey}, {Spreeuw}, {Steinmetz}, {Sterks},
  {Stiepel}, {Stuurwold}, {Tagger}, {Tang}, {Tasse}, {Thomas}, {Thoudam},
  {Toribio}, {van der Tol}, {Usov}, {van Veelen}, {van der Veen}, {ter Veen},
  {Verbiest}, {Vermeulen}, {Vermaas}, {Vocks}, {Vogt}, {de Vos}, {van der Wal},
  {van Weeren}, {Weggemans}, {Weltevrede}, {White}, {Wijnholds}, {Wilhelmsson},
  {Wucknitz}, {Yatawatta}, {Zarka}, {Zensus}, \& {van Zwieten}}]{haarlem13}
{van Haarlem}, M.~P., {Wise}, M.~W., {Gunst}, A.~W., {et~al.} 2013, \aap, 556,
  A2

\bibitem[{{White} {et~al.}(1999){White}, {Carlstrom}, {Dragovan}, \&
  {Holzapfel}}]{white99}
{White}, M., {Carlstrom}, J.~E., {Dragovan}, M., \& {Holzapfel}, W.~L. 1999,
  \apj, 514, 12

\bibitem[{{Wieringa} {et~al.}(1993){Wieringa}, {de Bruyn}, {Jansen}, {Brouw},
  \& {Katgert}}]{wieringa93}
{Wieringa}, M.~H., {de Bruyn}, A.~G., {Jansen}, D., {Brouw}, W.~N., \&
  {Katgert}, P. 1993, \aap, 268, 215

\bibitem[{{Yatawatta}(2014)}]{yatawatta14}
{Yatawatta}, S. 2014, \mnras, 444, 790

\bibitem[{{Yatawatta} {et~al.}(2008){Yatawatta}, {Zaroubi}, {de Bruyn},
  {Koopmans}, \& {Noordam}}]{yatawatta08}
{Yatawatta}, S., {Zaroubi}, S., {de Bruyn}, G., {Koopmans}, L., \& {Noordam},
  J. 2008, ArXiv e-prints [\eprint[arXiv]{0810.5751}]

\bibitem[{{Zaroubi} {et~al.}(2015){Zaroubi}, {Jeli{\'c}}, {de Bruyn},
  {Boulanger}, {Bracco}, {Kooistra}, {Alves}, {Brentjens}, {Ferri{\`e}re},
  {Ghosh}, {Koopmans}, {Levrier}, {Miville-Desch{\^e}nes}, {Montier}, {Pandey},
  \& {Soler}}]{zaroubi15}
{Zaroubi}, S., {Jeli{\'c}}, V., {de Bruyn}, A.~G., {et~al.} 2015, ArXiv
  e-prints [\eprint[arXiv]{1508.06652}]

\end{thebibliography}
\end{document}